%% file: paper.tex
\documentclass[acmtog,authorversion]{acmart} %
\acmSubmissionID{389}
\usepackage{booktabs} %
\usepackage[ruled]{algorithm2e} %
\usepackage{nicefrac}

\usepackage[normalem]{ulem}  %
\usepackage[utf8]{inputenc}
\usepackage{multirow}
\usepackage{enumerate}
\SetAlFnt{\small}
\SetAlCapFnt{\small}
\SetAlCapNameFnt{\small}
\SetAlCapHSkip{0pt}

\newcommand{\confnotice}{}

\setcopyright{acmcopyright}\acmJournal{TOG}
\acmYear{2020}\acmVolume{39}\acmNumber{6}\acmArticle{194}\acmMonth{12} \acmDOI{10.1145/3414685.3417861}

\newcommand{\res}[1]{\mbox{#1\hspace*{0.1em}$\times$\hspace*{0.1em}#1}}
\newcommand{\ress}[2]{\mbox{#1\hspace*{0.1em}$\times$\hspace*{0.1em}#2}}
\newcommand{\dL}{\partial{L}}
\newcommand{\diff}[1]{\partial\{{#1}\}}
\newcommand{\mymatrix}[1]{\left[{#1}\right]}
\newcommand{\noraise}[1]{\raisebox{0mm}[0mm][0mm]{#1}}
\DeclareMathOperator*{\filterop}{filter}
\DeclareMathOperator*{\shadeop}{shade}
\DeclareMathOperator*{\lightsop}{lights}

\newcommand{\Gparams}{\theta_G}
\newcommand{\Mparams}{\theta_M}
\newcommand{\Cparams}{\theta_C}
\newcommand{\Lparams}{\theta_L}

\usepackage{array}
\newcommand{\PreserveBackslash}[1]{\let\temp=\\#1\let\\=\temp}
\newcolumntype{C}[1]{>{\PreserveBackslash\centering}p{#1}}
\newcolumntype{R}[1]{>{\PreserveBackslash\raggedleft}p{#1}}
\newcolumntype{L}[1]{>{\PreserveBackslash\raggedright}p{#1}}
\newcommand{\ID}{\mathit{ID}}

\def\clap#1{\hbox to 0pt{\hss #1\hss}}%
\definecolor{olive}{rgb}{0.5, 0.5, 0.0}
\definecolor{maroon}{rgb}{0.69, 0.19, 0.38}
\definecolor{celestialblue}{rgb}{0.29, 0.59, 0.82}
\definecolor{darkgreen}{rgb}{0.0, 0.5, 0.0}
\definecolor{grey}{rgb}{0.5,0.5,0.5}

\newcommand{\rawurl}{github.com/NVlabs/nvdiffrast}

\newcommand{\codeurl}{\url{https://\rawurl}}

\usepackage{pifont}%
\usepackage{tabu}
\usepackage{colortbl}
\usepackage{relsize}

\begin{document}%
\input{figures}

\title{Modular Primitives for High-Performance Differentiable Rendering}

\author{Samuli Laine}\affiliation{\institution{NVIDIA}}\email{slaine@nvidia.com}
\author{Janne Hellsten}\affiliation{\institution{NVIDIA}}\email{jhellsten@nvidia.com}
\author{Tero Karras}\affiliation{\institution{NVIDIA}}\email{tkarras@nvidia.com}
\author{Yeongho Seol}\affiliation{\institution{NVIDIA}}\email{yseol@nvidia.com}
\author{Jaakko Lehtinen}\affiliation{\institution{NVIDIA}}\affiliation{\institution{Aalto University}}\email{jlehtinen@nvidia.com}
\author{Timo Aila}\affiliation{\institution{NVIDIA}}\email{taila@nvidia.com}

\begin{abstract}

We present a modular differentiable renderer design that yields performance superior to previous methods by leveraging existing, highly optimized hardware graphics pipelines.
Our design supports all crucial operations in a modern graphics pipeline: rasterizing large numbers of triangles, attribute interpolation, filtered texture lookups, as well as user-programmable shading and geometry processing, all in high resolutions. Our modular primitives allow custom, high-performance graphics pipelines to be built directly within automatic differentiation frameworks such as PyTorch or TensorFlow.
As a motivating application, we formulate facial performance capture as an inverse rendering problem and show that it can be solved efficiently using our tools.
Our results indicate that this simple and straightforward approach achieves excellent geometric correspondence between rendered results and reference imagery.

\end{abstract}



\maketitle
\confnotice

\section{Introduction}
\label{sec:intro}

Differentiable rendering is a fundamental building block in machine learning of 3D geometry.
Typically training data is available only as images, and finding a corresponding 3D representation requires \emph{analysis by synthesis}, i.e., rendering candidate images, computing the loss based on training and candidate images, and propagating the errors back to 3D positions and other scene attributes.
Many classical computer graphics and vision problems including the estimation of reflectance, geometry, lighting, and camera parameters can be cast into this \emph{inverse rendering}  framework \cite{Patow2003}.

Much of modern machine learning makes use of first order (gra\-di\-ent-based) optimization techniques implemented using  backpropagation. From a computational point of view, the explosive growth in model sizes and capabilities has, for a large part, relied on the availability of primitive operations that allow massively parallel and coherent execution of both the forward and backward (gradient) passes, allowing highly complex computation graphs to be built out of them.
However, the majority of effort in creating efficient primitive operations has focused on operating on densely-sampled data stored in multidimensional regular grids.
These kind of operations alone are not sufficient for 3D rendering because the mapping from a scene representation to pixel values is highly irregular and dynamic.
As such, specialized differentiable rendering algorithms have emerged for %
solving two problems:
\begin{enumerate}
\item \emph{Forward pass:} Given the factors that affect the shape and appearance of a 3D scene, render a 2D image; and
\item \emph{Backward pass:} Given the gradient of a loss function defined on the output image pixels, compute the gradient of the loss with respect to the input shape and appearance factors.
\end{enumerate}
This interface is universally used by autodifferentiation frameworks such as PyTorch and TensorFlow, and it allows 3D rendering to be used as a building block in a complex model that is trained by modern stochastic first order optimization techniques. %

Despite its long history, differentiable rendering can be considered a nascent field due to the recent proliferation of algorithms and applications. Most previous research is targeted towards a specific use case (e.g., pose or shape estimation), and is typically only evaluated on downstream tasks as part of a larger machine learning system \cite{softras,dibr,nmr}. These specific use cases and data sets allow optimizations and design choices that do not scale to other uses. For example, low geometric complexity may make it acceptable to not parallelize over triangles, but this quickly backfires on a larger scene; single objects viewed in a vacuum may enable one to disregard the effects of mutual occlusion between triangles when computing gradients, but this approximation is untenable in a scene with non-trivial depth complexity.

Another parallel line of research studies differentiable physically-based light transport simulation that models complex effects such as area light sources and indirect illumination \cite{redner,mitsuba2,Loubet2019}. Built on Monte Carlo sampling, these methods seek the best attainable image quality and accuracy at the cost of longer rendering times.

We build on the rich literature on real-time graphics systems that has long sought efficient solutions for managing the complex, dynamic mapping between world points and image pixels, and has delivered extremely efficient and practical hardware implementations. In particular, we seek to formulate and implement a differentiable rendering system that makes use of these pipelines to maximal extent, without sacrificing their desirable properties such as correct outputs, a high degree of user control through programmable shading and geometry processing, massive parallelization in all operations, and the ability to render high-resolution images of scenes consisting of millions of geometric primitives. %

Concretely, we describe a differentiable rendering system based on deferred shading \cite{deering}, and identify four primitive operations for which we provide custom, high-performance implementations: rasterization, attribute interpolation, texture filtering, and antialiasing. Our modular primitives enable rendering high-resolution images of complex scenes, using arbitrary user-specified shading, directly inside automatic differentiation (AD) frameworks such as TensorFlow or PyTorch.

As a motivating example, we cast facial performance capture as an inverse rendering problem and show that it can be efficiently solved using direct photometric optimization of shape and surface texture in megapixel resolutions. 
While our proof-of-concept solution does not aim to reconstruct the mouth, eyes, or complex material appearance, the high accuracy of the results in comparison to a state-of-the-art commercial solution demonstrates 
the viability of high-performance differentiable rendering in solving this problem.  %

Our differential rasterization primitives are publicly available at \mbox{\codeurl}.

\tblcaps %

\section{Related Work}
\label{sec:prevwork}

There is a large body of work on using rendering as part of an optimization process that infers properties of the world from images. These include inverse rendering algorithms that fit 3D and appearance models to photographs in an analysis-by-synthesis loop \cite{Patow2003}, as well as techniques that use a 3D renderer as part of a more complex machine learning model. %

By far the most common approach in previous work is to design a special-purpose differentiable image synthesis pipeline focusing on the particular requirements of the downstream task, with no particular emphasis on flexibility or generality \cite{softras,dibr,nmr}. A notable exception is OpenDR \cite{opendr} that, despite its limited shading model, explicitly sets out to develop a general-purpose differentiable rendering system.

Research on general-purpose differentiable rendering divides into two categories depending on whether the primary motivation is image quality or performance.
Table~\ref{tbl:caps} summarizes various characteristics of the previous methods analyzed below.

Li et al.~\shortcite{redner} introduce differentiable, physically-based rendering using Monte Carlo ray tracing with proper visibility gradients.
Light transport is integrated using random sampling, which leads to noise in the images that diminishes with more sampling.
Mitsuba~2 \cite{mitsuba2,Loubet2019} is a versatile rendering framework that can target a wide array of rendering problems.
The primary problem in using these renderers as parts of an intensive optimization or learning task is their lack of performance\,---\,%
to not become a bottleneck over millions of iterations, the rendering times in sufficiently high resolutions should be measured in milliseconds, instead of seconds or minutes needed for recursive light transport simulation.
These systems can also be configured to compute only primary visibility and local shading, which is sufficient for many applications. 
This boosts the performance of Li et al. \shortcite{redner} to $\sim$100ms in a simple scene in \ress{640}{480} resolution, which is still orders of magnitude too slow for many applications.

The second category of differentiable rendering aims at higher performance.
Primarily targeted at solving tasks such as shape or pose inference \cite{softras,dibr,handpose,opendr}, they render and shade 3D meshes using local shading only.
Strong emphasis is placed on obtaining useful visibility gradients to facilitate shape inference via gradient descent.

Soft Rasterizer \cite{softras} rasterizes each triangle as a probabilistic cloud with a configurable blur radius;
these clouds are combined heuristically based on other configurable parameters.
This blur makes coverage a continuous function of vertex positions, which is necessary for obtaining visibility gradients.
However, the blur also means that opaque surfaces become transparent around edges, leading to an incorrect image.
Optimization thus requires tuning these parameters to reach a balance between image correctness and gradient quality.
DIB-R \cite{dibr} renders colors without antialiasing and outputs an additional alpha channel that extends outside the covered pixels by a configurable blur radius.
The alpha channel can be used for approximating visibility gradients, but only if an alpha mask is available for reference images as well.
Also, these gradients are affected by all triangles regardless of occlusion.
Because color channels are point sampled, no visibility gradients are obtained for silhouettes that are in front of other geometry.
This is insufficient for, e.g., determining hand poses \cite{handpose}, and would also fail if one were to render, e.g., a skybox behind the mesh, so the method cannot be considered general-purpose.
As with Soft Rasterizer, it is not obvious how the blur parameter should be set.
Neural Mesh Rendering \cite{nmr} produces the image using point sampling and no antialiasing, and in the backward pass hallucinates image-based gradients on triangle edges based on the geometry.
The gradients are thus not consistent with the rendered image.

In contrast to these approximations, our aim is to differentiate the standard hardware graphics pipeline without altering its image formation principles. This places special emphasis on occlusion by opaque surfaces. For the gradient to be consistent with the forward imaging model, a 3D primitive that has no effect on the image\,---\,for instance, due to being off the screen or occluded by other primitives\,---\,should, by our premise, receive a zero gradient.

The reparameterization technique of Loubet et al.~\shortcite{Loubet2019} used in Mitsuba~2 \cite{mitsuba2} produces correct visibility gradients only when occluders and occludees can be inferred from four samples.
Although more complex occlusion scenarios are not handled correctly, Loubet et al. introduce a bandwidth parameter that can be adjusted to reduce the errors at the cost of increased noise.
It is argued that handling the most common, simple cases is sufficient for practical purposes, and our approach for visiblity gradients is founded on the same premise.
OpenDR \cite{opendr} approximates \emph{all} gradients based on the final image and knowledge of which triangle was rendered into each pixel.
This has the unfortunate effect of producing incorrect gradients for effects such as highlights, because all shading is assumed to be ``glued'' onto the surfaces.
As an example, OpenDR's gradients falsely indicate that moving a planar surface tangentially would move the highlights and reflections on it as well.
In addition, inferring gradients from the final pixels can be seen as equivalent to taking finite differences instead of analytic gradients.
Flexibility is limited because textures can modulate appearance only multiplicatively, and differentiation with respect to textures is not supported.

\figpipeline %

\section{Differentiable Rendering Primitives}
\label{sec:raster}

Given a 3D scene description in the form of geometric shapes, materials, and camera and lighting models, rendering 2D images boils down to two computational problems: figuring out the things that are visible in each pixel, and what color those things appear to be. %
A proper differentiable renderer has to provide gradients for all the parameters\,---\,e.g., lighting and material parameters, as well as the contents of texture maps\,---\,used in the process.

For what follows, it is useful to break the rendering process down into the following form, where the final color $I_i$ of the pixel at screen coordinates $(x_i, y_i)$ is given by
\begin{equation}
I_i = \filterop_{x,y} \bigg( \shadeop \Big( M \big( P(x,y) \big), \lightsop \Big) \bigg) \big( x_i, y_i \big).
\label{eq:rendering}
\end{equation}
Here, $P(x, y)$ denotes the world point visible at (continuous) screen coordinates $(x, y)$ after projection from 3D to 2D, and $M(P)$ denotes all the spatially-varying factors (texture maps, normal vectors, etc.) that live on the surfaces of the scene. The shade function typically models light-surface interactions. The 2D antialiasing filter, crucial for both image quality and differentiability, is applied to the shading results in continuous $(x, y)$, and the final color is obtained by sampling the result at the pixel center $(x_i, y_i)$. In real-time graphics, these steps are typically approximated by techniques like multisample antialiasing (MSAA).

The geometry, projection, and lights can all be considered as parametric functions.
The visible world point is affected by the geometry, parameterized by $\Gparams$, as well as the projection, parameterized by $\Cparams$. Similarly, the surface factors are parameterized by $\Mparams$, and light sources by $\Lparams$.\footnote{In the simplest case, $\Gparams$ and $\Mparams$, could describe, say, the vertex coordinates of a triangle mesh of a fixed topology and a diffuse albedo stored at the vertices and interpolated into the interiors of triangles; we use the abstract notation to allow for complex parameterizations fed into the renderer from within a deep learning model.} We follow the common view and take differentiable rendering to mean computing the gradients $\partial L(I)/\partial \{\Gparams, \Mparams, \Cparams, \Lparams\}$ of a scalar function $L(I)$ of the rendered image $I$ with respect to the scene parameters. Note that this does not require computing the (very large) Jacobian matrices $[\partial I/\partial \Gparams]$, etc., but rather only the ability to implement multiplication with the Jacobian transpose (``backpropagation''), yielding the final result through the chain rule:
\begin{equation*}
\left[\frac{\partial L(I)}{\partial \Gparams}\right] = \left[\frac{\partial I}{\partial \Gparams}\right] \left[\frac{\partial L}{\partial I}\right],
\end{equation*}
and similarly for the other parameter vectors.

Two main factors make the design of efficient rendering algorithms challenging. First, the mapping $P(x, y)$ between world points and screen coordinates is dynamic: it is affected by changes in both scene geometry and the 3D-to-2D projection. Furthermore, it is discontinuous due to occlusion boundaries. These two factors are also central points of difficulty in computing the gradients we seek. The following sections outline our approach to addressing them.

\subsection{Design Goals}
Our overall aim is to implement an efficient differentiable real-time graphics pipeline, with the following specific design goals:
\begin{enumerate}[G1]
\item \textbf{Efficiency}. Support modern graphics pipelines' ability to render, in high resolution, 3D scenes that are complex in terms of geometric detail, occlusion, and appearance.\label{goal:efficiency}
\item \textbf{Minimalism}. Easy integration with modern automatic differentiation (AD) frameworks, such as PyTorch and Tensorflow, without duplication of features.\label{goal:minimalism}
\item \textbf{Freedom}. Support arbitrary user-specified shading, as well as arbitrary parameterizations of input geometry, without committing to specific forms such as the Phong model or blendshapes.\label{goal:freedom}
\item \textbf{Quality}. Support the texture filtering operations required by shaders that implement complex appearance models, while making no assumptions about the contents of the textures. In addition to quality, this is also important for optimization dynamics.\label{goal:quality}
\end{enumerate}

\subsection{System Design}
Goal G\ref{goal:efficiency} immediately precludes algorithms that do not parallelize over both the geometric primitives and pixels, and those that do not properly account for occlusion of overlapping primitives. In the remaining space, we make the following design choices:
\begin{enumerate}[C1]
\item \textbf{Modularity}. We identify four modular primitive operations that implement crucial operations in a graphics pipeline. Each primitive is exposed as a backpropagation-capable operation with a fixed input/output interface to the host AD framework. Much like today's configurable and programmable hardware graphics pipelines, this non-monolithic design enables easy construction of potentially complex custom rendering pipelines.
\item \textbf{Positions and Textures are Tensors}. Our system takes the input geometry and texture maps in the form of tensors from the host AD system. This allows parameterizing both in a freely-chosen manner, and enables our rendering primitives to be used as building blocks of a complex learning system.
\item \textbf{Operate in Clip Space}. Contrary to common differentiable rendering systems, we place it on the user's responsibility to perform world, view, and homogeneous perspective transformations --- but not perspective division --- on the geometry using the host AD system. By this, we follow the separation between geometry and pixel processing made by all major graphics APIs. We feel this offers the cleanest possible interface between the host AD system and the renderer, further amplifying the benefits of their co-existence.
\item \textbf{Deferred Shading}. We build on the concept of deferred shading \cite{deering}. This entails first computing, for each pixel, the $M(P(x, y))$ terms from Equation~\eqref{eq:rendering} and storing the intermediate results in an image-space regular grid. The grid is subsequently consumed by the shading function. As shading is performed on a regular grid, it can be implemented entirely outside our rendering primitives using the efficient dense tensor operations in the host AD library, in line with G\ref{goal:minimalism} and G\ref{goal:freedom}.
\item \textbf{Image-space Antialiasing}. We approach differentiation of coverage in image space, approximating the inputs of the antialiasing filter in Equation~\eqref{eq:rendering} by the output grid of the deferred shading pass. 
Effectively, we assume shading to be constant with respect to the coverage effects at silhouette boundaries, but not with respect to other effects in appearance.
\item \textbf{Triangles}. We focus on triangle meshes as the basic geometric primitive, and seek to utilize the modern graphics pipelines' immensely optimized rasterization subsystem to maximal extent.
\end{enumerate}

We build pipelines out of the following four primitive operations customized for gradient computation. Figure~\ref{fig:pipeline} illustrates an example graphics pipeline built out of them.

\textbf{Rasterization} implements the dynamic mapping between world coordinates and discrete pixel coordinates. Leveraging the hardware rasterizer, we store per-pixel auxiliary data in the form of barycentric coordinates and triangle IDs in the forward pass. Using barycentrics as a base coordinate system allows easy coupling of shading and interpolation, as well as combining texture gradients with geometry gradients in the backward pass.

\textbf{Interpolation} is a pipeline operation that expands user-defined per-vertex data (i.e., vertex attributes) to pixel space. Making use of the barycentrics computed by the rasterizer, the interpolator module manages this mapping in both directions.

\textbf{Texture filtering} is a key operation in a shading system. Taking as inputs the interpolated texture coordinates and their screen-space derivatives for MIP-mapping, as well as texture data tensors, our texture filtering module performs trilinear MIP-mapping with gradients correctly propagated through both input texture coordinates as well as the contents of the (MIP-mapped) texture maps.

\textbf{Antialiasing} is performed on the output of the deferred shading operation, taking as additional inputs the barycentrics, triangle IDs, and vertex positions and indices.

We now proceed to describe each primitive operation in detail. For simplicity, the following discussion assumes that a single image is being rendered. However, a differentiable renderer is typically used in stochastic gradient descent -type schemes using minibatches of multiple rendered images. All our operations efficiently support minibatching. %

\subsection{Rasterization}

As per widely adopted graphics API standards, our rasterization module consumes triangles with vertex positions given as an array of clip-space homogeneous coordinates $(x_c, y_c, z_c, w_c)$. We leave it as the user's responsibility to compute clip space positions\,---\,often, this comprises only a few homogeneous \res{4} matrix multiplications. The backward pass then computes the gradient $\dL/\partial\{x_c, y_c, z_c, w_c\}$ of the loss $L$ with respect to the clip-space positions, leaving differentiation with respect to any higher-level parameterizations for the host AD library.

\paragraph{Forward pass.}
In the forward pass, the rasterizer outputs a 2D sample grid, with each position storing a tuple $(\ID, u, v, z_c/w_c)$, where $\ID$ identifies the triangle covering the sample, $(u, v)$ are barycentric coordinates specifying relative position along the triangle, and $z/w$ corresponds to the depth in normalized device coordinates (NDC). A special $\ID$ is reserved for blank pixels. Barycentrics serve as a convenient base domain for interpolation and texture mapping computations further down the pipeline. 
The NDC depth is utilized only by the subsequent antialiasing module, and does not propagate gradients. 
As a secondary output, the rasterizer outputs a buffer with the \res{2} Jacobian of the barycentrics w.r.t. the screen coordinates $J_\mathit{uv} = \diff{u, v}/\diff{x, y}$ for each pixel. These are later used for determining the footprint for filtered texture lookups.

Internally, the rasterization is performed through OpenGL, leveraging the hardware graphics pipeline.\footnote{
On compute clusters, we use OpenGL with EGL for displayless hardware-accelerated rendering.%
} Using the hardware graphics pipeline ensures that the rasterization is accurate and there are, e.g., no visibility leaks due to precision issues.
We also automatically get proper view frustum clipping as performed by the hardware.
The output values, including the per-pixel Jacobians between barycentrics and screen coordinates, are calculated using an OpenGL fragment shader.

Both TensorFlow and PyTorch implement GPU tensor operations in CUDA. To bridge them with OpenGL, we use the driver's OpenGL/CUDA interoperability API. The API minimizes data copies, using the same physical memory when possible, and never requires data to leave the GPU memory.

\paragraph{Backward pass.} The backward pass receives, for each pixel, the gradient $\dL/\diff{u, v}$ with respect to the barycentrics output by the rasterizer, and computes the gradients $\dL/\partial\{x_c, y_c, z_c, w_c\}$ for each input vertex. The perspective mapping between barycentrics and clip-space positions is readily differentiated analytically, and the necessary output is obtained through
\begin{equation}
\left[\frac{\dL}{\diff{x_c, y_c, z_c, w_c}}\right] = \left[\frac{\dL}{\diff{u, v}}\right] \left[\frac{\diff{u, v}}{\diff{x_c, y_c, z_c, w_c}}\right]\textrm{.}
\end{equation}
The gradients w.r.t. the screen-space derivatives of the barycentrics ($\dL/\partial J_\mathit{uv}$) are taken into account in a similar fashion.
The backward pass is implemented as a dense operation over output pixels, using a scatter-add operation to accumulate the gradients from the pixels to the correct vertices based on the triangle IDs. It can thus be trivially parallelized using a CUDA kernel.

\subsection{Interpolation}

Attribute interpolation is a standard part of the graphics pipeline. Specifically, it entails computing weighted sums of vertex attributes, with the weights given by the barycentrics, thereby creating a mapping between the pixels and the attributes.\footnote{
Note that with our rendering primitives, one can supply a different index buffer for attribute interpolation than was used for rasterization.
This is convenient when source data comes from a modeler such as Autodesk Maya that associates each vertex with a 3D position and a texture coordinate from separately indexed arrays.
If attribute interpolation were bundled with rasterization, such flexibility would not be possible.}

Generally, vertex attributes can be used for arbitrary purposes. One of their typical uses, however, is to provide 2D coordinates for texture mapping. 
Because of this, our interpolator module supports computing, in the forward pass, screen-space derivatives $J_A = \partial{A}/\diff{x,y}$ of all or a subset of attributes for later use in determining texture filter footprints and other purposes. 

\paragraph{Forward pass.}
Consider a single pixel at $(x, y)$. Denoting a vector of attributes associated with the $i$th vertex by $A_i$, the attribute indices of the triangle visible in the pixel $(x, y)$ by $i_{0, 1, 2}$, and the barycentrics generated by the rasterizer by $u = u(x, y)$ and $v = v(x, y)$, the interpolated vector $A$ is defined as
\begin{equation}
A = u\,A_{i_0} + v\,A_{i_1} + (1-u-v)\,A_{i_2}.
\label{eq:interpolation}
\end{equation}
Given the rasterizer's outputs (per-pixel triangle IDs and barycentrics), implementation of the forward pass is trivial.

The screen-space derivatives for attributes tagged as requiring them are computed using the barycenter Jacobians output by the rasterizer by $\partial{A}/\diff{x, y} = [\diff{u, v}/\diff{x, y}] [\partial{A}/\diff{u, v}]$, where the last Jacobian is simple to derive from Equation~\eqref{eq:interpolation}.

\paragraph{Backward pass.}
The inputs to the backward pass are the per-pixel gradients $\partial L/\partial A$ w.r.t. the interpolated attributes, as well as gradients w.r.t. the screen-space derivatives of the attributes. 
Much like the backward pass of the rasterizer, the gradients w.r.t. the attribute tensor are computed by a scatter-add into the tensor, applying the Jacobians $\partial A/\diff{A_{i_0, i_1, i_2}} = \{u, v, 1-u-v\}$ to the per-pixel input gradients. 
By simple differentiation, the gradients w.r.t. the input barycentrics are given by
\begin{equation}
\mymatrix{\frac{\dL}{\partial{u}}} = \mymatrix{A_{i_0} - A_{i_2}}^{\mathrm{T}} \mymatrix{\frac{\dL}{\partial A}}, \; \mymatrix{\frac{\dL}{\partial{v}}} = \mymatrix{A_{i_1} - A_{i_2}}^{\mathrm{T}} \mymatrix{\frac{\dL}{\partial A}}.
\end{equation}
In the same vein, the gradients w.r.t. the screen-space derivatives of the input barycentrics $\dL/\partial J_\mathit{uv}$ are computed based on the incoming gradients w.r.t. the screen-space derivatives of the attributes $\dL/\partial{J_\mathit{A}}$.

\subsection{Texture Mapping}

We perform texture mapping using trilinear MIP-mapped texture fetches. 
In the general case, this entails picking a fractional MIP-map pyramid level (i.e., level-of-detail, LOD) based on the incoming screen-space derivatives of the attributes used as texture coordinates, and performing a trilinear interpolation from the eight texels on the appropriate MIP pyramid levels. 
Figure~\ref{fig:texture} illustrates our implementation. 
We choose the MIP level based on the texture-space length of the major axis of the sample footprint as defined by the screen-space derivatives of the texture coordinates. 
This is conservative in the sense that grazing angles result in blurring instead of aliasing. 

\figtexture %

Once a pair of MIP-map levels has been picked, operation of the forward and backward passes closely resemble attribute interpolation: 
On each level, the four closest texels take the place of the three triangle vertices, and the two sub-texel coordinates that determine exact position within the four-pixel ensemble take the place of the barycentrics. 
Consequently, our implementation is highly similar, with the forward pass requiring a gather, and the backward pass requiring a scatter-add, with the related Jacobians computed with equal simplicity from the bilinear basis functions and texture contents. 
As the derivations are highly similar, we omit them for space. 
Note, however, that gradients are computed also for the texture coordinate attributes, as well as for the screen-space derivatives for the texture coordinates.

MIP-mapped texturing differs from attribute interpolation by its multiscale nature: gradients are accumulated on various levels of the MIP-map pyramid in the backward pass. 
As all MIP-map levels are obtained from the finest-level texture during the construction in the forward pass, the backward pass needs to finish by transposing the construction operation and flattening the gradient pyramid so that the gradient is specified densely at the finest level. 
Fortunately, this is implemented easily by starting at the coarsest level, recursively up-sampling the result and adding gradients from the next level, precisely like collapsing a Laplacian pyramid.

To the best of our knowledge, no previous differentiable renderer except for Li et al.~\shortcite{redner} has supported differentiable, filtered texture sampling. While we currently do not do so, we note that it would be possible to utilize the hardware texture unit in the forward pass, and retain the CUDA kernel only for the backward pass.
The key challenge is that we cannot be certain of the implementation (e.g., numerical precision) of the hardware texture unit, and thus the gradients might not match the forward pass. This will be especially true for anisotropic texture fetches, where considerable freedom exists in covering the footprint \cite{Schilling1996}. Hardware texture units also require specific memory layouts.

\subsection{Analytic Antialiasing for Visibility Gradients}

As usual in real-time graphics, we expect shading to be band-limited via filtered texture lookups and other means, and thus not exhibit aliasing within surfaces.
However, point-sampled visibility causes aliasing at visibility discontinuities, and more crucially, cannot produce visibility-related gradients for vertex positions.
Antialiasing converts the discontinuities to smooth changes, from which the gradients can be computed \cite{redner}.
Note that antialiasing can only be performed after shading, and therefore must be implemented as a separate stage instead of bundling it into rasterization.%

We follow the same approach as several previous methods \cite{opendr,handpose} and approach the problem by analytic post-process edge antialiasing. Image-based post-process antialiasing is an old and widely-used technique in real-time graphics, with famous techniques such as FXAA being recently superseded by deep learning algorithms \cite{Turing}. %
For an overview, see Jimenez et al.~\shortcite{aacourse}.
Our method is a variant of distance-to-edge anti-aliasing (DEAA) \cite{deaa} and geometric post-process antialiasing (GPAA) \cite{gpaa}.
The main differences are in how visibility discontinuities are detected and attributed to vertex positions, as required for computing gradients.

\figantialias %

\paragraph{Forward pass.}
Figure~\ref{fig:antialias} illustrates our antialiasing method. 
We first detect potential visibility discontinuities by finding all neighboring horizontal and vertical pixel pairs with mismatching triangle IDs.
For each potential discontinuity, we fetch the triangle associated with the surface closer to camera, as determined from the NDC depths computed during rasterization.
We then examine the edges of the triangle to see if any of them are silhouettes%
\footnote{
We consider an edge to be on a silhouette if it has only one connecting triangle, or if it connects two triangles that lie on the same side of it after projection.
This includes silhouettes that have another surface behind them, unlike DIB-R \cite{dibr} that only considers silhouettes against the background.
}
and pass between the neighboring pixel centers.
For horizontal pixel pairs, we consider only vertically oriented edges (\mbox{$|w_{c,1} \cdot y_{c,2} - w_{c,2} \cdot y_{c,1}| > |w_{c,1} \cdot x_{c,2} - w_{c,2} \cdot x_{c,1}|$}), and vice versa.

If a silhouette edge crosses the segment between pixel centers, we compute a blend weight by examining where this crossing happens.
Pixel colors are then adjusted to reflect the approximated coverage of either surface in the pixels.
Essentially this approach approximates the exact surface coverage per pixel \cite{Jalobeanu2004} using an axis-aligned slab. Consequently the coverage estimate is exact for only perfectly vertical and horizontal edges that extend beyond the pixel. For a diagonal long edge that passes exactly between the pixel centers, the error in coverage is $\frac{1}{8}$th of a pixel.

Finely tessellated surfaces reveal two further approximations. Theoretically, the silhouette between two pixel centers can take any poly-line shape, and the axis-aligned slab approximation can be arbitrarily poor. However, typically additional tessellation manifests itself on the pixel-scale as slightly rounded silhouettes, and for these the approximation accuracy is only slightly worse than for long edges, although exact error bounds cannot be given.

A potentially more serious approximation results from the assumption that all triangles that contain silhouette edges overlap pixel centers and are thus stored during rasterization.  
Clearly, if we tessellate a surface enough, it is rare for a triangle with a silhouette edge to get rasterized.
In this situation, some silhouette edges are not found during antialiasing and no visibility gradient is obtained for these pixels.
Occasionally missing a gradient can slow down optimization but rarely prevents it from succeeding, as we show in Section~\ref{sec:analysis}.

\paragraph{Backward pass.}
To prepare for the gradient computation, we store the results of the discontinuity analysis in the forward pass so that we do not have to repeat it in the backward pass.
Gradient computation with the stored data is then easy\,---\,%
for each pixel pair that was antialiased in the forward pass, we determine how both vertex positions influence the blending coefficient, and transfer incoming pixel gradients to vertex positions accordingly using scatter-add operations.

\subsection{Discussion}

Our design aims to do as little as possible apart from managing the complex and dynamic mappings between pixels and the input vertices and textures, leaving the field open for utilizing novel parameterizations for geometry, textures, and lighting models. In particular, the modular design allows the units to be chained many times in a single rendering pipeline.
The deferred shading design also leaves many options open. For example, it is easy to perform texture-space shading instead of shading in the pixel domain: after computing shading results using array operations in texture space,
one would simply look up the surface colors from the texture in the deferred shading pass.

\section{Analysis}
\label{sec:analysis}

In this section, we validate our design principles via targeted, synthetic tests.
We first examine the properties of the visibility gradients resulting from our antialiasing, as well as the effects of filtered texture lookups on texture convergence.
Then, we construct a rendering pipeline with nontrivial shading and use it to demonstrate an optimization task that involves indirect texture lookups.
We also examine a pose fitting task with a difficult optimization landscape.
Finally, we measure the performance of our system and compare it to previous differentiable rasterizers.

\subsection{Visibility Gradients}

To examine the validity of the gradients, we perform a synthetic test where we attempt to infer vertex positions and colors of a simple unit cube.
We initialize the solution by taking the true vertex positions and perturbing them randomly in range $[-\frac{1}{2},\frac{1}{2}]^3$.
The vertex colors are initialized to random RGB values in $[0,1]^3$. 
We then run Adam optimizer \cite{adam} ($\beta_1=0.9$, $\beta_2=0.999$) for 5000 iterations, where in each iteration we render the reference mesh and the optimized mesh from same, random viewpoint, and take the image-space $L_2$ loss between the images.
Based on this loss, we learn both vertex positions and colors simultaneously.
The learning rate was ramped down exponentially from $10^{-2}$ to $10^{-4}$ over the course of the optimization.

We implemented two modes for coloring the vertices. 
In the \emph{continuous} coloring mode, the vertex colors at each corner are shared.
This yields a coloring that is continuous across the surface of the cube and has~8 unique colors to optimize.
In the \emph{discontinuous} coloring mode, each face has four unique colors at the corners, i.e., a total of~24 unique colors.
Consequently, the coloring is not continuous across the edges or vertices of the cube.
It can be expected that the latter mode is more difficult to optimize because of the larger number of unknowns and presumably less smooth gradients due to color discontinuities.

Figure~\ref{fig:cube} illustrates the results in the continuous coloring mode.
To our surprise, the optimization succeeds even at \res{4} resolution, where the average size of a rendered triangle is approximately half a pixel.
This indicates that our antialiasing-based visibility gradients offer enough information even for small triangles such as those seen in finely tessellated meshes.
However, rendering in higher resolution offers faster convergence, highlighting that rendering performance in high resolutions is crucial.

\figcube %
\figcubeplot %

Figure~\ref{fig:cubeplot} shows the average convergence curves for both coloring modes in the three resolutions tested.
Each curve is an average over 10 successful optimization runs. We manually excluded the cases where early optimization steps produced an irrecoverable self-intersecting mesh, which happened in approximately 25\% of runs in \res{4} resolution, and less often in higher resolutions.
This concurs with the observation from Mitsuba~2 \cite{mitsuba2} that some rendering-related components require careful initialization.
We could have made these configurations less likely by lowering the learning rate, 
initializing the mesh to further away from self-intersecting states, 
or by using a suitable regularization term that pushes apart geometry that is in danger of folding over itself.
We chose not to use any regularizers in this test because we explicitly wanted our gradients to be based on image-space loss only.

\subsection{Texture Filtering}
 
\figearth %

To measure the importance of texture filtering via mipmaps, we constructed a test where we attempt to learn a texture based on synthetic, high-quality reference images that exhibit large variations in scale.
We then measure how well the texture is learned with and without mipmapping.

Figure~\ref{fig:earth}a shows example reference images of this task.
The reference images are rendered first in \res{4096} resolution and then downsampled to \res{512} using a high-quality downsampling filter.
The reference images are thus well band-limited and display no aliasing, blurring, or other artifacts.

The goal of the optimization is to learn a cube map -parameterized texture with \res{512} pixel faces, mapped onto a unit sphere, based on the reference images.
We again use Adam \cite{adam} as the optimizer ($\beta_1=0.9$, $\beta_2=0.99$) and run it for 20\,000 iterations, ramping the learning rate from $10^{-2}$ to $10^{-3}$ during the course of optimization.
This learning rate schedule was chosen to be optimal for the case without mipmapping.
The training images are rendered directly in \res{512} resolution, from the same, random viewpoints as the reference images, and the optimizer attempts to minimize $L_2$ loss between training and reference images.
The same mesh and texture parameterization are used for all images.

Learning the texture is made difficult by the sphere being placed randomly at distance $[1.5, 50]$ from the camera.
Hence some reference images view a close-up patch of the surface, whereas most are too distant to infer texel-level details.
This replicates the effects of highly variable pixel-to-texel ratio in reference imagery, which we expect to be present in many kinds of real-world data such as street view images or sets of in-the-wild photographs.

Figures~\ref{fig:earth}b,c illustrate that with mipmapped texture filtering, the learned texture converges to a solution much closer to the reference (32.9\,dB vs 25.6\,dB).
The convergence failure of non-mipmapped version can be explained by a simple thought experiment.
When the reference image has a faraway pixel with a large texture footprint, its value is determined by a weighted mean of the reference texture over that footprint.
Without mipmapping, we will sample whichever full-resolution texel quad lands under that pixel center.
If this value deviates from the large-area average in the reference image, the gradients will pull the texels in the learned texture towards this average.
Over many such updates, this pull towards the mean leads to attenuated high-contrast details, which can be seen in Figure~\ref{fig:earth}b where even the converged non-mipmapped solution has less visible contrast than the solution obtained via proper texture filtering.

\figenv %

\subsection{Indirect Texturing, BRDF Optimization}

To demonstrate the flexibility of our modules, we construct a rendering pipeline that computes reflections via environment mapping \cite{envmap} and adds a highlight from an additional light source using a Phong BRDF \cite{phong}.
Figure~\ref{fig:env} illustrates the use of this rendering pipeline for solving the environment map contents and Phong BRDF parameters based on the reflections from an irregular object with known geometry and pose.

At the beginning of optimization, the BRDF parameters are initialized to random values, whereas the environment map is initialized to uniform gray.
In each iteration, the camera angle and light direction are randomized.
Optimization is done using Adam with a fixed learning rate of $10^{-2}$ and a simple image-space $L_2$ loss.
In this synthetic test, the unknown environment texture and BRDF parameters rapidly converge to the reference solutions.

The rendering pipeline is constructed as follows.
We start by rasterizing the geometry as usual, obtaining a frame buffer with per-triangle barycentrics and their screen-space derivatives.
We also calculate a normalized reflection vector for each vertex.
These reflection vectors are then used as attributes for interpolation, which yields per-pixel reflection vectors and screen-space derivatives for each of their components.
We represent the environment map as a cube map, so for each per-pixel reflection vector we determine the corresponding cube map face and 2D texture coordinates within it.
The same calculation also yields the screen-space derivatives of the texture coordinates, and we perform a trilinear texture fetch to the environment map.
This is important because reflections from curved surfaces introduce highly variable distortions and texture footprint sizes.
The cosine between the reflection vector and light direction vector required by the Phong BRDF model is computed based on the per-pixel reflection vectors.

The shading computation involves~18 lines of Python code using standard TensorFlow operations.
In our opinion, this is a small price to pay for the complete freedom to tailor shading, data representations, etc., to the needs of the application, compared to incorporating a fixed set of shading models into the differentiable rendering system itself.
It would not be possible to implement a similar setup in previous rasterization-based differentiable renderers without modifying their internals.

\paragraph{Limitations}
Local shading models, such as the one demonstrated here, cannot accurately model global phenomena such as interreflections.
If such fidelity is required, a path tracing based differentiable renderer will be necessary \cite{redner,mitsuba2,Loubet2019}.
However, there is no inherent limit on the complexity of the local shading model, so e.g. microfacet \cite{cook82} or Gaussian mixture model \cite{herholz2016} BRDFs could be used to seek a better fit to data.
Ultimately it depends on the intended use how physically accurate the shading should be\,---\,even a crude approximation of appearance may be sufficient for inferring other unknowns such as pose or geometry.
In Section~\ref{sec:perfcap}, we demonstrate that in the context of facial performance capture, the per-frame geometry of skin areas can be accurately recovered without any shading at all.
In situations like this, striving for physical fidelity would unnecessarily slow down the optimization.

\subsection{Pose Optimization}

\figpose %

In the SoftRas paper, Liu et al.~\shortcite{softras} investigate the problem of resolving the pose of a rendered cube using gradient-based optimization of image-space loss.
The task is made difficult by an optimization landscape with many local minima (Figure~\ref{fig:pose}).
The image synthesis model of SoftRas allows turning all surfaces partially transparent and blurring them by an arbitrary amount.
This results in a smoother loss function and a modest improvement in the resolved poses.

To demonstrate that the nonstandard image synthesis model of SoftRas is not necessary for solving this task, we focus on the optimization process instead of manipulating the rendering model.
A simple and efficient way to discourage local minima in a stochastic fashion is to add noise to the unknown parameters\footnote{%
We represent the pose as a quaternion. Noise is applied by constructing a random quaternion and mixing it with the pose using spherical interpolation \protect\cite{slerp}.
}
during optimization.
Indeed, running the optimization for 10k iterations using Adam and ramping down the noise strength from $1$ to $0.003$ over the course of optimization yields an average pose error of 48.62$^\circ$ measured over 100 random trials.
This is an improvement over the best result of 63.57$^\circ$ reported by Liu et al.~\shortcite{softras}, indicating that noise-based regularization is at least as effective as their approach based on transparency and blur.

However, we note that gradient descent from a random initial state is an ineffective way to solve this problem.
Splitting the optimization into two phases\,---\,first greedily seeking for a good initial pose by applying ramped-down noise in a gradient-free fashion, and then continuing with Adam from the pose with the smallest image-space loss\,---\,lowers the average error to 22.49$^\circ$.
As a further task-specific optimization, we can take the symmetries of the cube into account and customize the noise to incorporate random symmetry-preserving rotations.
This effectively bridges the local minima with similar pose but different color combinations and lowers the average error to 2.61$^\circ$.

\subsection{Performance}
\label{sec:performance}

\tblperf %

To assess the performance of our method, we selected 14 meshes of varying triangle counts from the ShapeNet database \cite{shapenet}.
We rendered these meshes using both our method and two comparison methods in multiple resolutions.
As comparison methods we used the official implementation of Soft Rasterizer \cite{softras}, and PyTorch3D~\cite{pytorch3d}, a more recent differentiable rasterization library for PyTorch.
The test was set up to include both forward and gradient evaluations, reflecting the total cost of including a rendering operation in an optimization task.

Default $\gamma,\sigma$ parameter values were used for Soft Rasterizer.
PyTorch3D was set up to render one pixel blur radius, one face per pixel, and soft compositing (SoftGouraudShader).
Default bin size heuristic was enabled.
We originally intended to include interior scenes in our tests, but Soft Rasterizer could not render ``in-scene'' viewpoints due to lack of clipping, making triangles behind the camera render erroneously in front of the camera.
Therefore, we limited our test to individual objects rendered in front of the camera.
All tests were run on a single NVIDIA TITAN V GPU with 12\,GB of memory.

Results of the test are summarized in Table~\ref{tbl:perf}.
We can see that our method is much less sensitive to triangle and pixel counts than the comparison methods.
Soft Rasterizer slows down quickly because it tests each triangle for every pixel, and consequently loses to our method by several orders of magnitude with nontrivial triangle counts and resolutions.
PyTorch3D fares better than Soft Rasterizer thanks to its coarse-to-fine rasterization architecture.
Still, the performance difference is more than an order of magnitude in our favor and grows with high resolutions and triangle counts, highlighting the better scalability of our method.

\figoccluplot %

\paragraph{Occluded vs visible geometry}
It is generally desirable that rendering performance is not affected by the amount of geometry that is not visible in the rendered image.
Our method employs deferred shading, and is therefore mostly oblivious to occluded or out-of-view geometry except at the rasterization step.
The same holds for PyTorch3D when storing just one face per pixel, but its rasterization step is expensive so it is not obvious how hidden geometry affects the overall performance.

To quantify the effects of depth complexity, we constructed pairs of synthetic scenes where the number of triangles and covered pixels are held constant but the depth complexity is varied.
Specifically, we repeat a simple base mesh either so that all copies are visible and together cover the image, or so that only one is visible and covers the image while all other copies are hidden behind it.
Figure~\ref{fig:occluplot} shows the measured performance as function of geometric complexity and geometric setup.
Resolution is fixed to \res{1024} and the settings are otherwise the same as above.
PyTorch3D, disregarding the constant cost of $\sim$50\,ms, scales strongly with the total area of geometry, occluded or not.
This is due to its software rasterizer resolving visibility so late that practically no work is saved if the tested fragment is found to be occluded.
The performance of our method is mostly unaffected by the depth complexity, as it uses the hardware rasterizer with efficient hierarchical depth tests.
SoftRas scales linearly with geometric complexity regardless of the geometric setup, and does not compare favorably to the other two methods.

\section{Application: Facial Performance Capture}
\label{sec:perfcap}

To illustrate the performance and utility of the design of our differentiable rendering pipeline, we examine how to use it to solve markerless facial performance capture, i.e., inferring time-varying facial geometry based on multiple camera streams.
This is a non-trivial classical computer graphics problem that has been approached in several ways in the past.
Many methods utilize morphable 3D models \cite{Blanz1999} that enable approximating the facial geometry even from monocular data.
The downside of this class of methods is that the obtained geometry is approximate and cannot fully reproduce intricate motion.
High-quality markerless capture often requires complex capture setups involving structured light or special cameras \cite{Oleg2009,Bradley2010}.
The passive capture method of Beeler et al.~\shortcite{Beeler2011} first reconstructs each frame using a single-shot method \cite{Beeler2010} and then builds frame-to-frame correspondences iteratively.

While these methods yield great results, they are fairly complex and consequently difficult to implement.
As a result, the state of the art in many cases is using commercial capture systems such as DI4D PRO~\shortcite{di4d} or commercial software such as Agisoft Metashape~\shortcite{metashape} and R3DS Wrap~\shortcite{wrap}.
Our goal is not to attempt to surpass this state of the art, but to illustrate how far we can get with a near-trivial formulation as an inverse rendering problem.
In particular, we will not attempt to reconstruct tricky regions such as mouth, eyes, or hair, but instead focus on skin areas only.

Our test material consists of three performances captured in a DI4D PRO rig at 29.97~frames per second.
There are synchronized 9 camera feeds with 3 in color and 6 in monochrome\,---\,for simplicity, we convert the color reference images to monochrome as well prior to processing.
Resolution of the reference images is \ress{3008}{4112} pixels, and the camera intrinsics and extrinsics are known.
Lengths of the three performances (``Neutral'', ``Disgust'', and ``Anger'') are 89, 123, and 207 frames, respectively.

\subsection{Solution via Inverse Rendering}

Our goal is to find a global texture and a per-frame mesh so that when rendered from the known camera positions, the textured meshes match the reference footage as closely as possible measured using image-space $L_2$ loss.
We learn the geometry as per-frame deformation of a fixed-topology base mesh that has 16\,521 vertices and 16\,472 original faces that were triangulated into 32\,916 triangles for rendering.
The base mesh has texture coordinates referencing a 5:1 aspect ratio texture atlas, and our learned texture is a single-channel \ress{10240}{2048} texture initialized to zeros.
The texture coordinates of the base mesh are not modified during optimization.

We do not consider material properties or lighting in our image synthesis model, and assume skin to be Lambertian and lighting to be uniform.
Neither assumption is valid \cite{Marschner1999} but we can still expect the geometry to be reconstructed correctly if this produces the best achievable images.
Capturing material properties and incident lighting, along with geometry, should be possible with a more complex rendering pipeline \cite{Liu2017}.
Occasional convergence problems were caused by variable shadowing under the nose.
We alleviate the problem by passing both rendered and reference images through a high-pass filter so that all low-frequency effects are attenuated before computing the image-space loss.
This has the added benefit of making the optimization more resilient to illumination and shading effects that occur due to changes in head orientation.

\subsection{Parameterization and Regularization}

In principle, the vertex positions for every frame could be encoded in one large matrix to be optimized.
However, we decompose the vertex positions into a matrix product in order to make the optimization landscape more tractable.
One could see our model as a three-layer dense neural network that maps a one-hot animation frame index vector into a vector of per-vertex deltas from the base mesh.
Even though using a chain of matrices without nonlinearities does not increase the expressiveness of the representation, it accelerates optimization similar to how overparameterization accelerates deep learning \cite{Arora2018}.

\newcommand{\Vbase}{\boldsymbol{V}_\mathit{\!base}}
\newcommand{\Vi}   {\boldsymbol{V}_{\!i}}
\newcommand{\Rj}   {\boldsymbol{R}_{\!\hspace*{.05em}j}}
\newcommand{\vj}   {\boldsymbol{v}_{\!\hspace*{.05em}j}}
\newcommand{\vk}   {\boldsymbol{v}_k}
\newcommand{\M}[1] {\boldsymbol{M}_{#1}}
\newcommand{\wi}   {\boldsymbol{w}_i}
\renewcommand{\dj} {\boldsymbol{\delta}_{\!j}}
\newcommand{\djb}  {\boldsymbol{\delta}^\mathit{base}_{\!j}}
\newcommand{\cmdot}{{\cdot}}

Let us denote the vertex count $n$ and the number of reference frames $m$.
Denoting the $3n$ column vector encoding base mesh vertex positions as \mbox{$\Vbase = [x_1,y_1,z_1,x_2,\cdots]$}, the vertex positions for frame $i$ are computed as
$\Vi = \Vbase + \M{3}\M{2}\M{1}\wi$,
where $\wi$ is a one-hot column vector of length $m$ with entry at index $i$ being one.
Matrices $\M{1}$ and $\M{2}$ are of size $m \times m$, and $\M{3}$ has size $3n \times m$.
The square matrices $\M{1}$ and $\M{2}$ are initialized to identity, whereas $\M{3}$ is initialized to zero.
$\M{3}$ can be seen as a learned basis for the mesh deltas, and $\M{1}$ and $\M{2}$ acting as a mapping from frame index to this basis.
Finding the geometry thus corresponds to finding values for $\M{1,2,3}$ via optimization.

This formulation has no inherent propensity to, e.g., keep the surface tessellation of the mesh intact.
Because of this, we apply a mesh Laplacian regularization term that penalizes local curvature changes compared to the base mesh.
Sorkine~\shortcite{Sorkine2005} gives an overview of Laplacian-based methods for mesh processing.
Using their notation, the uniformly-weighted differential $\dj$ of vertex $\vj$ is \noraise{$\dj = \vj - \frac{1}{|N_j|}\sum_{k\in N_j}\vk$}, where
$N_j$ is the set of one-ring neighbors of vertex $\vj$.
In other words, $\dj$ is the difference between position of vertex $\vj$ and the average position of its neighbors.
Our Laplacian regularization term is \noraise{$L_{\boldsymbol{\delta}} = \frac{1}{n}\sum_{1 \le j \le n}||\dj-\djb||^2$}, i.e.,
the average square Euclidean difference between the vertex differentials of the base mesh and those of the deformed mesh.
Although this representation is not rotation-invariant \cite{Lipman2005}, we have not found this to be a problem in practice.
Similar regularization has been used in earlier work on shape inference \cite{softras}.

In addition to texture and geometry, we learn global, per-camera brightness and contrast adjustment values that are applied to the rendered images during training.
This accounts for differences between reference images originating from color vs monochrome cameras.
We do not use any form of temporal regularization, i.e., there are no terms that would prefer nearby frames to be similar to each other.
Regardless of this, the solution is temporally stable as can be seen in the accompanying video.

\subsection{Optimization}

To resolve geometry and texture for a sequence of frames, we initialize the geometry representation as explained above.
In each iteration, we choose random frame and camera indices, and render the corresponding mesh using a pipeline similar to one shown in Figure~\ref{fig:pipeline}.
We perform rendering in the same resolution as reference images, i.e., \ress{3008}{4112} pixels, which limits our minibatch size to one in practice.

In our primary configuration, optimization is run for 100\,000 iterations using Adam optimizer \cite{adam} ($\beta_1=0.9, \beta_2=0.999$) with a base learning rate of \noraise{$\lambda=10^{-3}$}.
The learning rate is decayed to $10^{-4}$ during the last 25\% of optimization.
High-pass filtering is computed as $x' = x - 0.3\cmdot\mathit{blur}(x)$ where $x$ is the rendered/reference image, and $\mathit{blur}(x)$ downsamples the image by a factor of \res{32} and upsamples it back using an approximate Gaussian filter.
Images are compared using $L_2$ loss, so the overall loss function with the Laplacian term included is \noraise{$L = ||x' - y'||_2^2 + 3 L_{\boldsymbol{\delta}}$} where $x'$ and $y'$ are the high-pass filtered rendered and reference images, respectively.
A typical optimization run takes 60--70 minutes on a single NVIDIA Tesla V100 GPU, with each optimization iteration taking approximately 40 milliseconds.

In an additional test, we resolved all three capture sequences in a single optimization run.
This task is complicated by the actor being positioned slightly differently in each sequence\,---\,with just one base mesh, the alignment is off and large motion of the mesh is required.
To circumvent this problem, we also learn a $3\times4$ rigid transformation matrix $\Rj$ for each sequence $j$, and apply it before the vertex deltas, i.e., $\Vi = \Rj\Vbase + \M{3}\M{2}\M{1}\wi$.
In addition we initialize the texture with one previously solved for the ``Neutral'' performance, limited $\M{2}$ to \res{100} elements, and adjusted the dimensions of $\M{1}$ and $\M{3}$ accordingly.
With these modifications and extending the computation to 800\,000 iterations, the optimization successfully found a rigid transformation for each sequence to handle the misalignments, and subsequently solved the vertex deltas for every frame of the combined set along with a texture that best fits all three sequences.
As a consequence of optimizing a single texture for the entire material, the mapping between surface and texture-space points becomes automatically consistent between all sequences.

\subsection{Results}

\figmesh %

Figure~\ref{fig:mesh} shows an example result of the geometry and texture optimization in the ``Neutral'' sequence.
For clarity, the wireframes in Figure~\ref{fig:mesh}{c--e} include only the edges of original base mesh instead of the triangulated version.
See the accompanying video for the sequences and our reconstructions, as well as the progression of geometry and texture during training.

In our base mesh, the holes for mouth and eyes are simply covered with triangles in order to avoid spurious visibility leaks to the opposite side of the mesh.
Obviously, this does not allow faithful reconstruction of mouth and eyes, because eyes have strong view-dependent reflections, and mouth has complex internal geometry.
As such, the optimization ends up texturing these covering triangles only somewhat believably.
Hair becomes similarly approximated by the learned texture.

\figdiff %

Figure~\ref{fig:diff} shows closeups of the nose region in five frames selected from sequences ``Neutral'' and ``Disgust''.
There is a surprising amount of fine-grained motion, and our method captures this very accurately as illustrated in the figure.
We obtained a 3D reconstruction from DI4D~\cite{di4d} for comparison purposes, and it shows markedly less deformation and does not align properly with the camera images.
Their optical flow based reconstruction does not attempt to track areas that lack high-quality multi-view observations such as ears and nostrils, whereas our solution automatically reconstructs this motion as well.

\figcombo %

Figure~\ref{fig:combo} shows three example frames from the test where all three performances were resolved at once.
In these selected frames, the mouth region looks reasonable, but the ``Anger'' sequence exhibits artifacts around the mouth in many frames.
This is not surprising\,---\,our model is unable to render the mouth adequately, so the optimum may be far from correct.
Nonetheless, the solution is otherwise temporally stable, and the motion and texture of skin areas are reconstructed well.
As the same texture and texture coordinates are used for all frames, our method provides highly consistent vertex to skin correspondence across all sequences.

\section{Discussion and Future Work}
\label{sec:discussion}

We have demonstrated a modular differentiable renderer design capable of rendering high-resolution images of complex 3D scenes up to several orders of magnitude faster than prior approaches, while supporting crucial features such as filtered texture mapping with correct gradients. 
We believe that a high-performance differentiable renderer enables countless uses in inverse graphics, generative modeling, and other computer vision and AI problems,
and to help this development, have made our library publicly available at \codeurl.

As a practical use case whose success hinges on high-performance differentiable rendering performance, we have demonstrated that multi-view facial performance capture from synchronized high-resolution video cameras can be solved accurately by casting it as a simple inverse rendering problem.
It will be interesting to extend this solution to joint material appearance capture, dynamic textures, as well as custom solutions for the eyes, mouth, and hair, integrated as a single optimization problem.

There may exist specific circumstances and applications (e.g., discovery of occluded geometry) where the all-transparent image formation model used by several earlier differentiable renderers may be beneficial for optimization. However, we believe that such problems can also be approached in a principled way via careful choices for mesh parameterization, regularization, and optimization methods, while following the standard occlusion model of the modern graphics hardware pipeline.

\begin{acks}
We thank
Simon Yuen for providing input and comparison data for the facial performance capture experiment,
David Luebke for comments,
and
Sanja Fidler and Wenzheng Chen for discussions on previous work.
\end{acks}

\bibliographystyle{ACM-Reference-Format}
\bibliography{paper}

\end{document}

%% file: figures.tex
\newcommand{\figpipeline}{
\begin{figure*}
\centering
\hfill\includegraphics[width=0.99\linewidth]{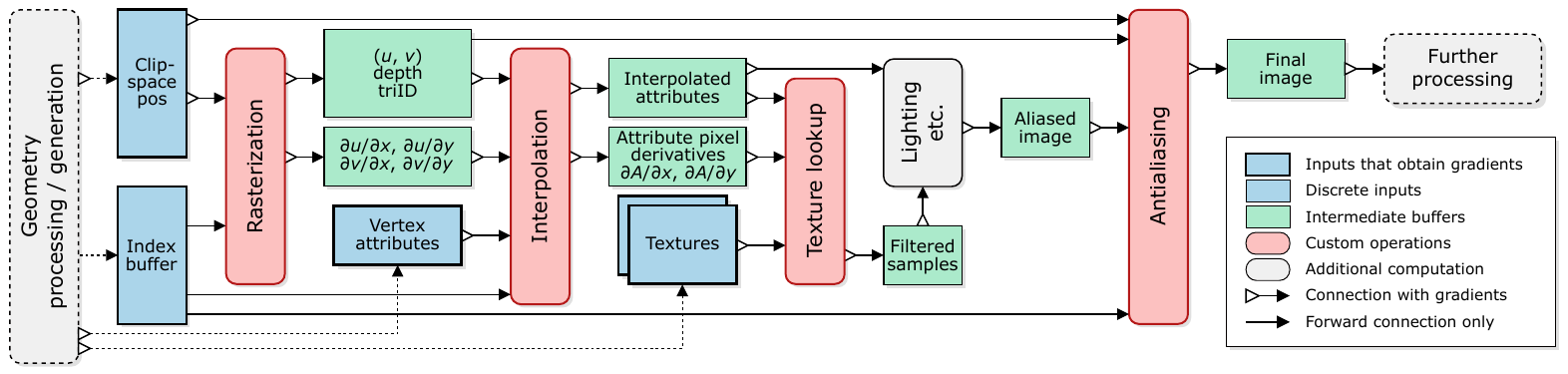}\vspace*{-1em}%
\caption{\label{fig:pipeline}%
A simple differentiable rendering pipeline with our proposed primitive operations highlighted in red.
The input data for rendering (blue) may be generated by, e.g., a neural network if the pipeline is part of a larger computation graph.
In simpler setups the geometry processing might include only the model/view/perspective transformations for vertex positions with other inputs being constants or learnable parameters.
All intermediate buffers (green) are in image space.
Connections with gradients are denoted by a white triangle.
Channel counts are fixed only for vertex positions and indices, and in the intermediate buffers produced by the rasterization operation.
There are no restrictions on the channel counts for vertex attributes, textures, related intermediate data, or the output image.
}
\end{figure*}
}

\newcommand{\figantialias}{
\begin{figure}
\centering
\includegraphics[width=0.85\linewidth]{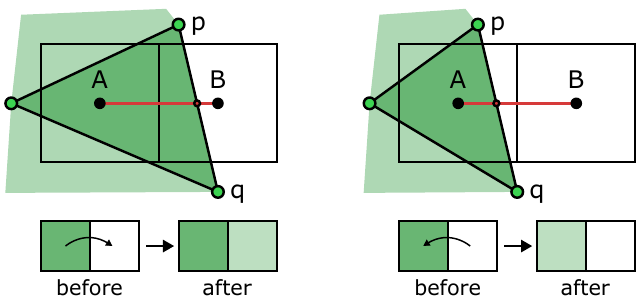}\hphantom{xxx}\\
(a)\hspace*{37mm}(b)%
\caption{\label{fig:antialias}%
Illustration of our analytic antialiasing method.
A vertical silhouette edge $p,q$ passes between centers of horizontally adjacent pixels $A$ and $B$.
This is detected by the pixels having a different triangle ID rasterized into them.
Pixel pair $A,B$ is processed together, and one of the following cases may occur.
(a) The edge crosses the segment connecting pixel centers inside pixel $B$, causing color of $A$ to blend into $B$.
(b) The crossing happens inside pixel $A$, so blending is done in the opposite direction.
To approximate the geometric coverage between surfaces, the blending factor is a linear function of the location of the crossing point\,---\,from zero at midpoint to 50\% at pixel center.
This antialiasing method is differentiable because the resulting pixel colors are continuous functions of positions of $p$ and $q$.
}
\end{figure}
}

\newcommand{\clod}{c_\mathit{lod}}
\newcommand{\figtexture}{
\begin{figure}
\centering
\includegraphics[width=1.0\linewidth]{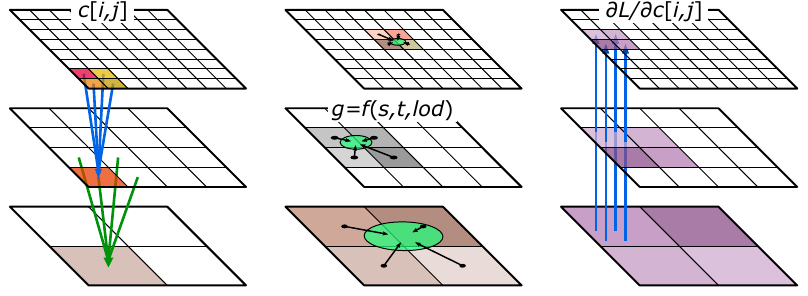}\\
\makebox[0.32\linewidth][c]{(a)}\hfill%
\makebox[0.32\linewidth][c]{(b)}\hfill%
\makebox[0.32\linewidth][c]{(c)}%
\caption{\label{fig:texture}%
Filtered differentiable texture lookup with a non-constant texture.
(a) In the beginning of forward pass, prefiltered MIP levels $\clod$ are constructed from the full-resolution texture $c[i,j]$ by repeated downsampling using a $2\times2$ box filter.
(b) In forward pass, each lookup $g=f(s,t,\mathit{lod})$ interpolates prefiltered values on the appropriate MIP level as determined by the size of sample footprint.
In backward pass, we receive incoming gradients $\partial{L}/\partial{g}$.
Texture coordinate gradients $\partial{L}/\partial{s}$ and $\partial{L}/\partial{t}$ for each lookup are computed based on these and contents of texels that were used in interpolation.
Simultaneously, texture image gradients $\partial{L}/\partial\clod[i,j]$ are accumulated into each MIP level.
In a trilinear lookup, these calculations are performed on two adjacent levels and weighted according to the fractional part of $\mathit{lod}$.
(c) To produce outgoing full-resolution texture image gradients $\partial{L}/\partial{c}[i,j]$, we sum the accumulated gradients from all MIP levels.
}
\end{figure}
}

\newcommand{\rr}[1]{\multirow{2}{*}{#1}}
\definecolor{goodcolor}{rgb}{0.1, 0.7, 0.0}
\definecolor{mehcolor}{rgb}{0.8, 0.6, 0.0}
\definecolor{badcolor}{rgb}{0.8, 0.0, 0.0}
\newcommand{\markbox}[1]{\makebox[10mm][c]{#1}}
\newcommand{\goodmark}{\markbox{\ding{52}}}%
\newcommand{\mehmark}{\markbox{\ding{55}}}%
\newcommand{\badmark}{\markbox{\ding{54}}}%
\newcommand{\GOOD}{\color{goodcolor}{\goodmark}}
\newcommand{\MEH}{\color{mehcolor}{\mehmark}}
\newcommand{\BAD}{\color{badcolor}{\badmark}}
\definecolor{lightgrey}{rgb}{0.9, 0.9, 0.9}

\newcommand{\tblcaps}{
\begin{table}
\footnotesize
\caption{\label{tbl:caps}Comparison of characteristics of selected differentiable rendering systems.}
\tabulinesep=0.494mm%
\centering%
\vspace*{-4mm}%
\begin{tabu}{@{\hspace*{1mm}}l@{\hspace*{1mm}}c@{}c@{}c@{}c@{}c@{}c@{}c@{}}%
\vspace*{-.5mm} & OpenDR & NMR & SoftRas & DIB-R & Li et al. & Mitsuba2 & \rr{\bf Our} \\
&\shortcite{opendr}&\shortcite{nmr}&\shortcite{softras}&\shortcite{dibr}&\shortcite{redner}&\shortcite{mitsuba2}&\\
\hline\arrayrulecolor{lightgrey}%
Performance%
\makebox[0mm][l]{$^*$}
             & \GOOD & \MEH  & \MEH  & \MEH  & \BAD  & \BAD  & \GOOD\\\hline
Scalability  & \GOOD & \BAD  & \BAD  & \BAD  & \GOOD & \GOOD & \GOOD\\\hline
Flexibility  & \BAD  & \BAD  & \BAD  & \BAD  & \GOOD & \GOOD & \GOOD\\\hline
Antialiasing & \BAD  & \BAD  & \MEH  & \BAD  & \GOOD & \GOOD & \GOOD\\\hline
Occlusion    & \GOOD & \GOOD & \BAD  & \MEH  & \GOOD & \GOOD & \GOOD\\\hline
Gradients    & \BAD  & \BAD  & \GOOD & \MEH  & \GOOD & \MEH  & \GOOD\\\hline
Noise-free   & \GOOD & \GOOD & \GOOD & \GOOD & \BAD  & \BAD  & \GOOD\\\hline
No tuning    & \GOOD & \GOOD & \BAD  & \BAD  & \GOOD & \MEH  & \GOOD\\\hline
GI           & \BAD  & \BAD  & \BAD  & \BAD  & \GOOD & \GOOD & \BAD\\\arrayrulecolor{black}\hline
\end{tabu}
\parbox{\linewidth}{\small%
$^*$%
\emph{Performance} considers suitability to intensive optimization such as in deep learning;
\emph{Scalability} refers to performance with respect to surface tessellation and image resolution;
\emph{Flexibility} is whether the system is designed to support arbitrary shading;
\emph{Antialiasing} requires that geometric edges are smoothed in the result image;
\emph{Occlusion} considers if geometrically obscured surfaces are guaranteed to not affect the resulting image;
\emph{Gradients} refers to the correctness of gradients with respect to rendered image;
\emph{Noise-free} systems do not rely on random sampling;
\emph{No tuning} refers to lack of tunable parameters that affect the rendered image or gradients.
\emph{GI} denotes global illumination, support for physically-based illumination and shadowing including indirect effects.
See text for detailed analysis.}
\end{table}
}

\newcommand{\diffimg}[1]{\includegraphics[width=0.158\linewidth]{figures/diff/#1.jpg}}
\newcommand{\diffhlabel}[1]{\hspace*{2mm}\raisebox{0.107\linewidth}{\rotatebox[origin=b]{90}{\smash{\footnotesize{#1}}}}\hfill\hfill}
\newcommand{\figdiff}{
\begin{figure}
\centering
\diffhlabel{Reference}%
\diffimg{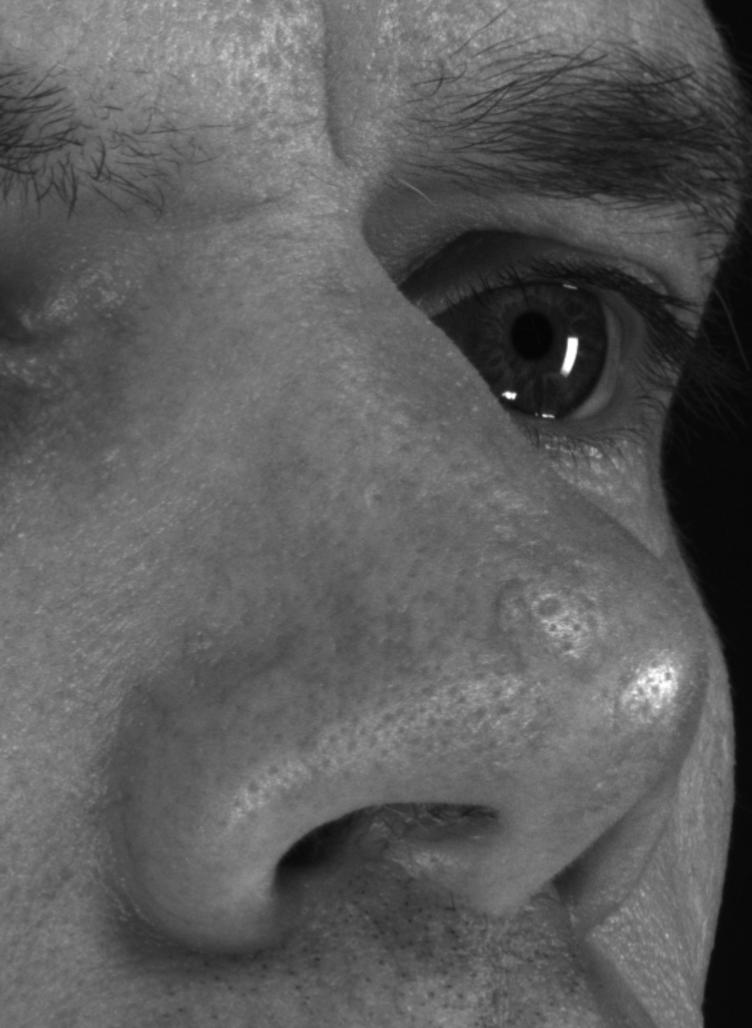}\hfill%
\diffimg{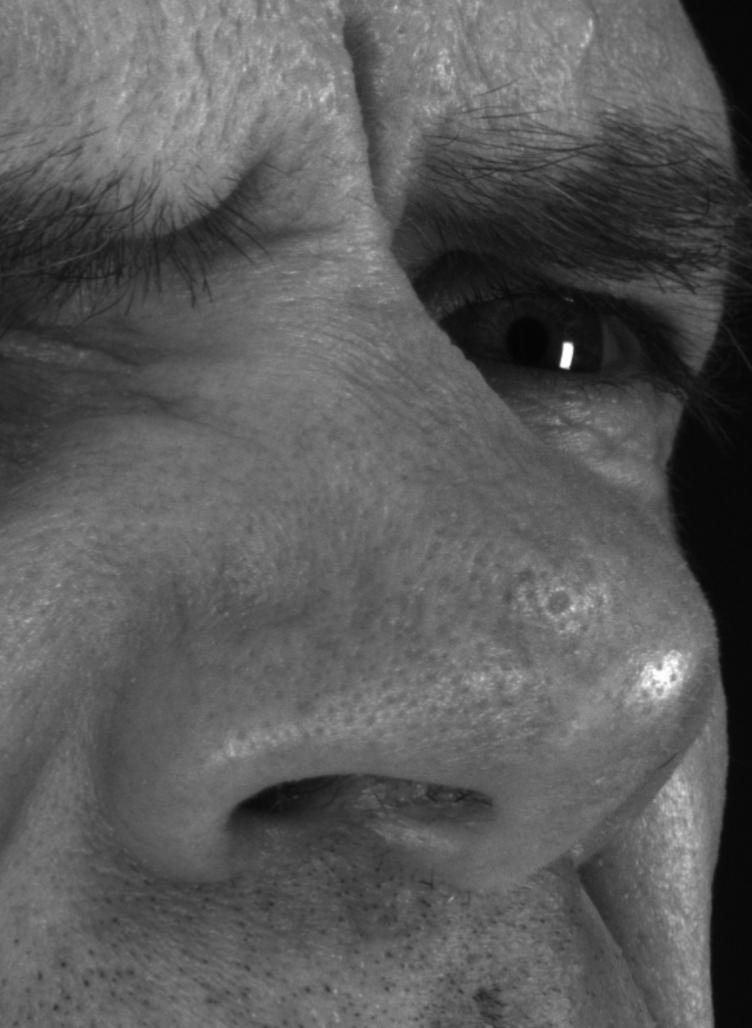}\hfill%
\diffimg{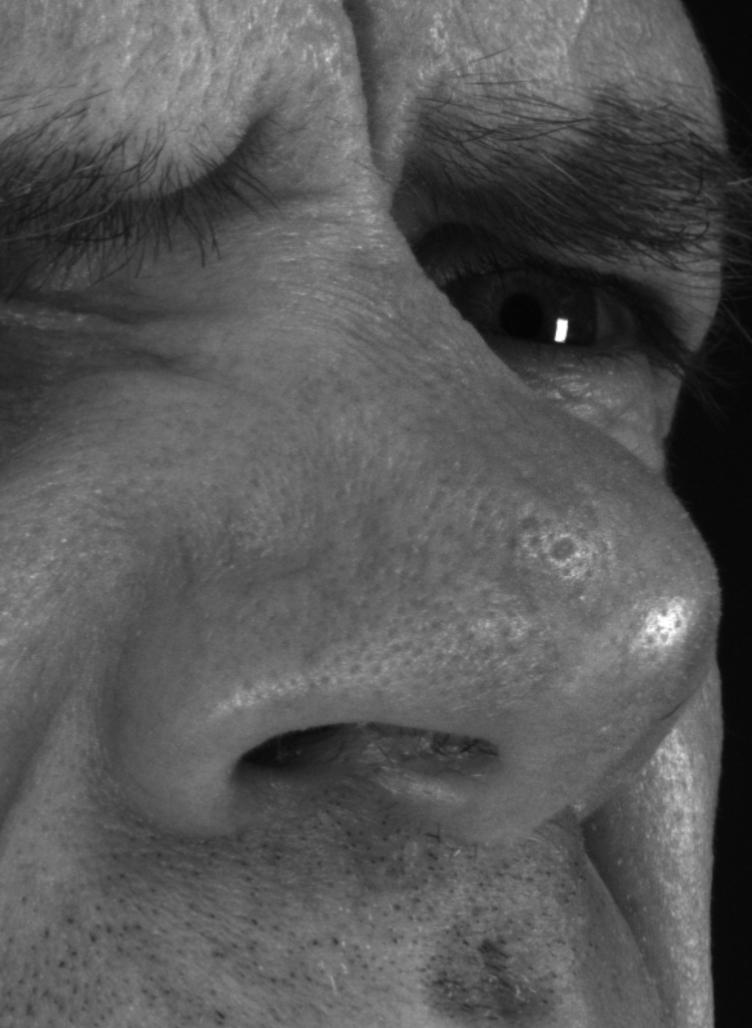}\hfill%
\diffimg{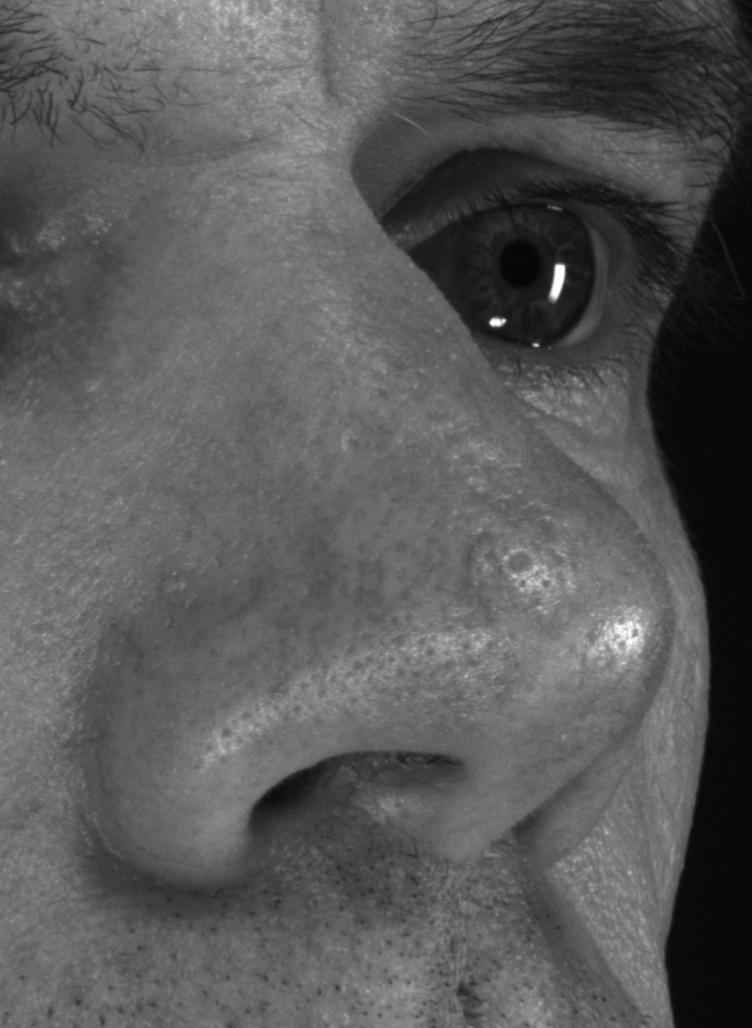}\hfill%
\diffimg{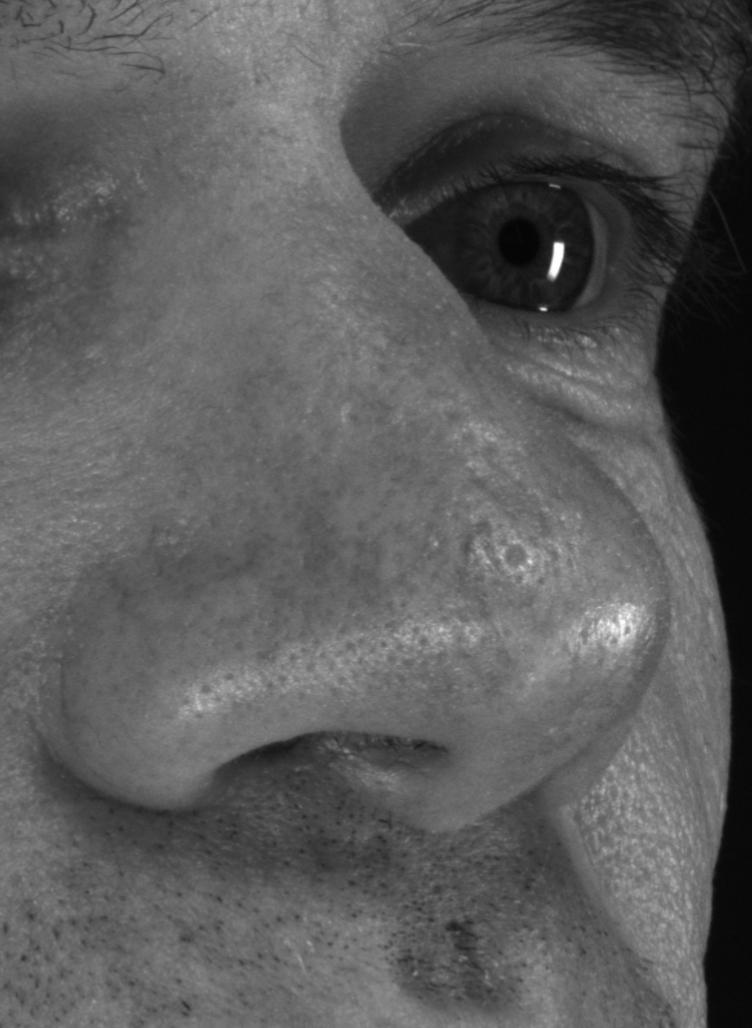}\hfill%
\diffimg{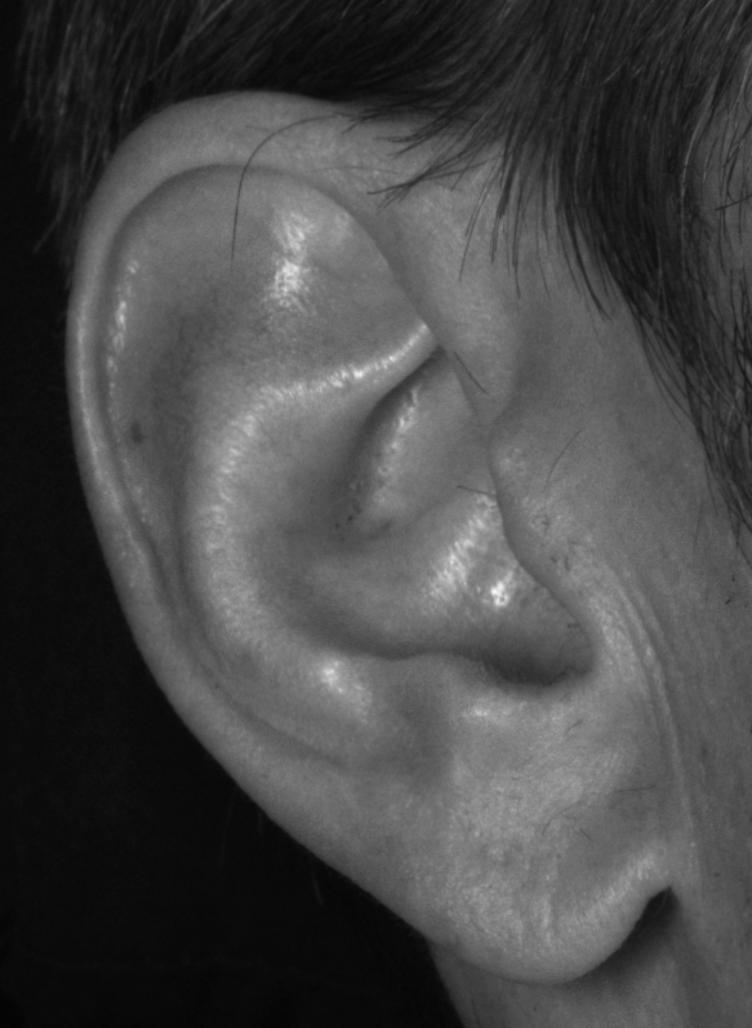}\\
\diffhlabel{Our rendering}%
\diffimg{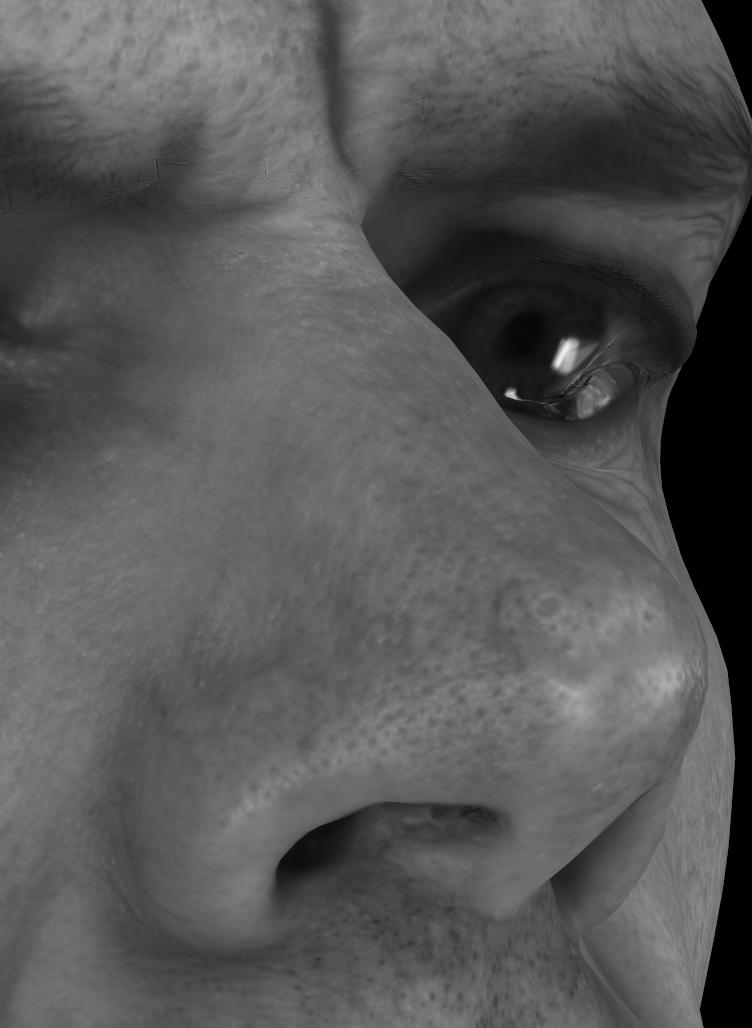}\hfill%
\diffimg{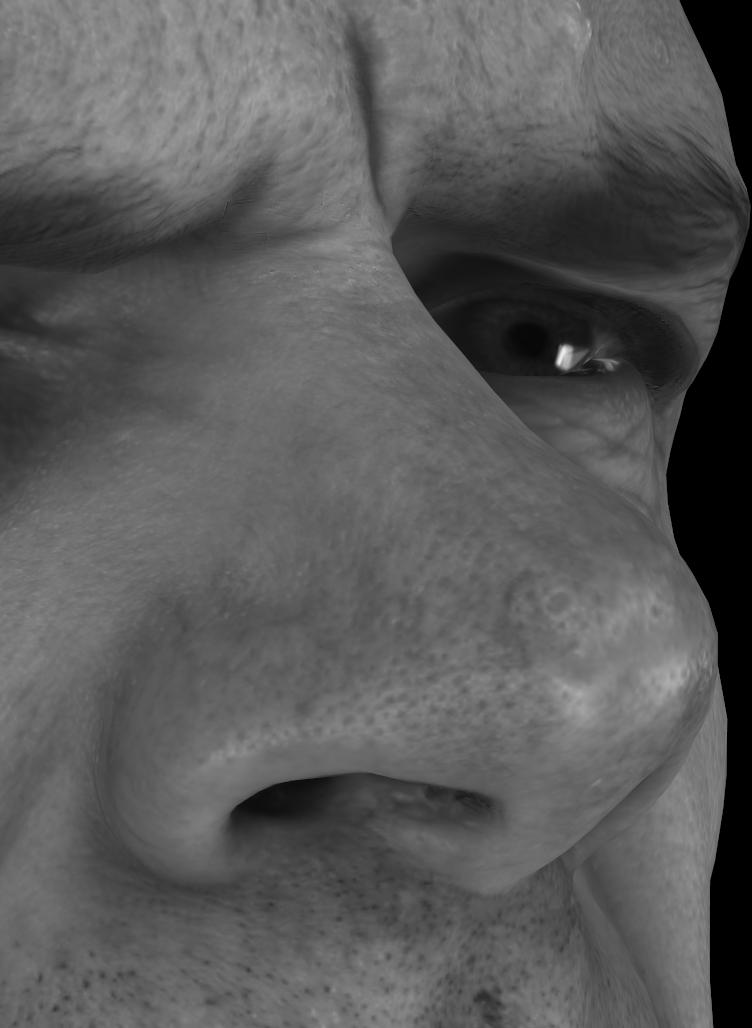}\hfill%
\diffimg{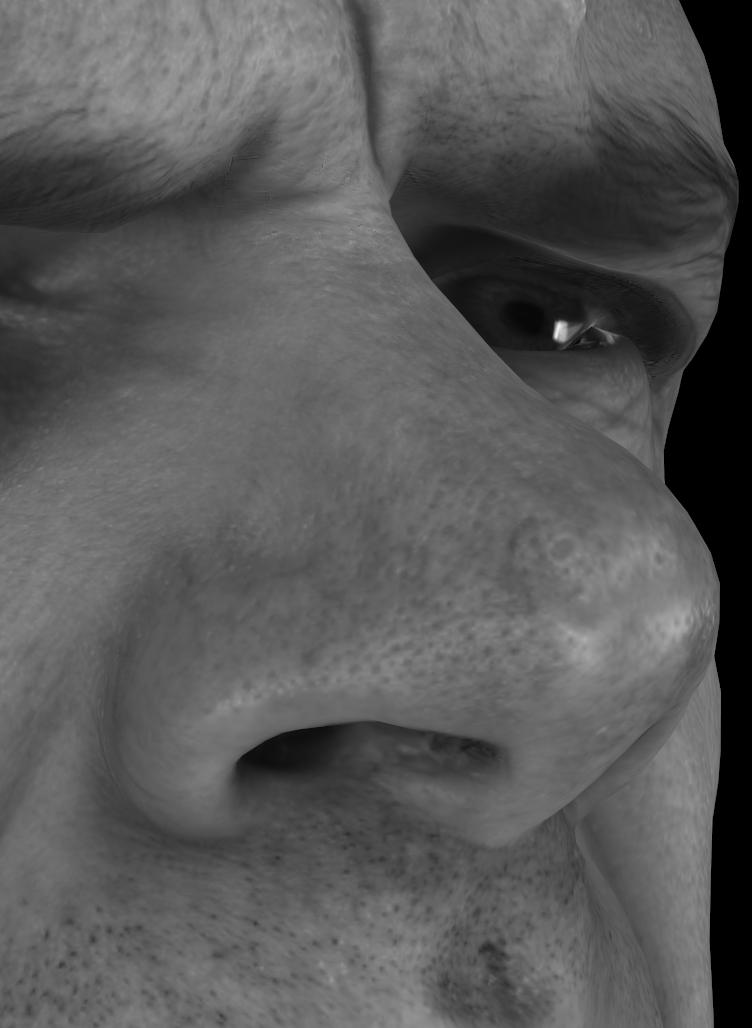}\hfill%
\diffimg{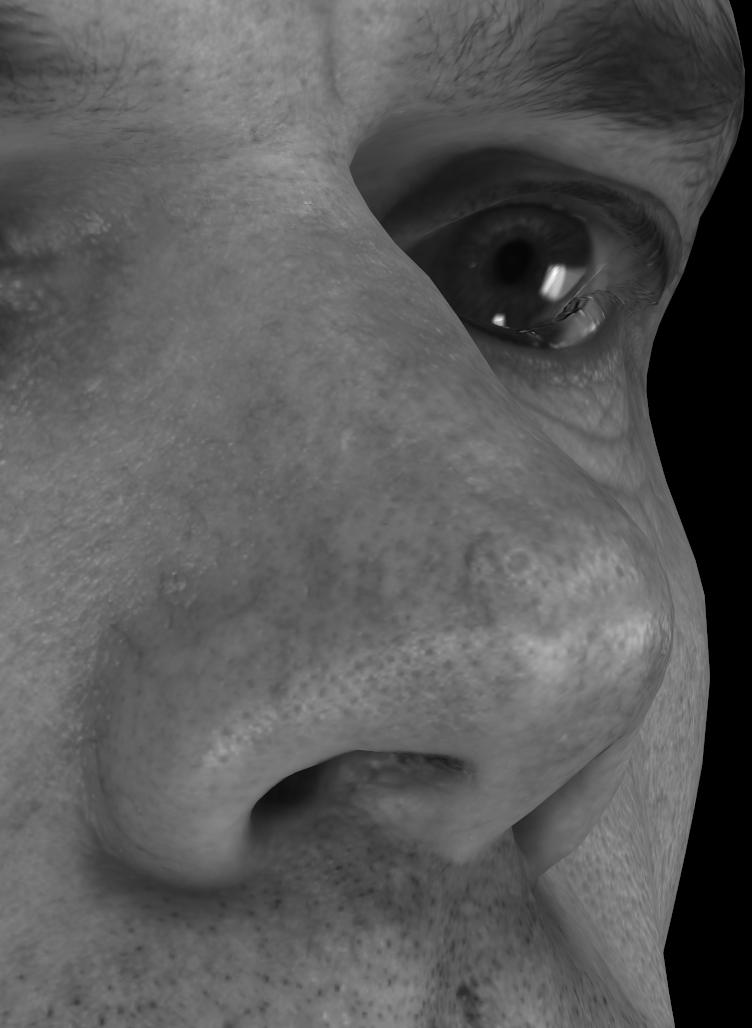}\hfill%
\diffimg{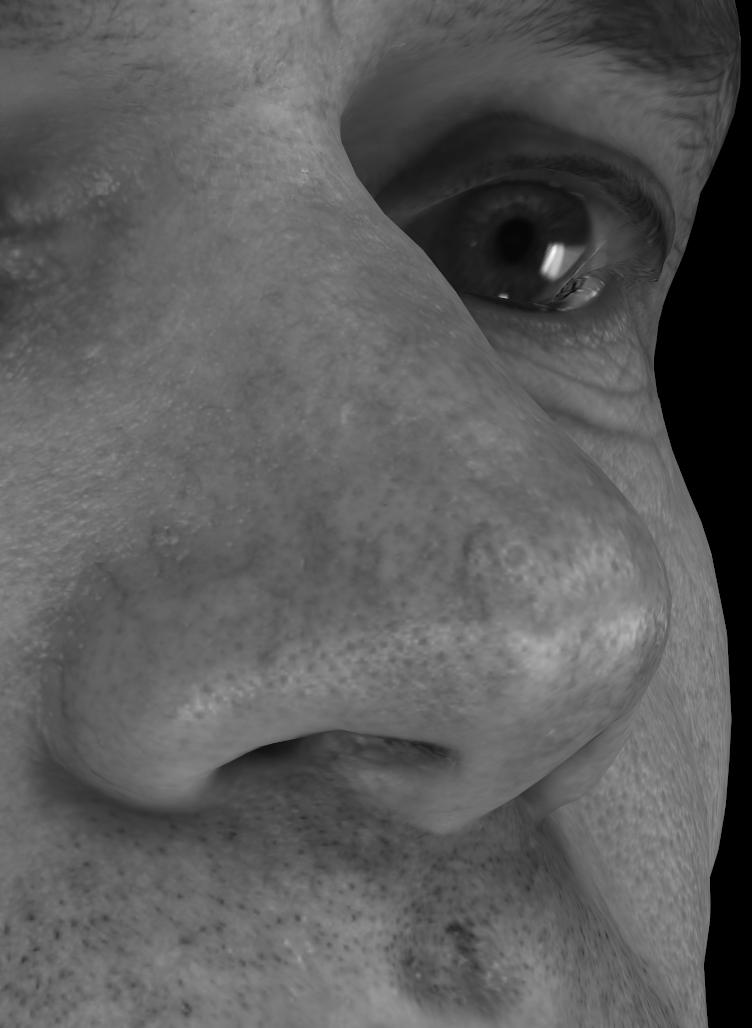}\hfill%
\diffimg{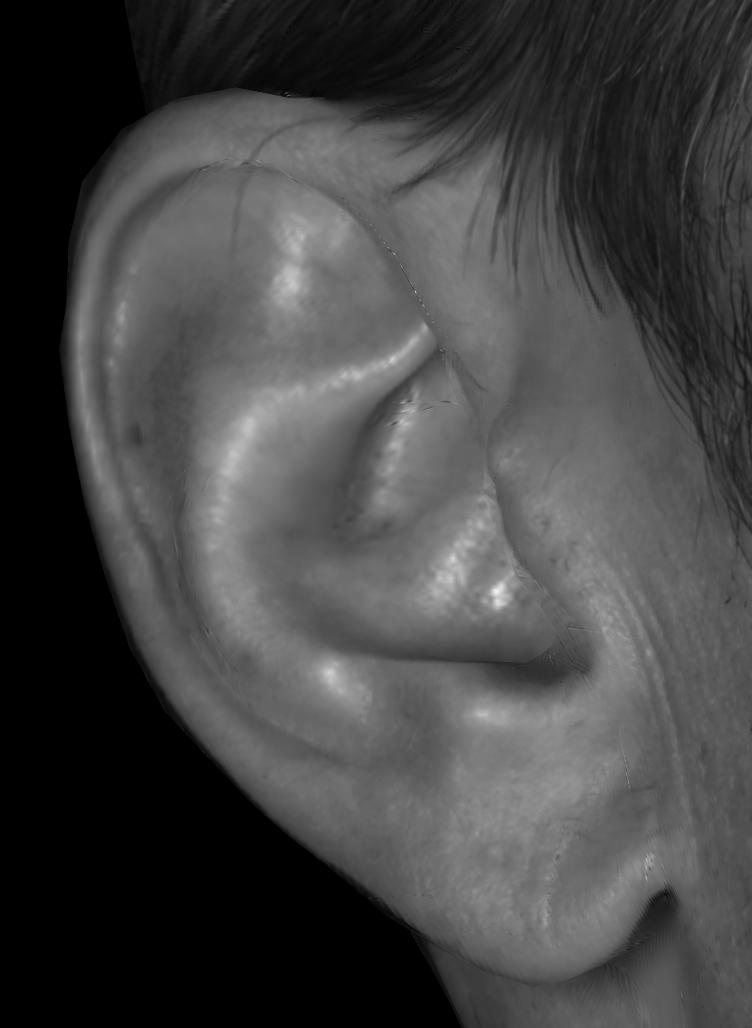}\\
\diffhlabel{DI4D~\protect\shortcite{di4d}}%
\diffimg{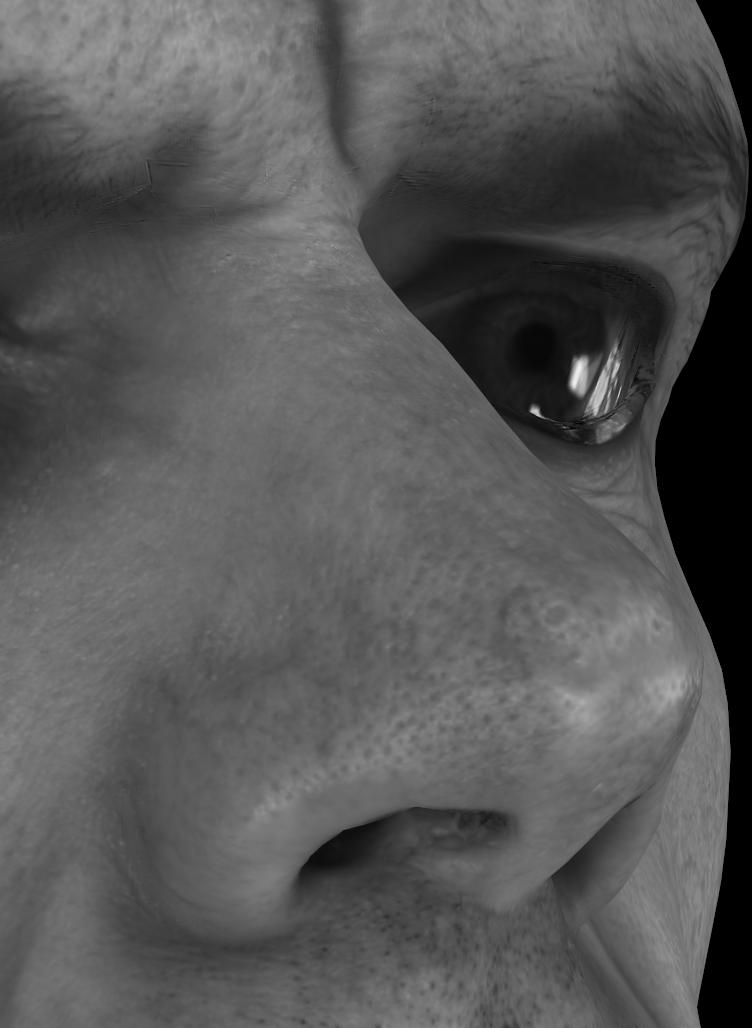}\hfill%
\diffimg{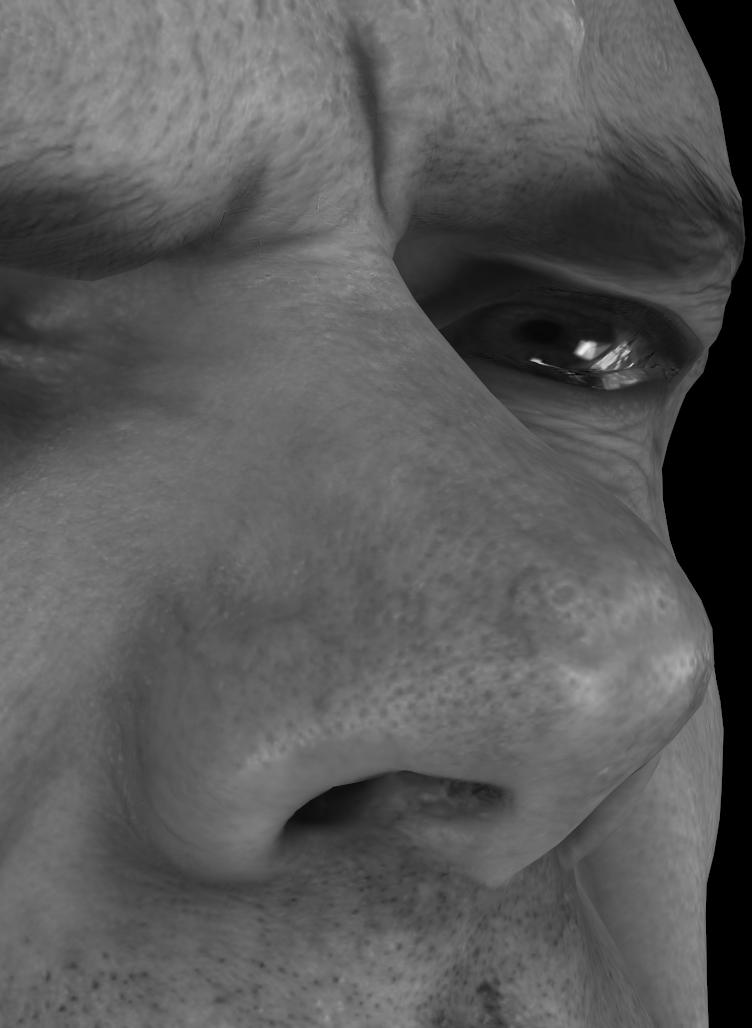}\hfill%
\diffimg{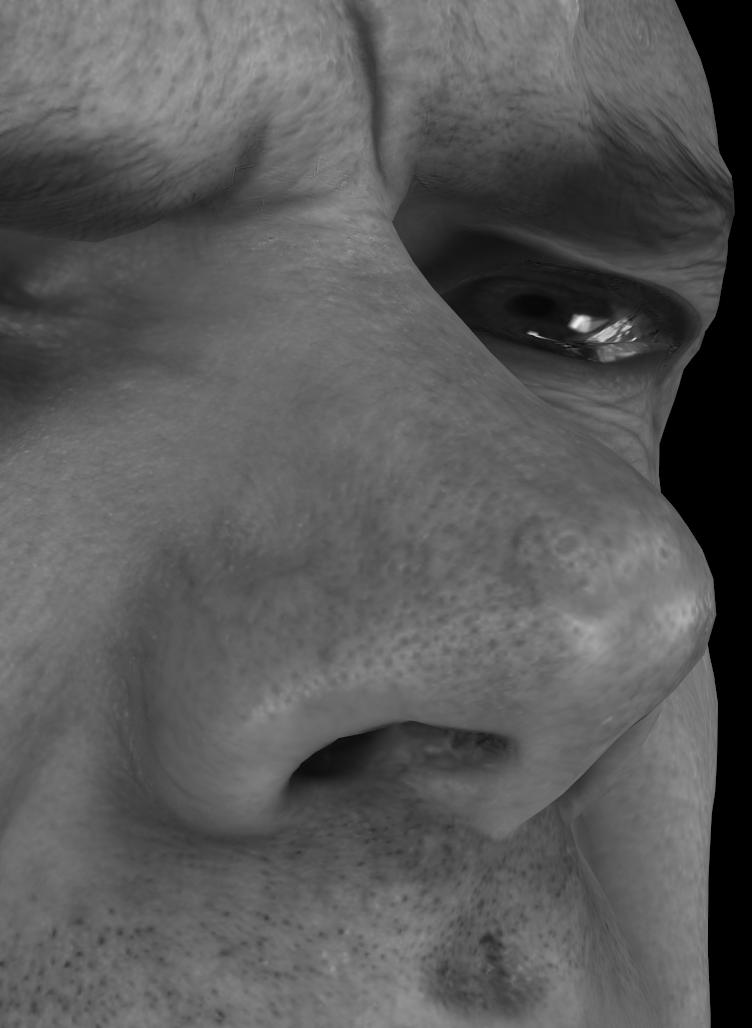}\hfill%
\diffimg{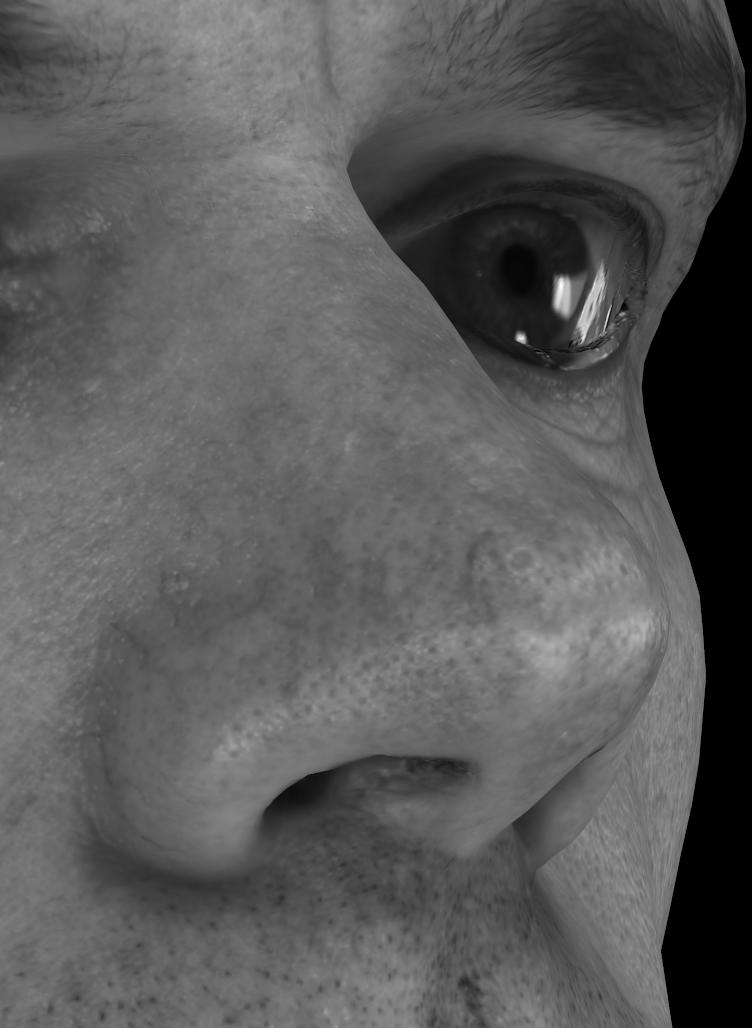}\hfill%
\diffimg{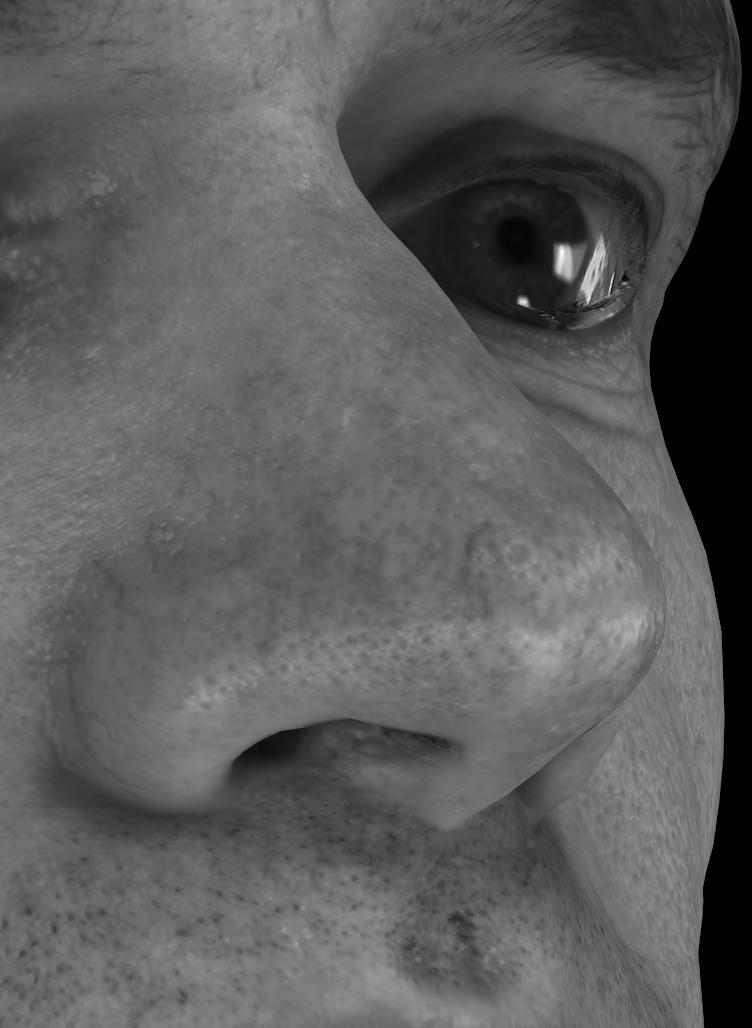}\hfill%
\diffimg{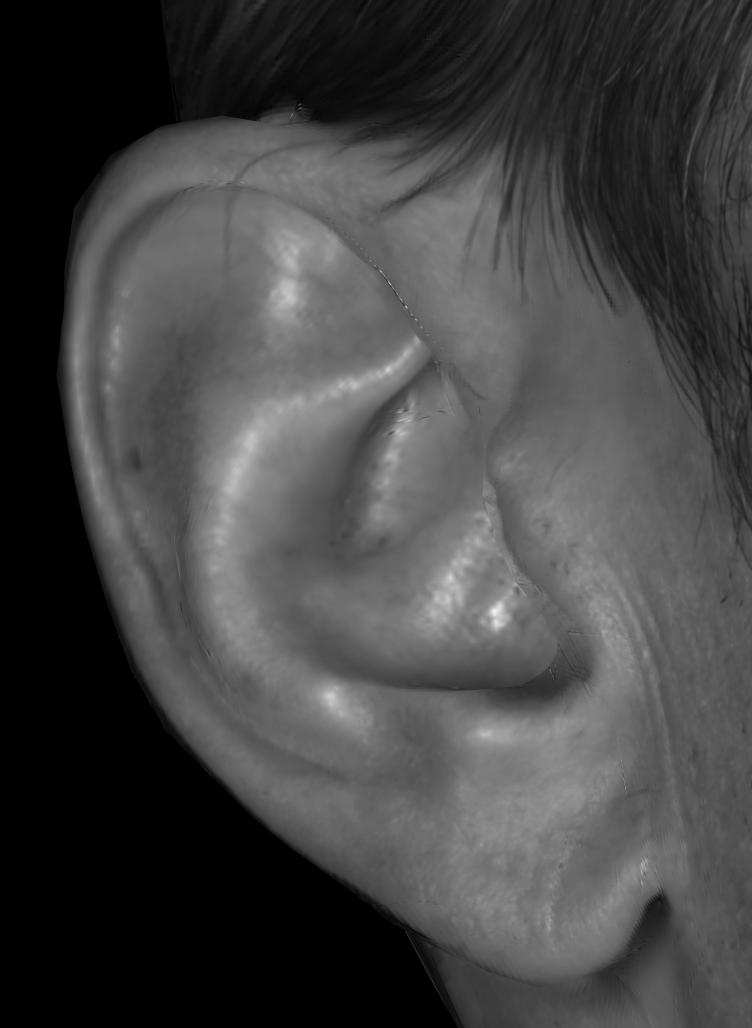}\\
\diffhlabel{Difference, Our}%
\diffimg{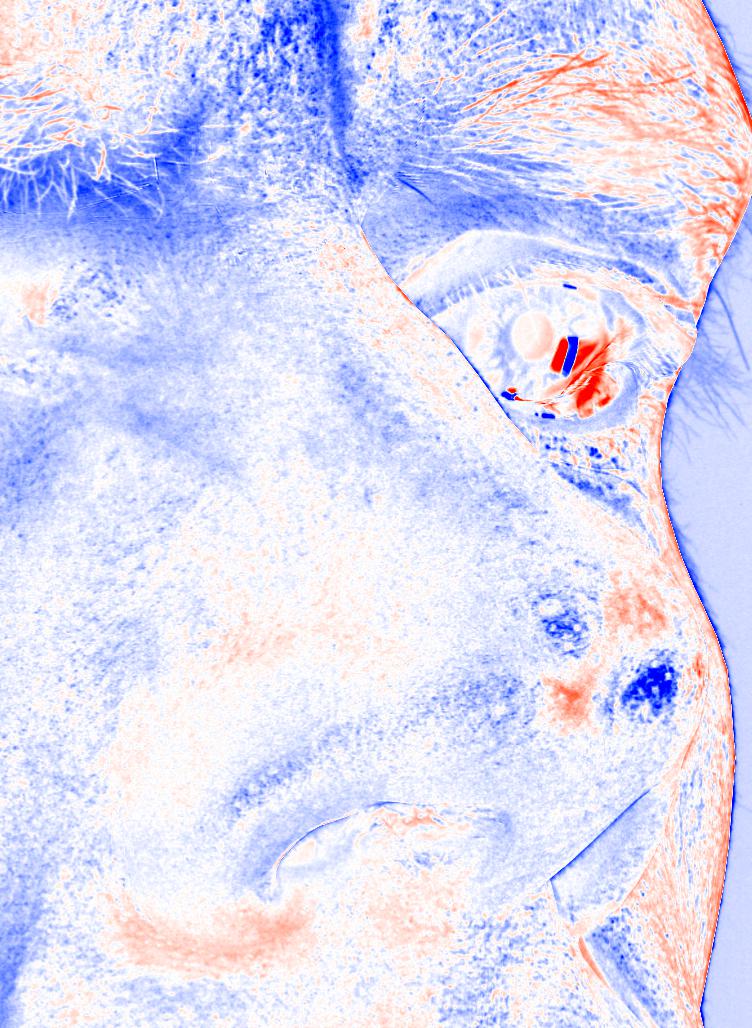}\hfill%
\diffimg{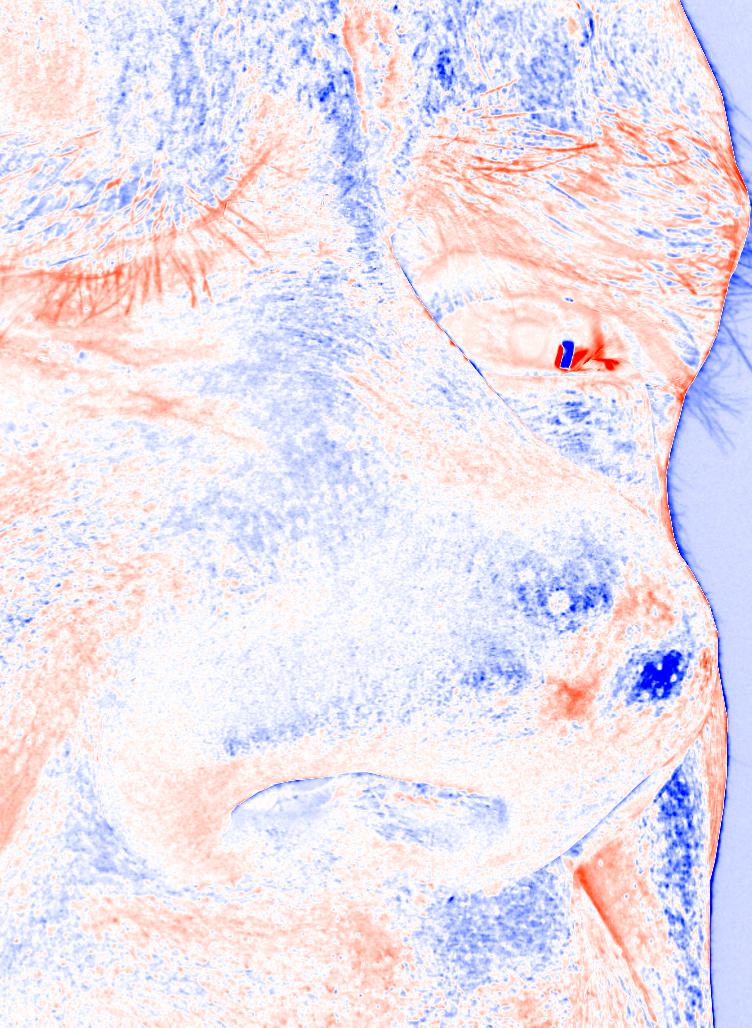}\hfill%
\diffimg{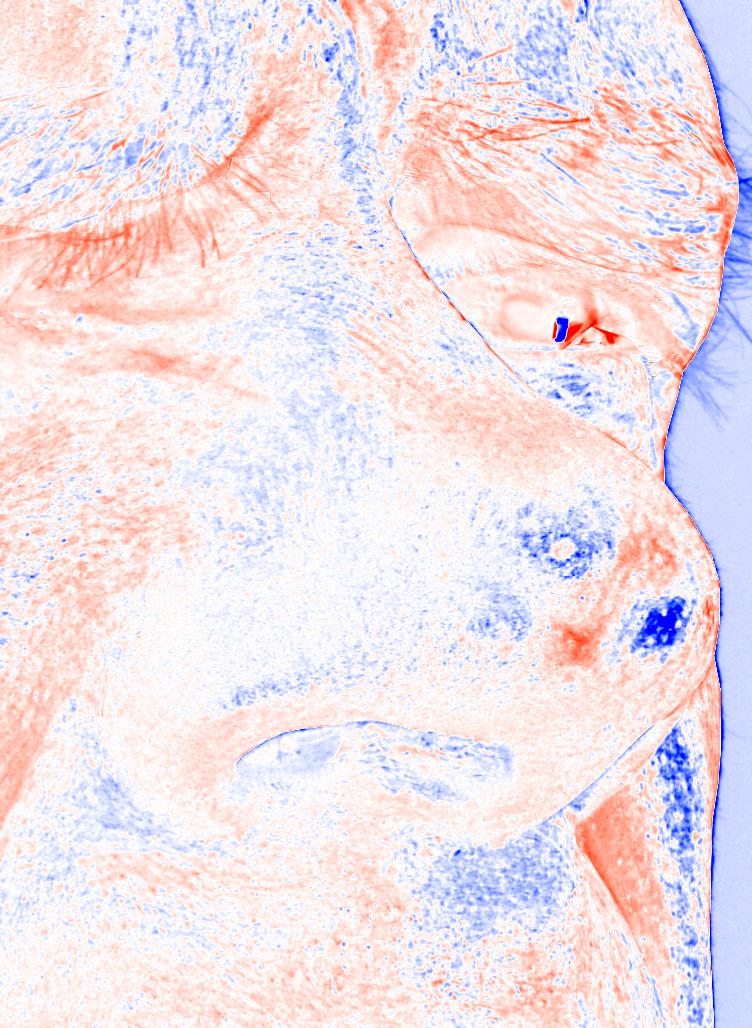}\hfill%
\diffimg{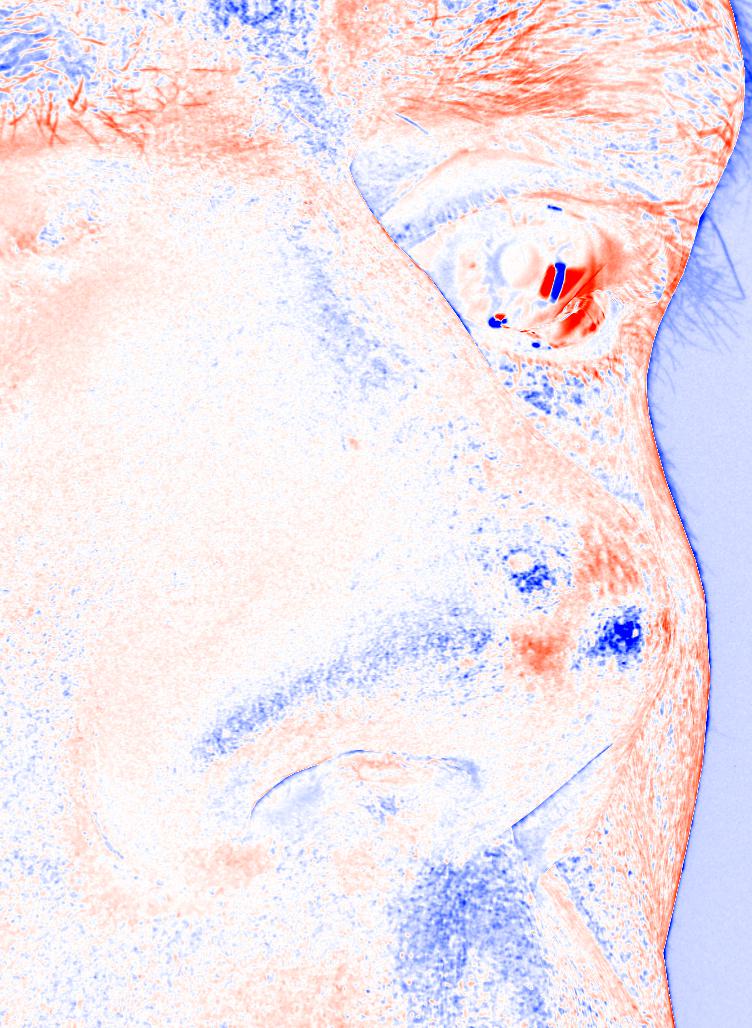}\hfill%
\diffimg{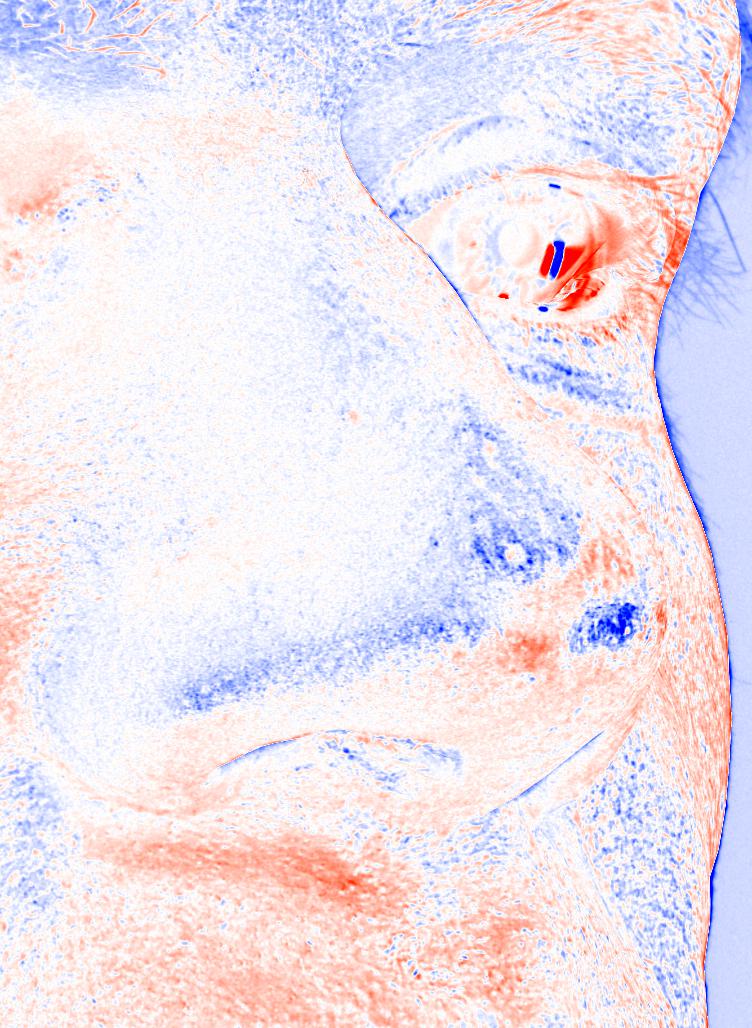}\hfill%
\diffimg{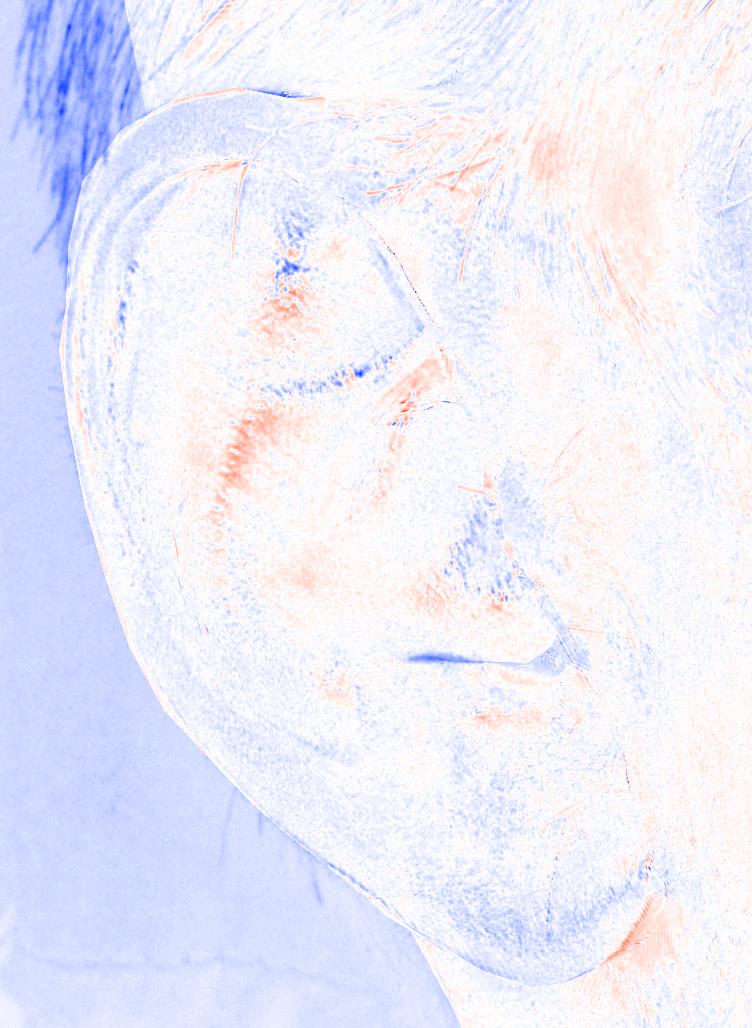}\\
\diffhlabel{Difference, DI4D}%
\diffimg{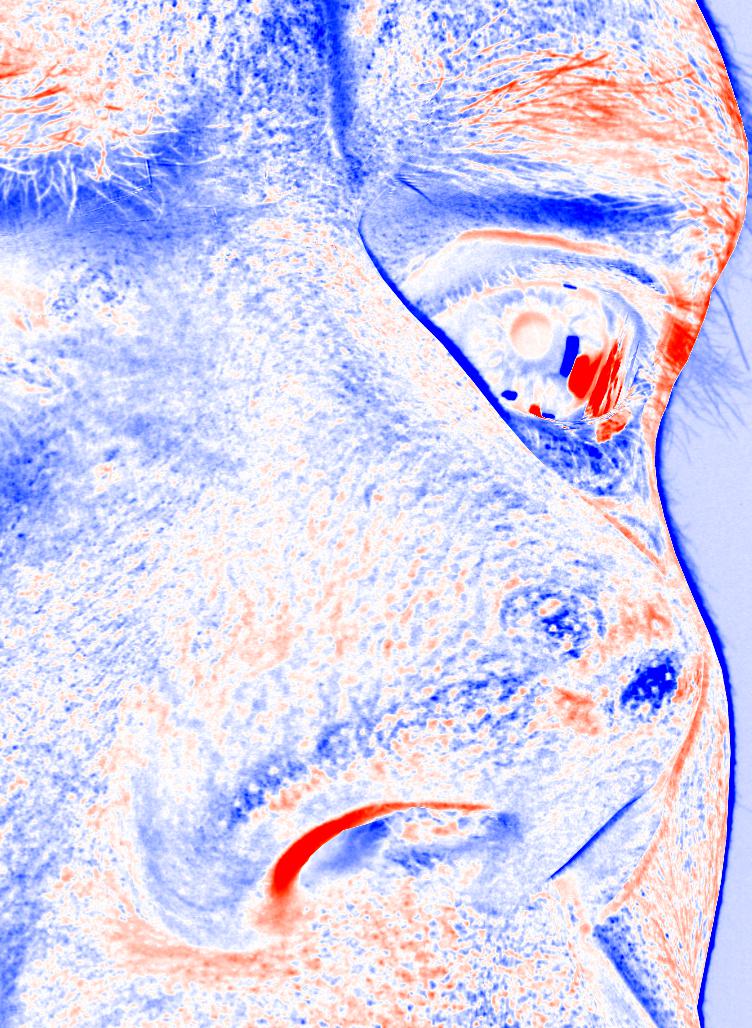}\hfill%
\diffimg{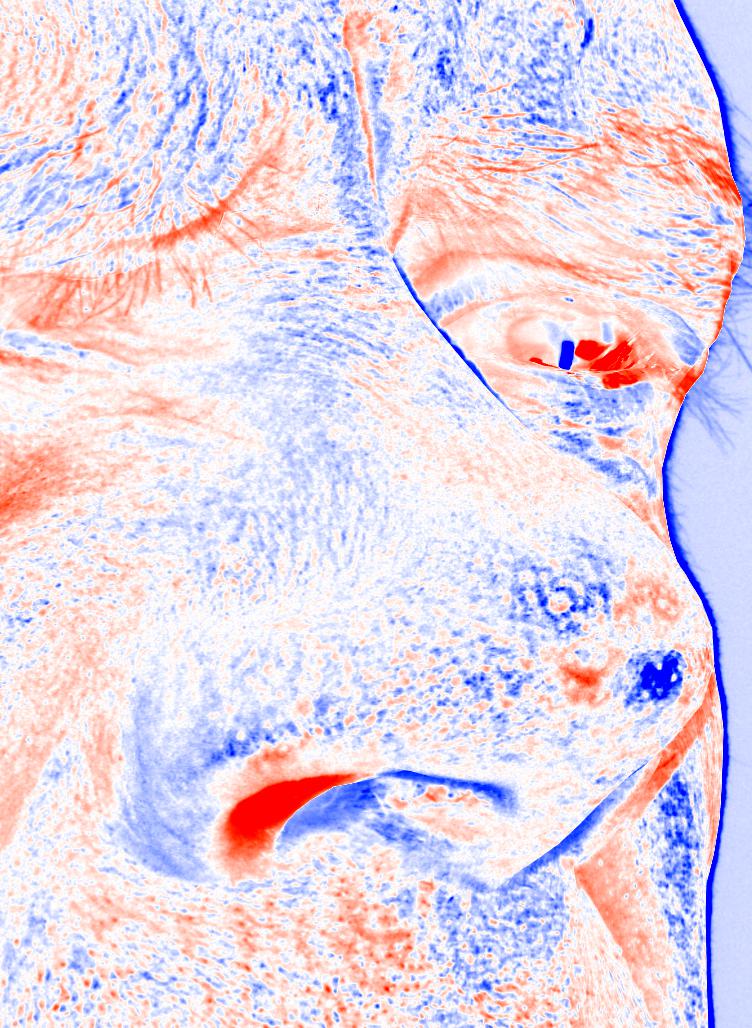}\hfill%
\diffimg{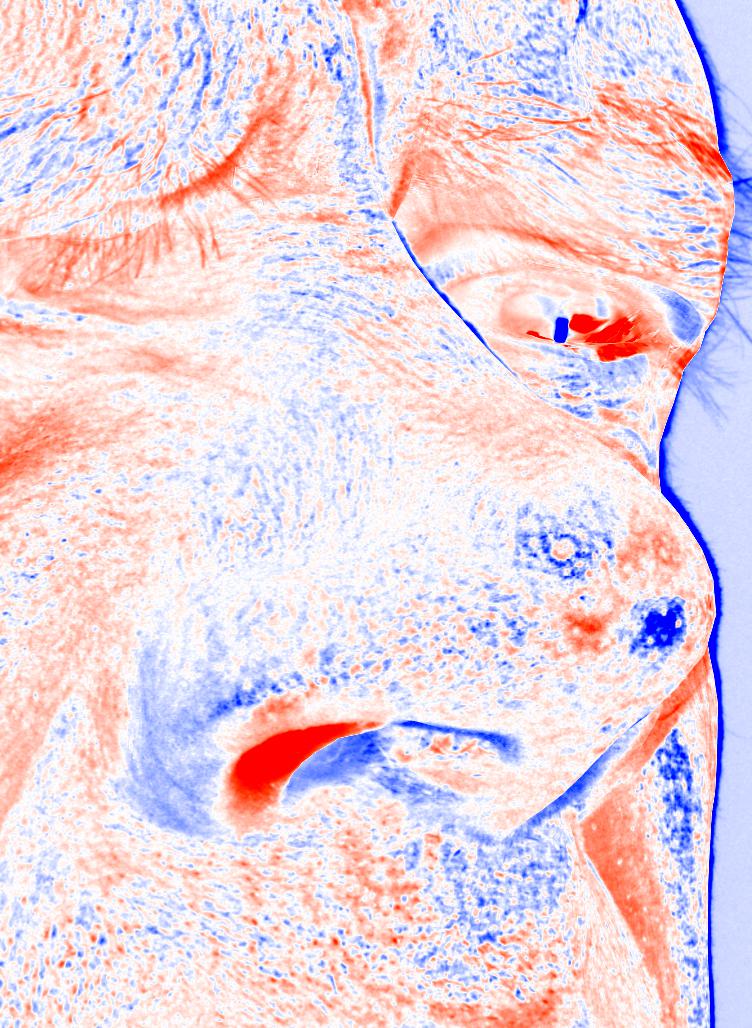}\hfill%
\diffimg{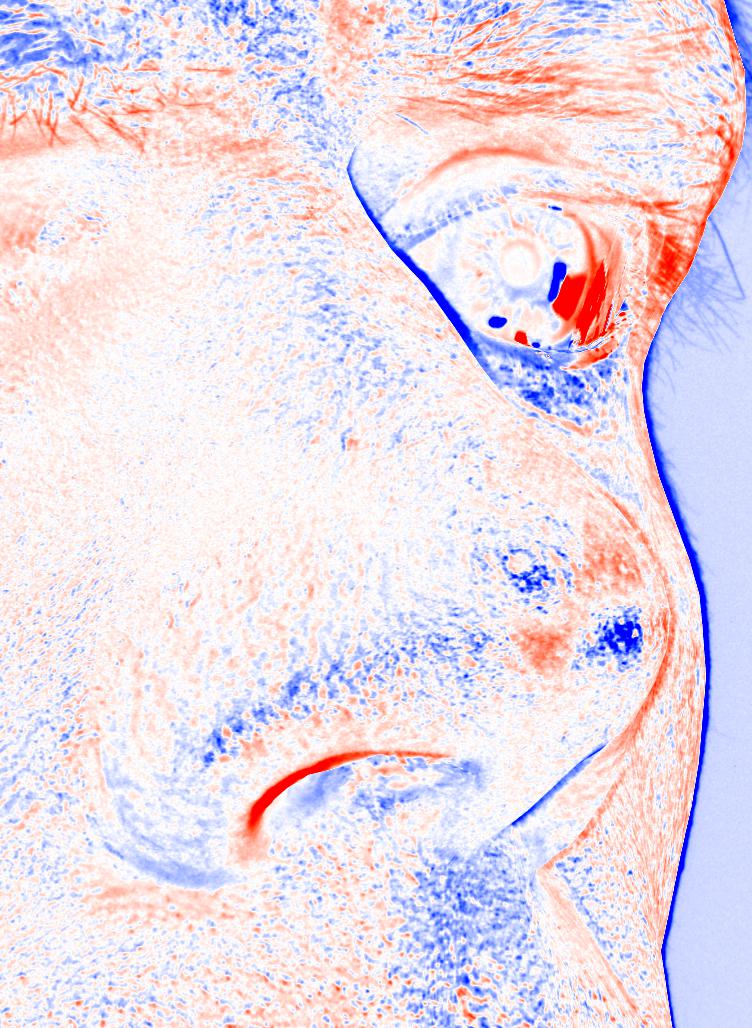}\hfill%
\diffimg{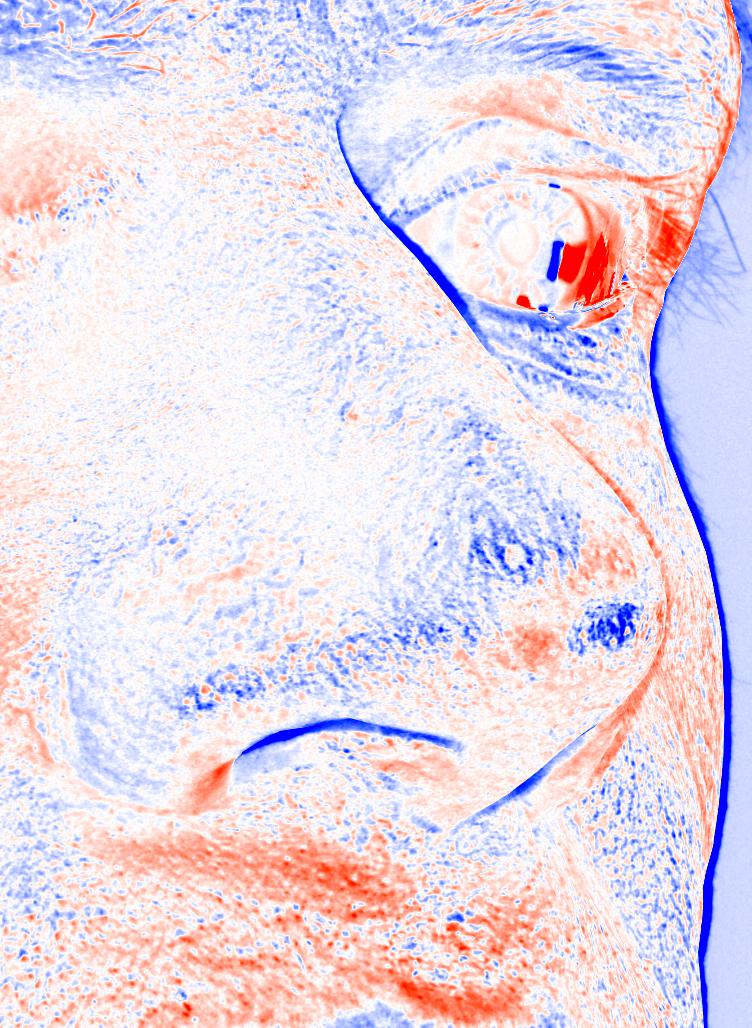}\hfill%
\diffimg{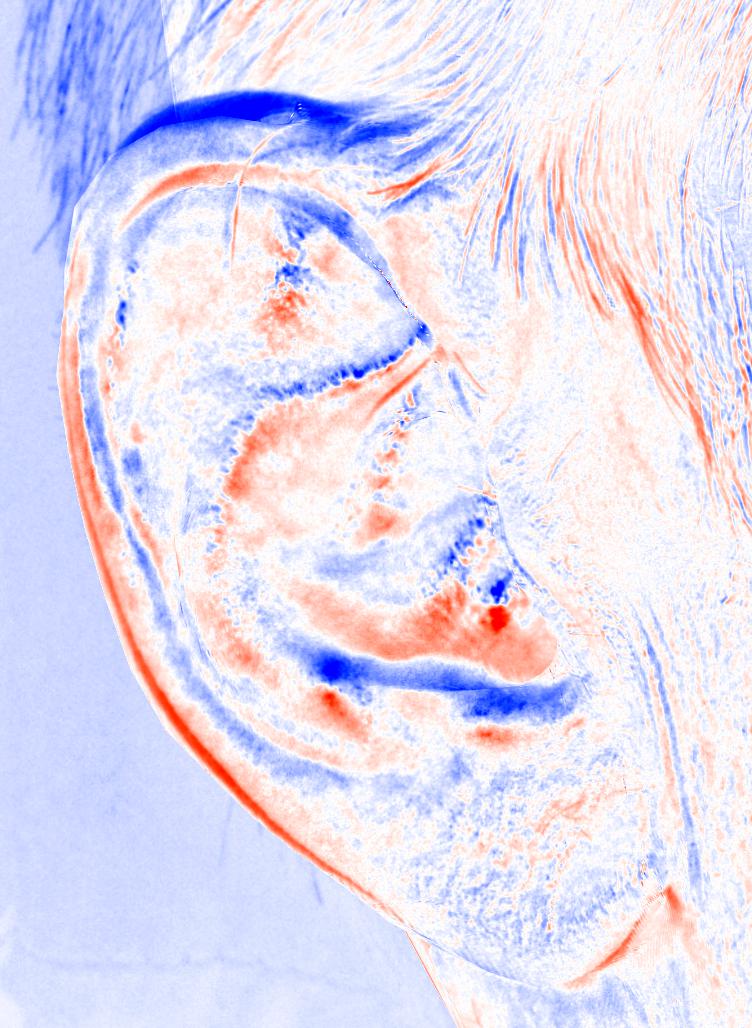}\vspace*{-.5em}\\
\caption{\label{fig:diff}%
Closeup of the nose reveals significant motion and deformation.
Our solution reproduces the changes in geometry faithfully, while the solution from DI4D~\protect\shortcite{di4d} severely attenuates the deformations.
False-color difference images (red/blue = brighter/darker than reference) highlight the geometric discrepancies around the nostrils.
In our solution, the geometric silhouettes are located correctly and the differences are only due to defocus blur in the reference images that is lacking in our renderings.
Right: The DI4D solution did not attempt to recover the motion of ears. 
We placed no constraints on which vertices are allowed to move during optimization, and thus also captured this motion.
}
\end{figure}
}

\newcommand{\meshbox}[1]{\makebox[0.195\linewidth][c]{#1}}
\newcommand{\meshimg}[1]{\includegraphics[width=0.195\linewidth]{figures/mesh/#1}}
\newcommand{\figmesh}{
\begin{figure*}
\centering\small%
\meshimg{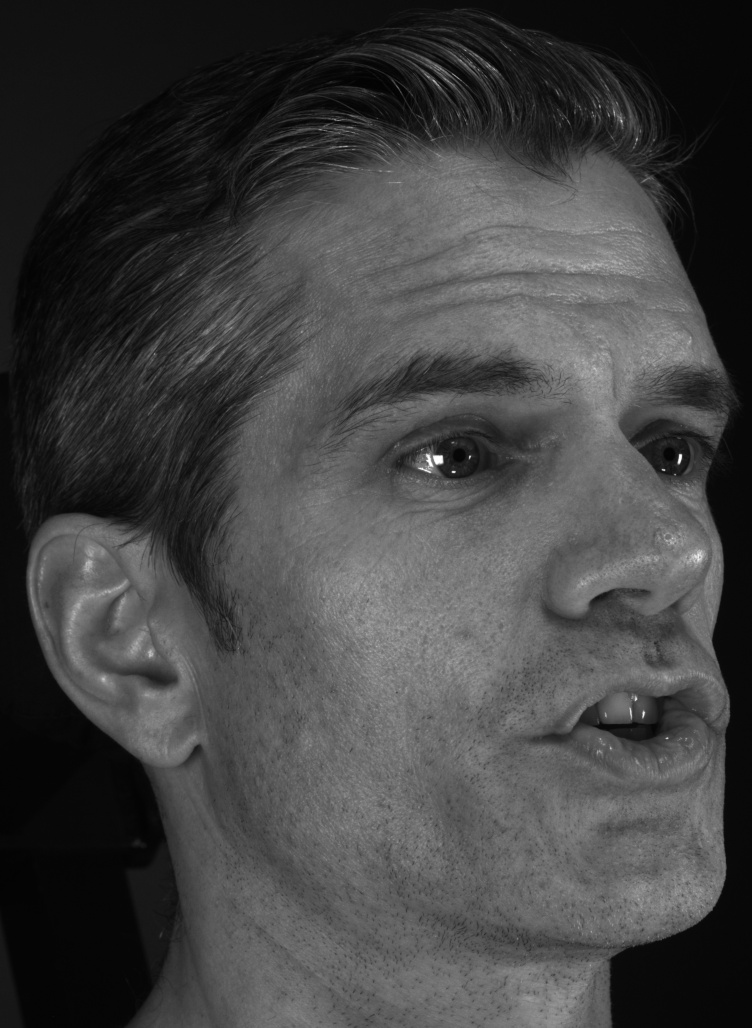}\hfill%
\meshimg{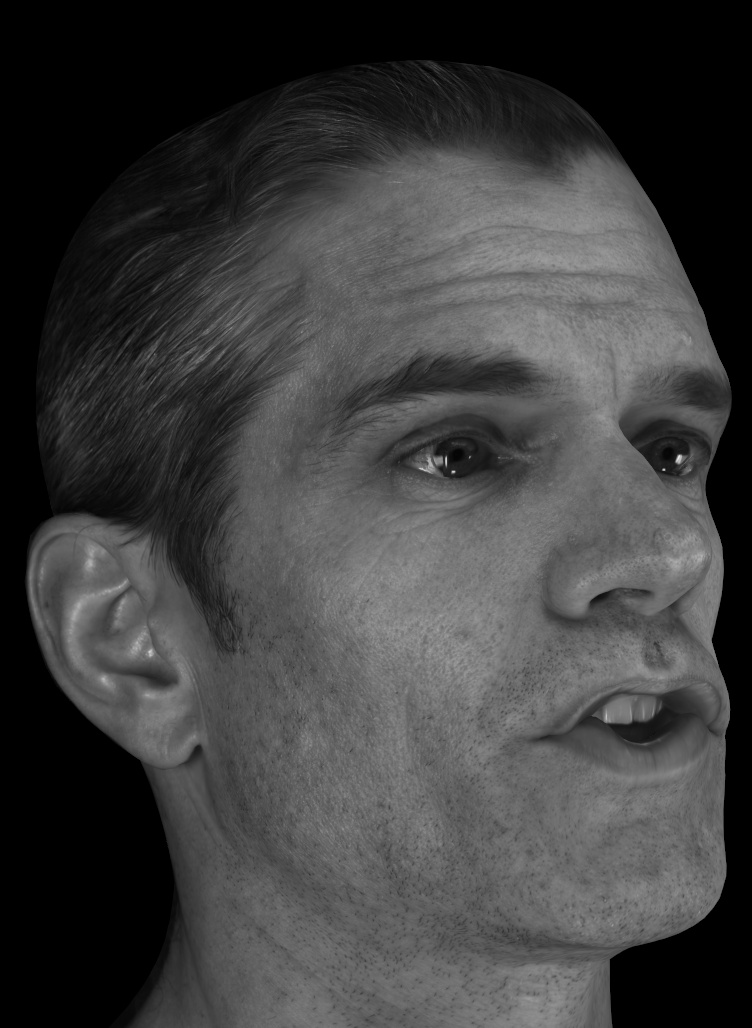}\hfill%
\meshimg{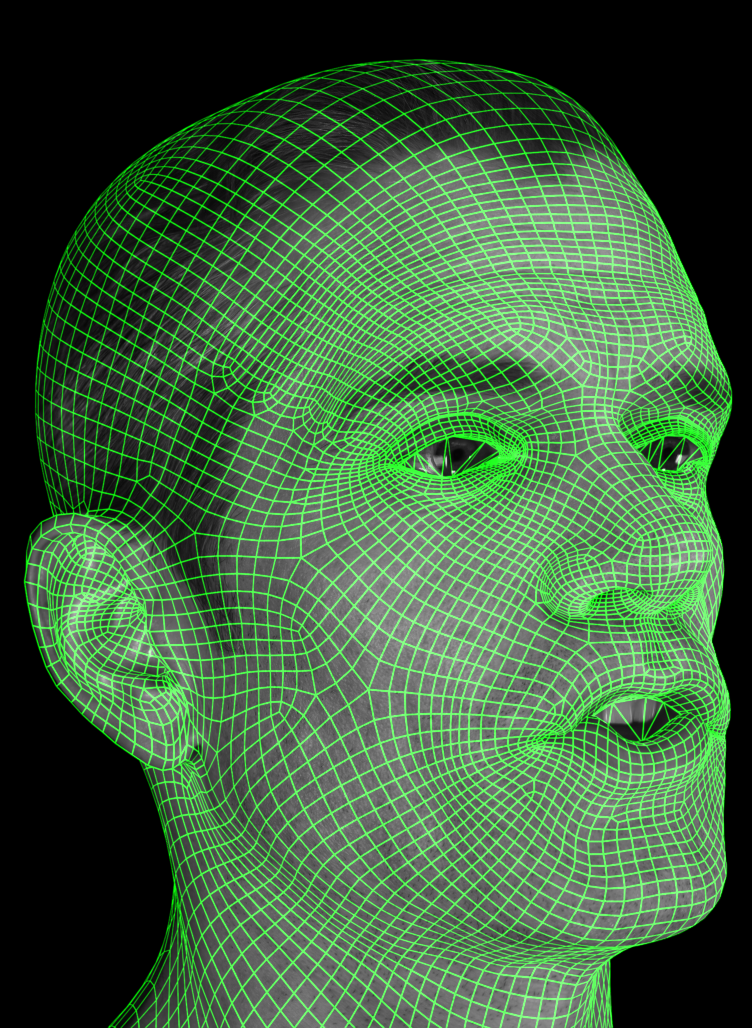}\hfill%
\meshimg{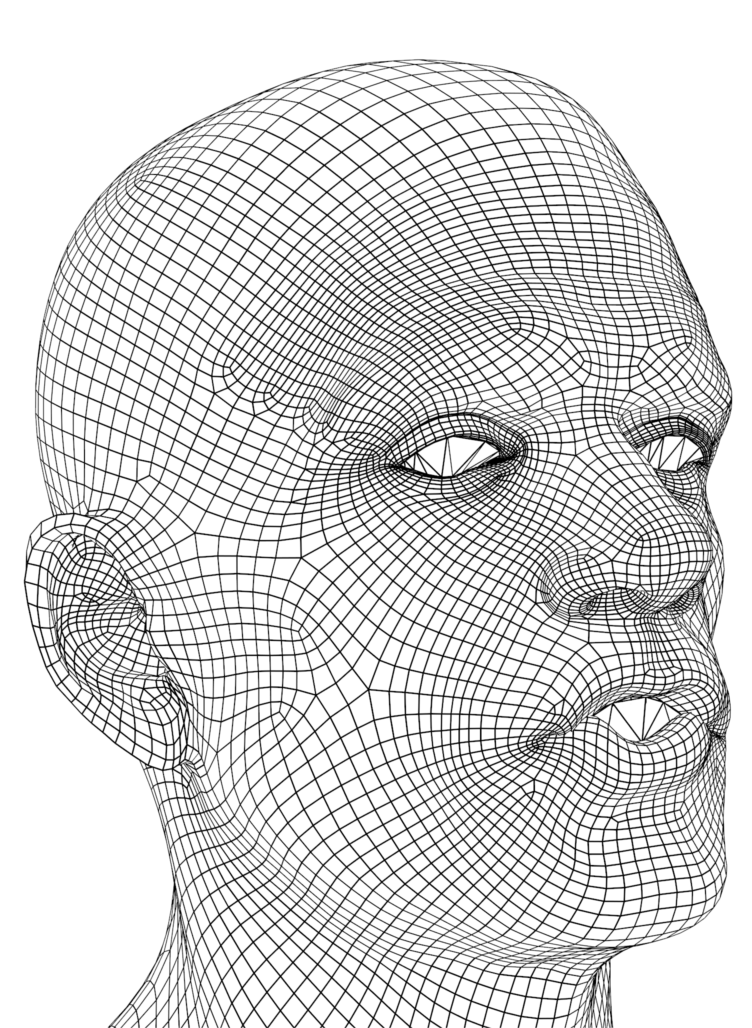}\hfill%
\meshimg{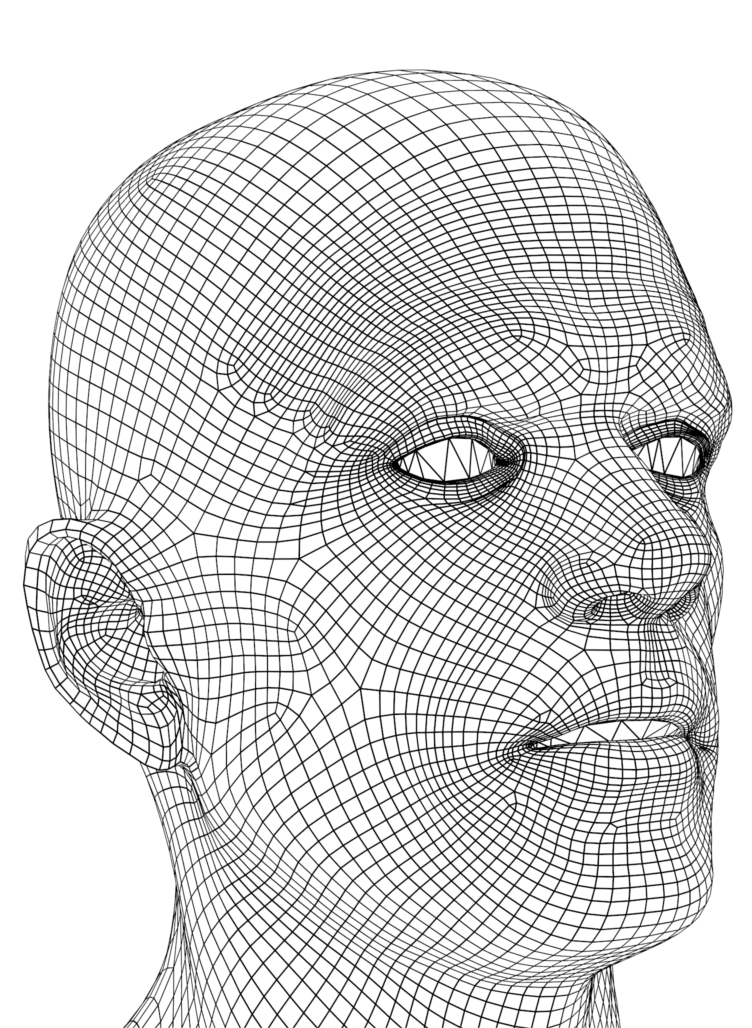}\\
\meshbox{(a) Reference}\hfill%
\meshbox{(b) Our reconstruction}\hfill%
\meshbox{(c) With wireframe}\hfill%
\meshbox{(d) Wireframe only}\hfill%
\meshbox{(e) Base mesh}\\
\caption{\label{fig:mesh}%
An example frame from the reconstruction of a 89-frame sequence.
(a) One of the 9 camera images for the frame.
(b--d) Our reconstructed texture and geometry, rendered from the same viewpoint using the known camera parameters.
(e) Base mesh $\Vbase$ used as the starting point for optimization.
}
\end{figure*}
}

\newcommand{\comboimg}[1]{\includegraphics[width=0.321\linewidth]{figures/mesh/#1.jpg}}
\newcommand{\combohlabel}[1]{\hspace*{1.5mm}\raisebox{0.22\linewidth}{\rotatebox[origin=b]{90}{\smash{\footnotesize{#1}}}\hspace*{0.5mm}}}
\newcommand{\figcombo}{
\begin{figure}
\centering\footnotesize%
\noraise{\combohlabel{}}\hfill%
\makebox[0.321\linewidth][c]{``Neutral'' (89 frames)}\hfill%
\makebox[0.321\linewidth][c]{``Disgust'' (123 frames)}\hfill%
\makebox[0.321\linewidth][c]{``Anger'' (207 frames)}\\
\combohlabel{Reference}\hfill%
\comboimg{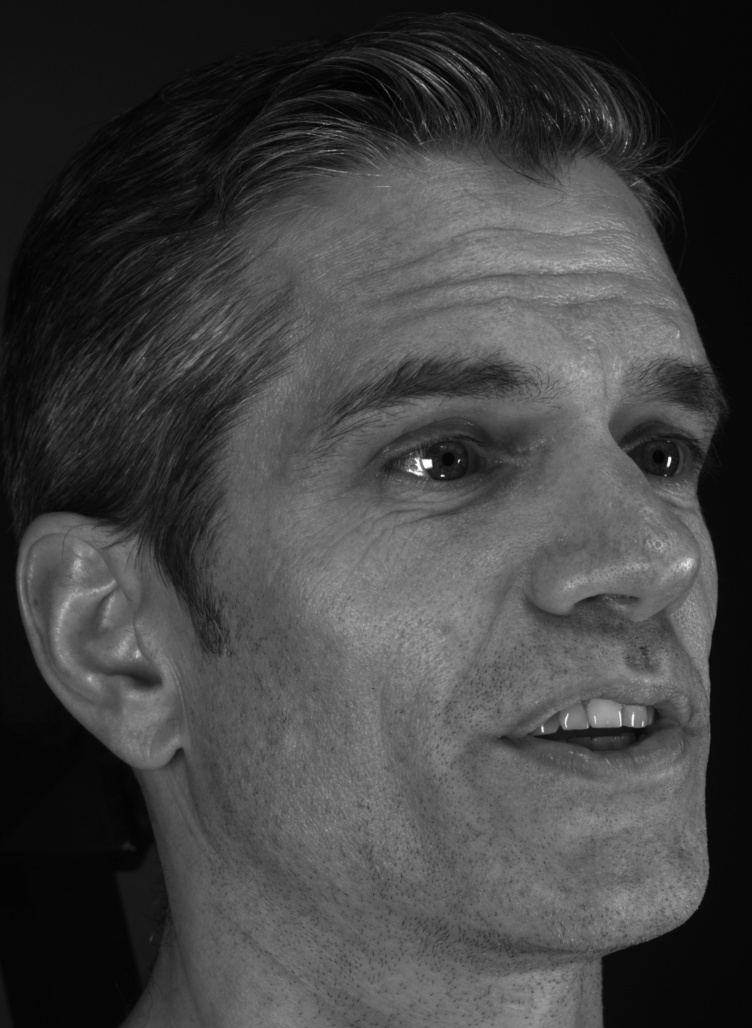}\hfill%
\comboimg{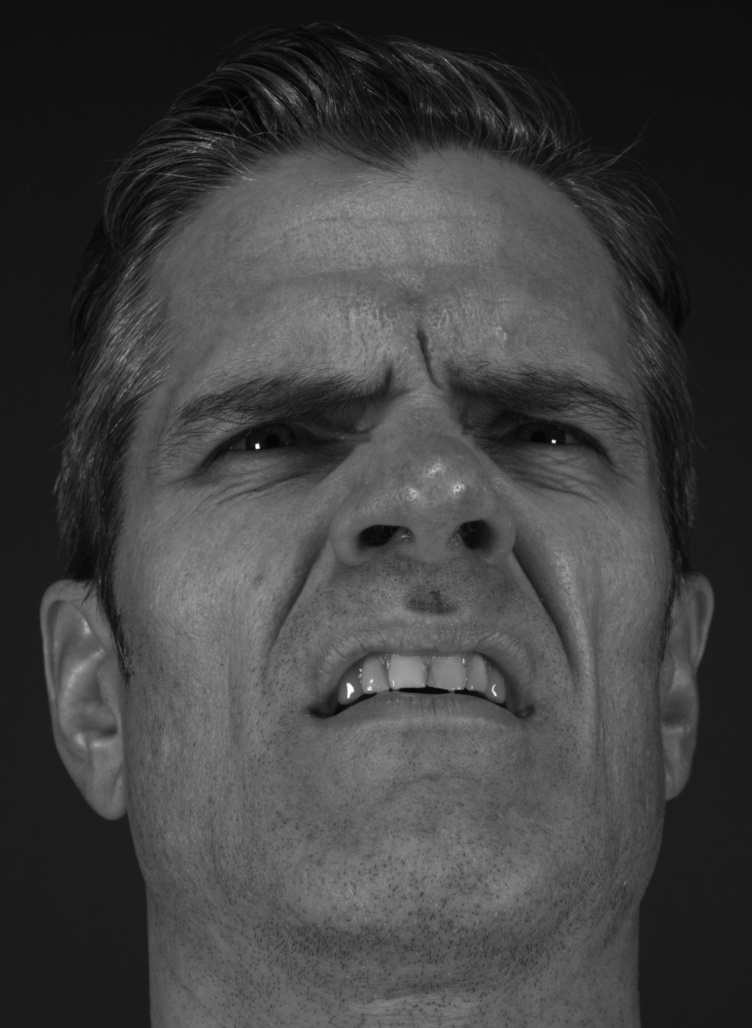}\hfill%
\comboimg{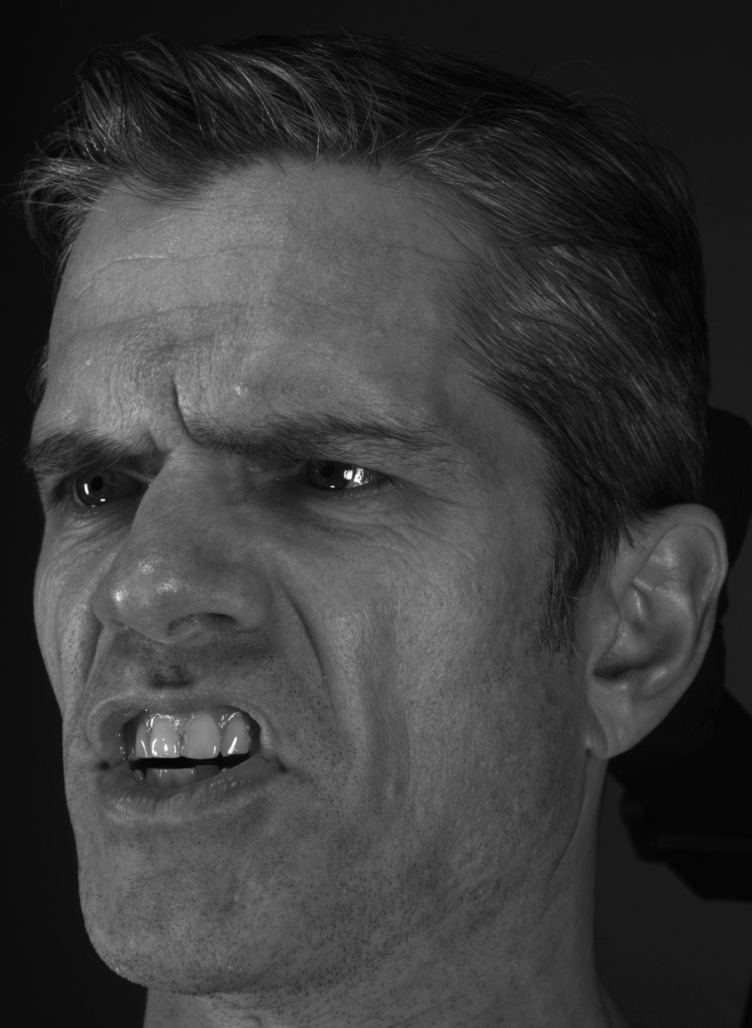}\\
\combohlabel{Our reconstruction}\hfill%
\comboimg{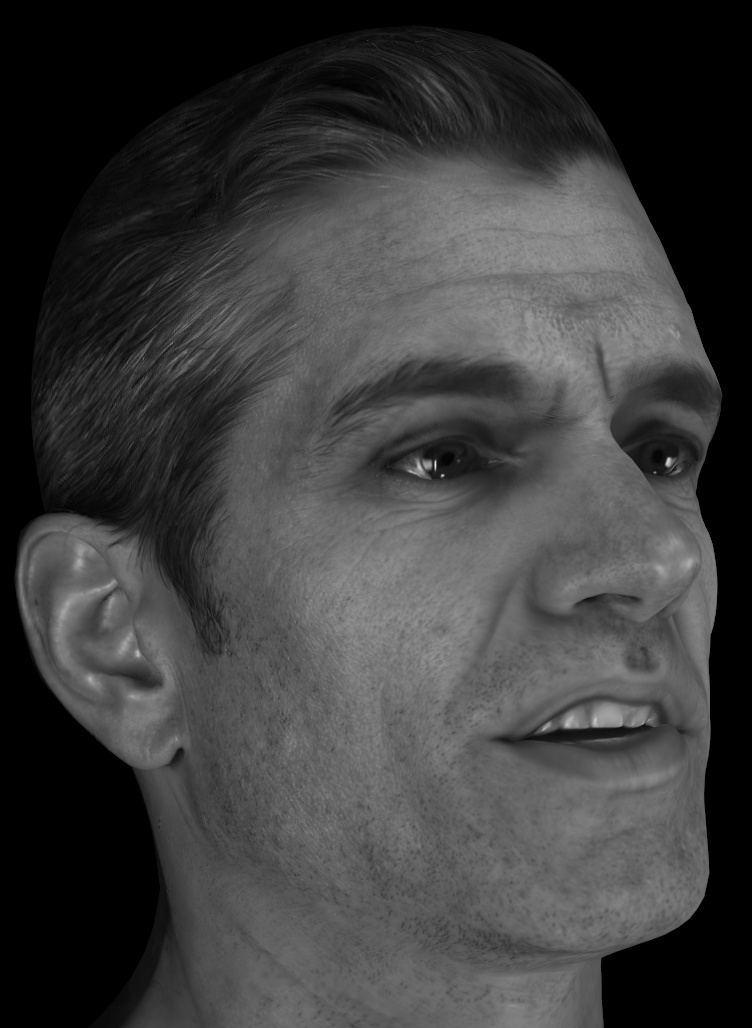}\hfill%
\comboimg{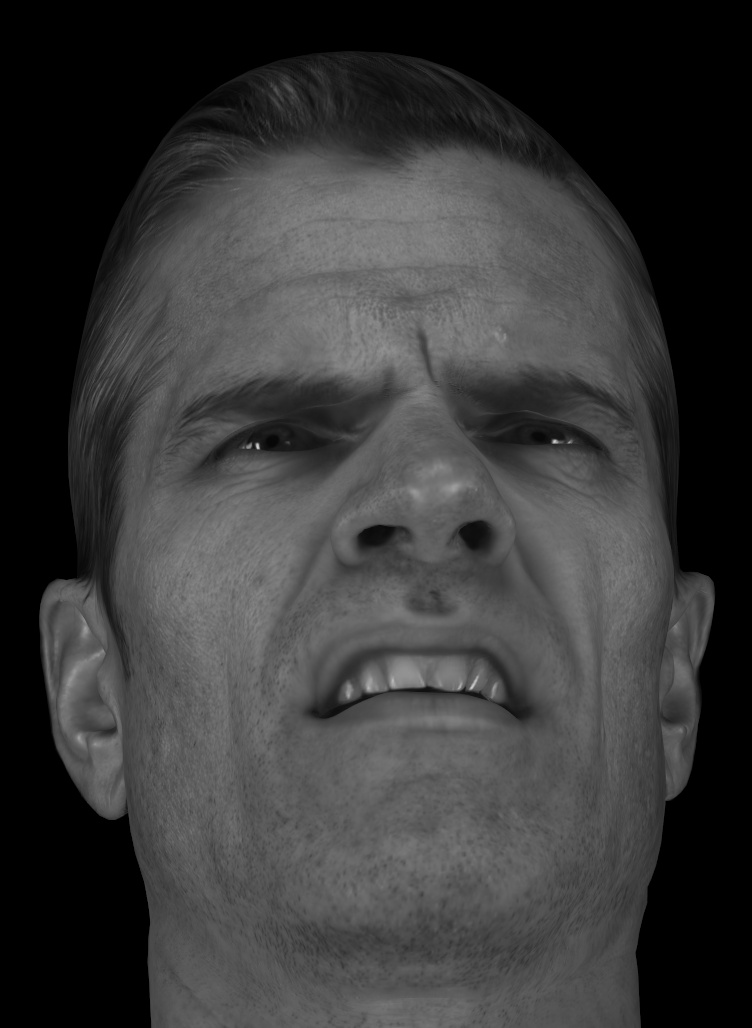}\hfill%
\comboimg{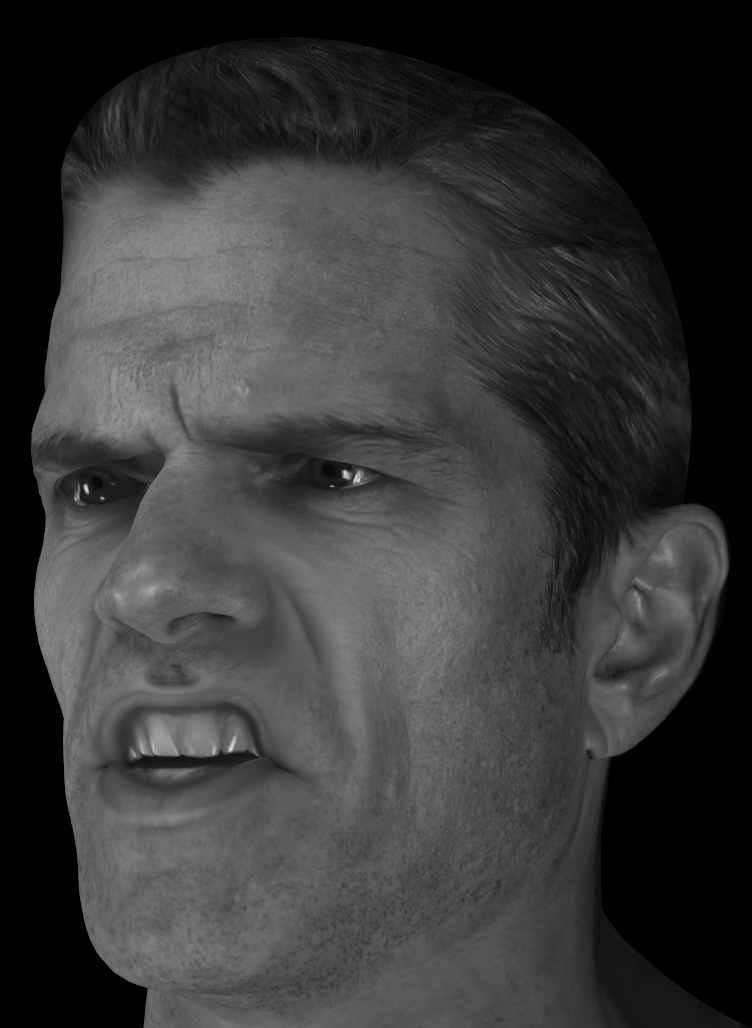}\\
\caption{\label{fig:combo}%
Example reconstructions from the optimization of three sequences as a single 419-frame sequence, starting from the same base mesh as shown in Figure~\protect\ref{fig:mesh}e.
Because a single texture is solved for all frames, the vertex to skin correspondence is consistent across all sequences.
}
\end{figure}
}

\newcommand{\cubeimg}[1]{\includegraphics[width=0.239\linewidth]{figures/cube/#1}}
\newcommand{\cubebox}[1]              {\makebox[0.239\linewidth][c]{#1}}
\newcommand{\cubehlabel}[1]{\hspace*{2mm}\raisebox{0.120\linewidth}{\rotatebox[origin=b]{90}{\smash{\footnotesize{#1}}}}\hfill\hfill}
\newcommand{\figcube}{
\begin{figure}
\centering\footnotesize%
\raisebox{0mm}[0mm][.5mm]{\cubehlabel{}}\hfill%
\cubebox{Iteration 100}\hfill%
\cubebox{Iteration 1000}\hfill%
\cubebox{Iteration 5000}\hfill%
\cubebox{Final mesh}\\
\cubehlabel{\res{16} pixels}\hfill%
\cubeimg{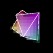}\hfill%
\cubeimg{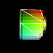}\hfill%
\cubeimg{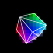}\hfill%
\cubeimg{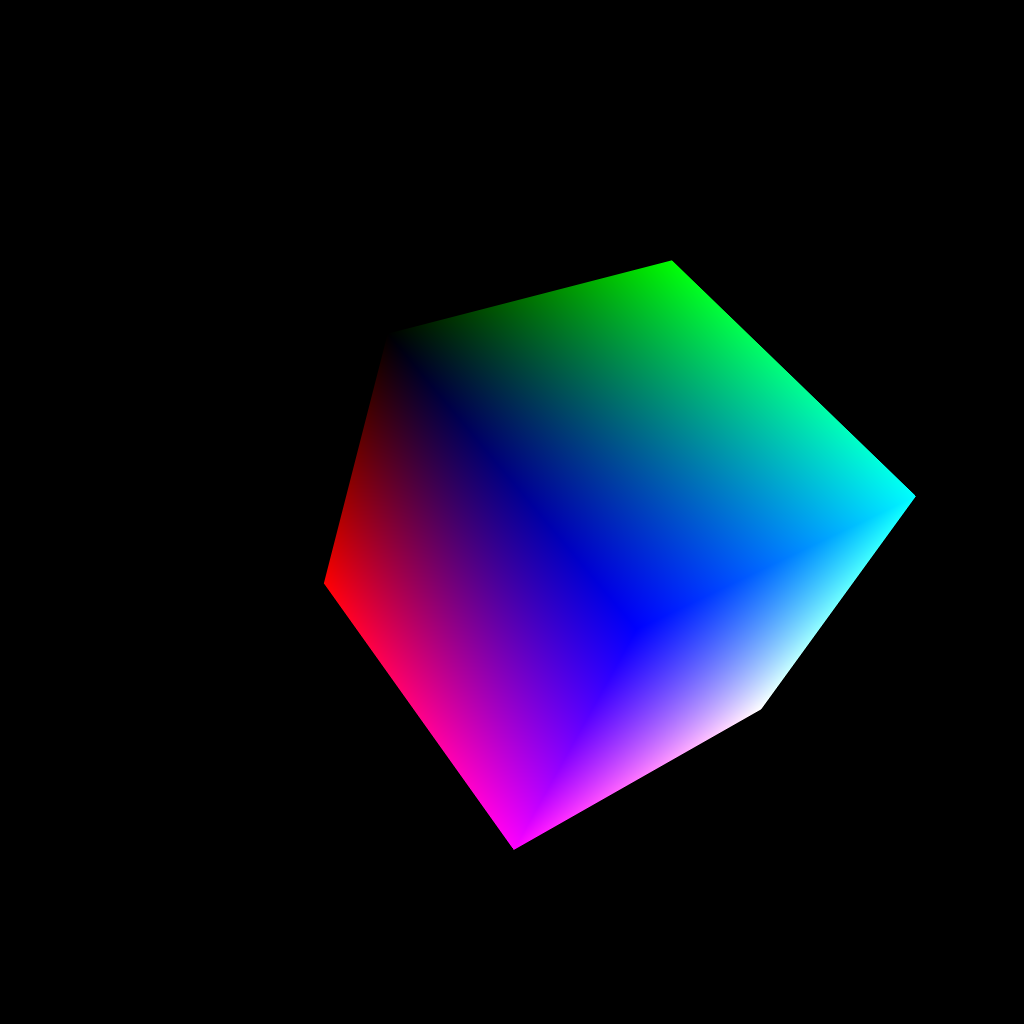}\\%
\cubehlabel{\res{8} pixels}\hfill%
\cubeimg{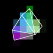}\hfill%
\cubeimg{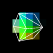}\hfill%
\cubeimg{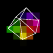}\hfill%
\cubeimg{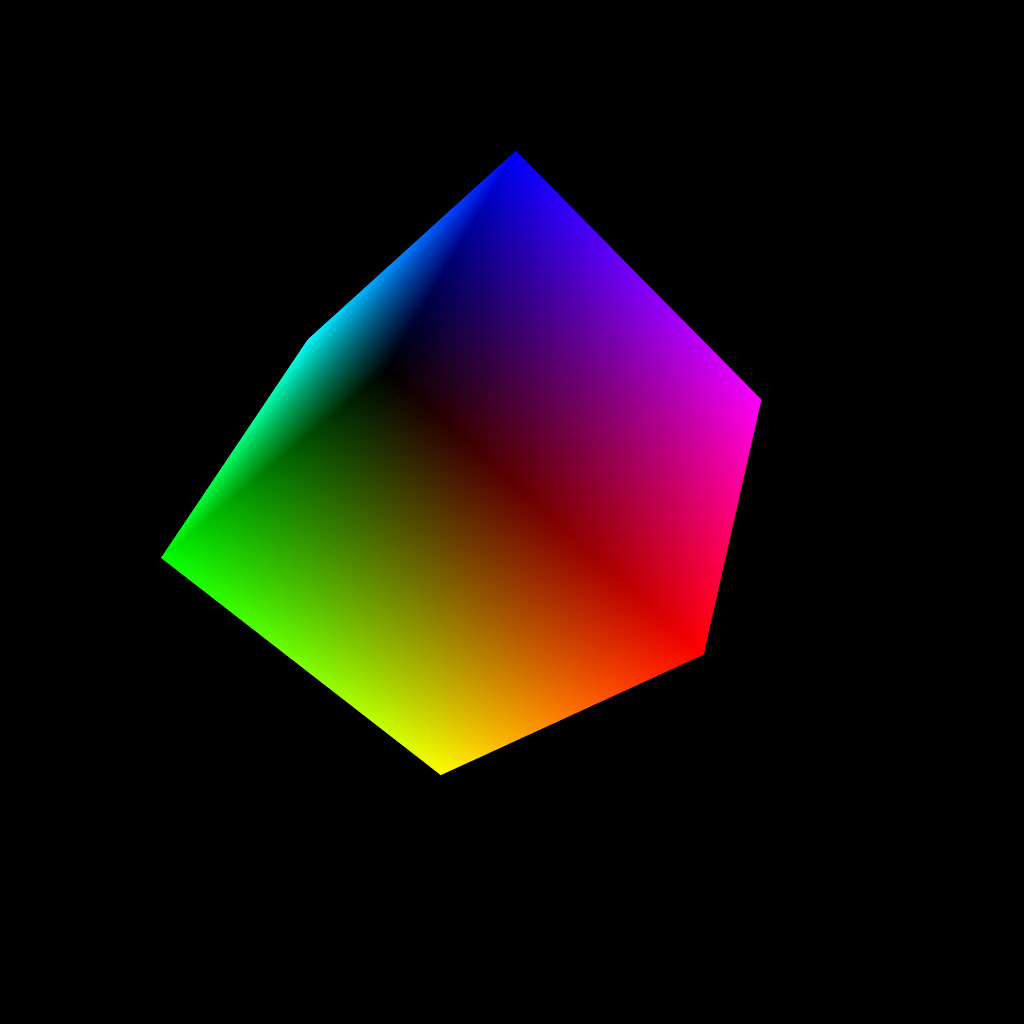}\\%
\cubehlabel{\res{4} pixels}\hfill%
\cubeimg{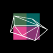}\hfill%
\cubeimg{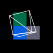}\hfill%
\cubeimg{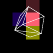}\hfill%
\cubeimg{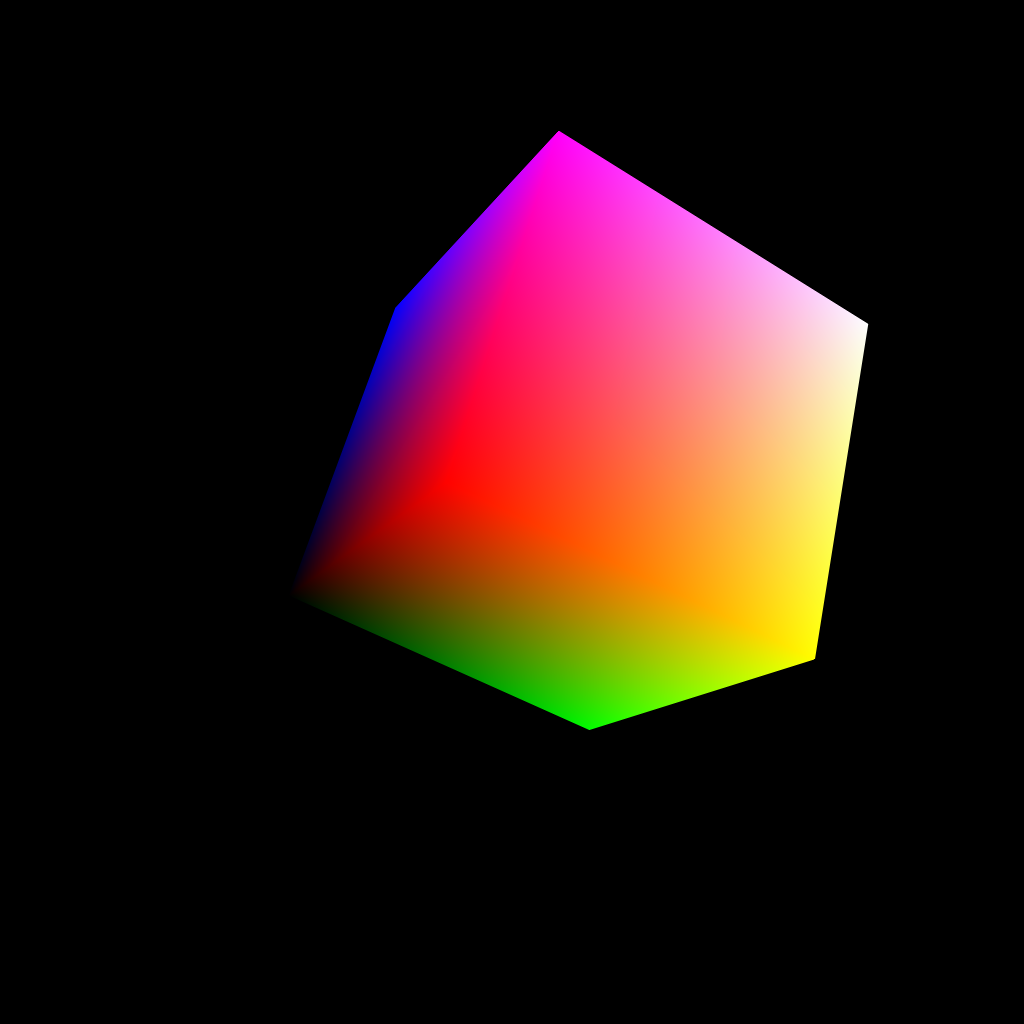}\vspace*{-.5em}\\%
\caption{\label{fig:cube}%
To validate that our visibility gradients provide useful information even for small triangles, we infer vertex positions and colors of a simple mesh in extremely small resolutions.
The geometry of the current solution is superimposed on the rasterized images for illustration purposes only.
Rightmost column shows the final, optimized mesh rendered in high resolution.
In \res{4} resolution, the average triangle area is only 0.54 pixels.
The optimization nonetheless converges to the correct solution, albeit slower than in higher resolutions.
In \res{2} resolution the optimization fails to converge.
}
\end{figure}
}

\newcommand{\figcubeplot}{
\begin{figure}
\centering\footnotesize
\includegraphics[width=0.9\linewidth]{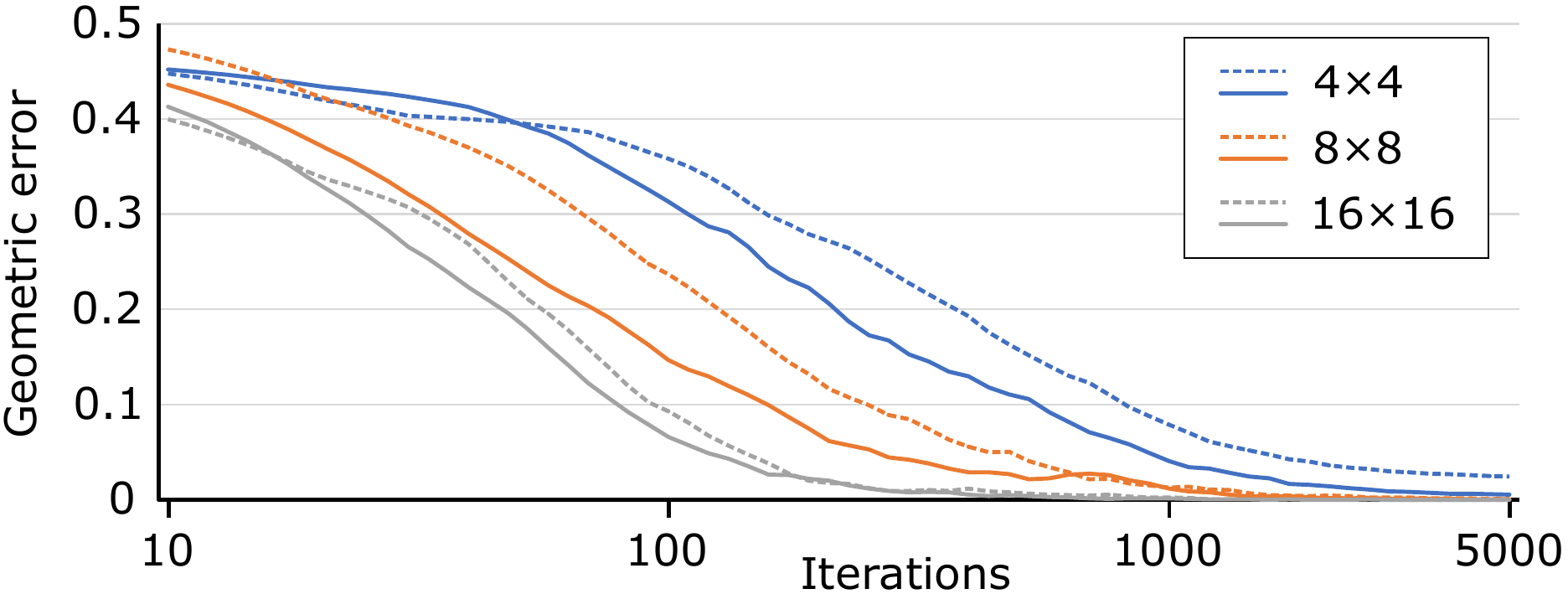}\hphantom{xxxx}\vspace*{-.5em}\\
\caption{\label{fig:cubeplot}%
Convergence of the cube shape and color optimization test (average of 10 successful optimizations).
Vertical axis shows the average distance between vertices and their true positions in the unit cube.
The solid curves indicate convergence in the continuous coloring mode (Figure~\protect\ref{fig:cube}), and the dashed curves correspond to the discontinuous coloring mode.
As expected, the latter is somewhat more difficult to optimize.
Note the logarithmic horizontal axis.
}
\end{figure}
}

\newcommand{\esp}{\hspace*{2.6mm}}
\newcommand{\figenv}{
\begin{figure}
\centering\footnotesize
\begin{tabular}{@{}c@{\esp}c@{\esp}c@{\esp}c@{\esp}c@{}}
\includegraphics[width=0.18\linewidth]{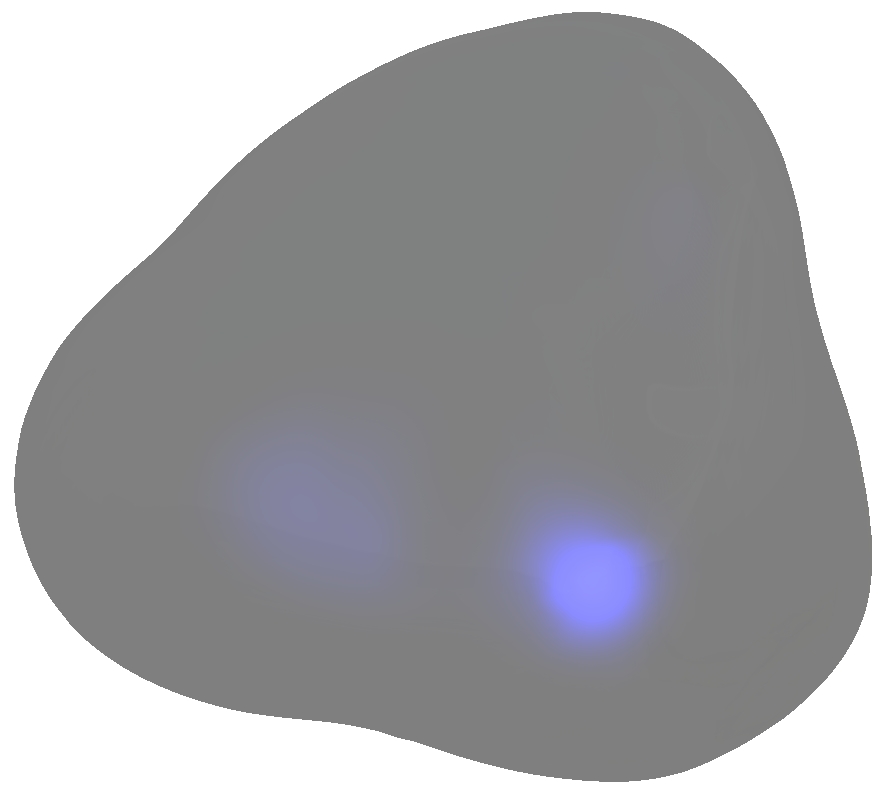}&%
\includegraphics[width=0.17\linewidth]{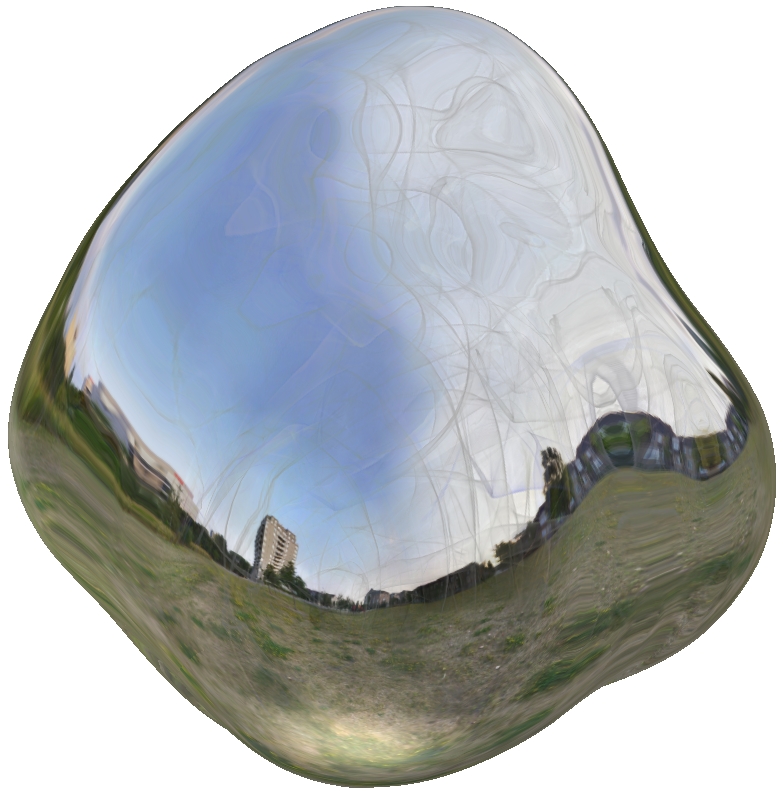}&%
\includegraphics[width=0.16\linewidth]{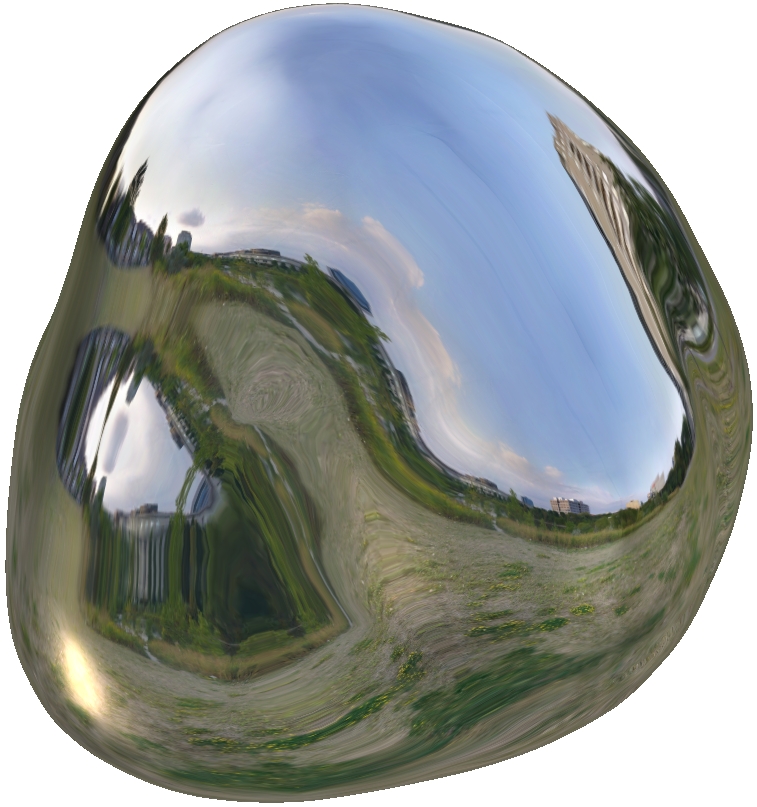}&%
\includegraphics[width=0.17\linewidth]{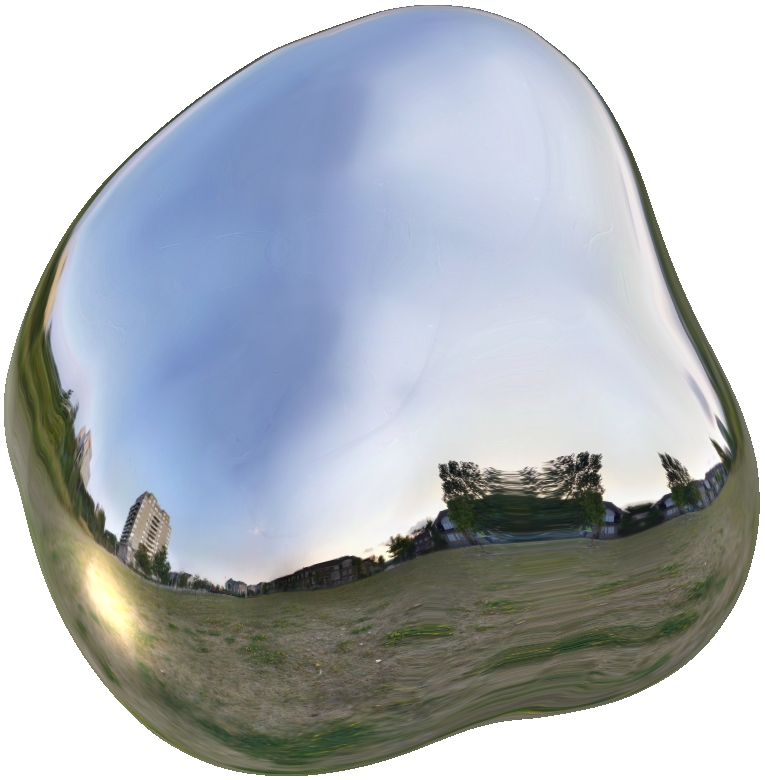}&%
\includegraphics[width=0.17\linewidth]{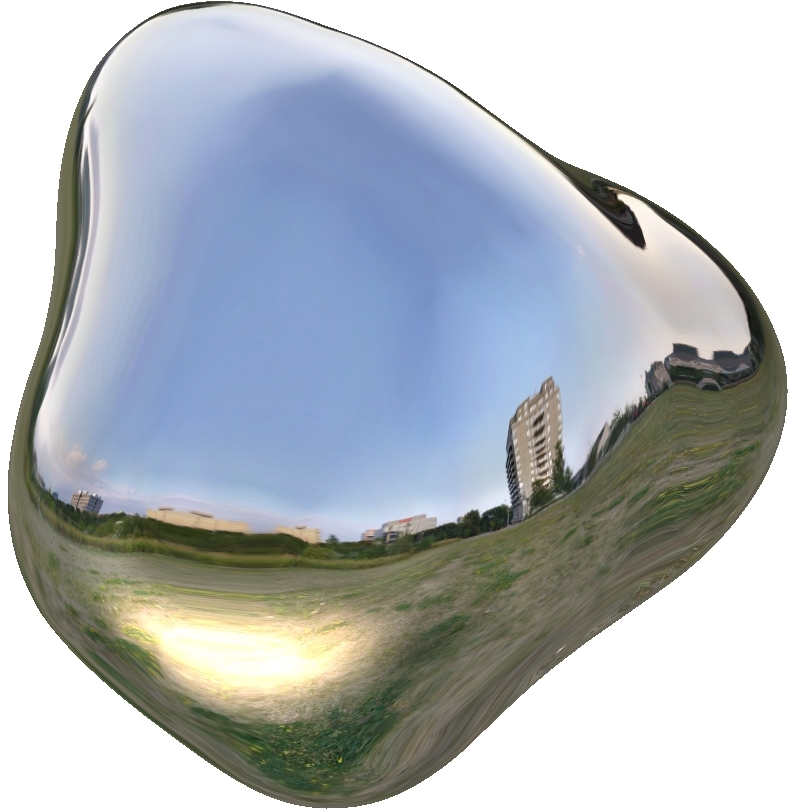}\\
Initialization&100 iter.&400 iter.&700 iter.&1000 iter.\\
\end{tabular}\\
\vphantom{1mm}%
\includegraphics[width=1.0\linewidth]{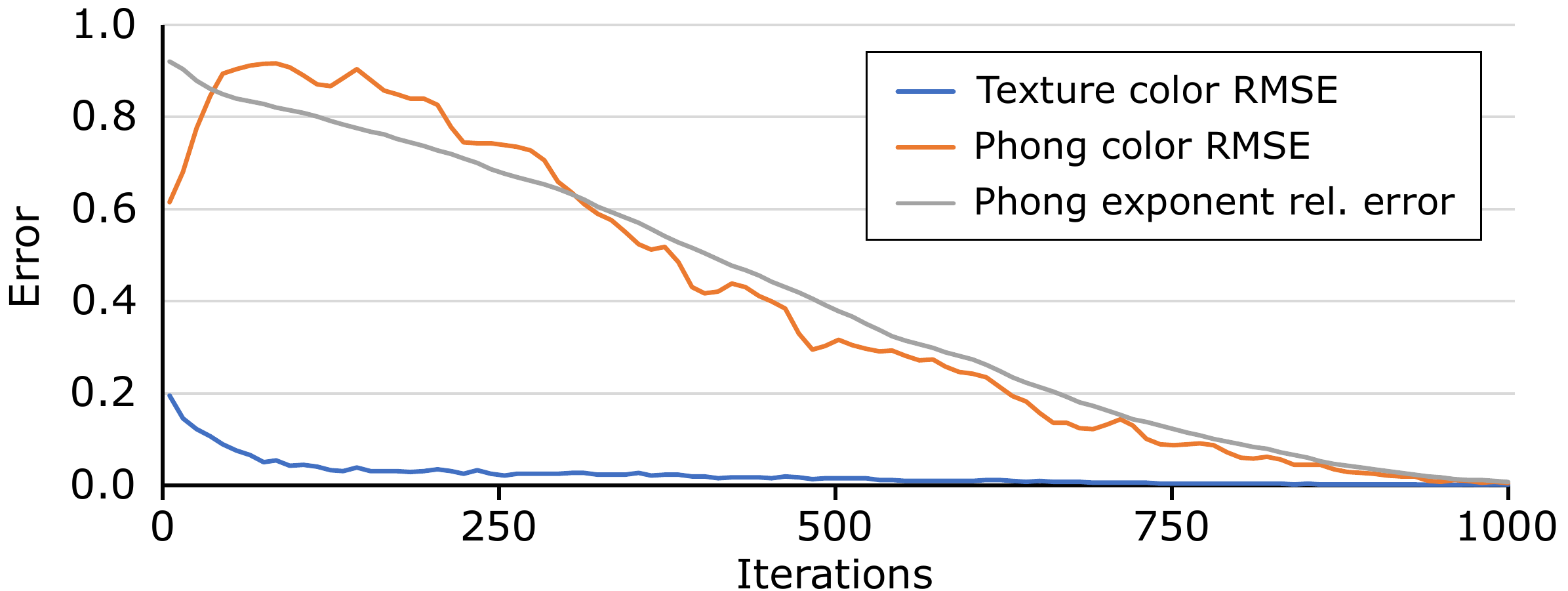}\\
\caption{\label{fig:env}%
Optimizing environment map texture and Phong BRDF parameters in a synthetic test case.
Top: Example renderings at various iteration counts.
The texture converges slightly unevenly due to the distribution of indirect texture lookups, as seen at 100 iterations.
Bottom: Convergence of the learned parameters over the course of optimization.
}
\end{figure}
}

\newcommand{\figpose}{
\begin{figure}
\centering\footnotesize
\begin{tabular}{@{}c@{}c@{}c@{\hspace*{0.016\linewidth}}c@{}c@{}c@{}}
\includegraphics[width=0.164\linewidth]{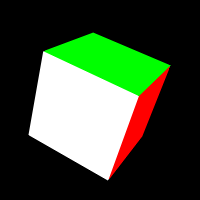}&%
\includegraphics[width=0.164\linewidth]{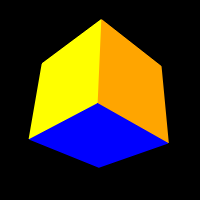}&%
\includegraphics[width=0.164\linewidth]{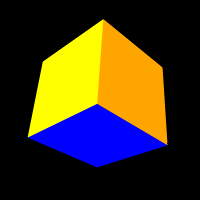}&%
\includegraphics[width=0.164\linewidth]{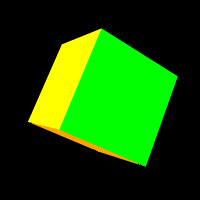}&%
\includegraphics[width=0.164\linewidth]{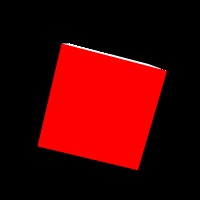}&%
\includegraphics[width=0.164\linewidth]{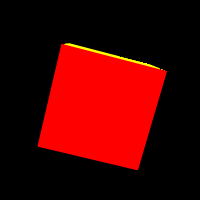}\\
Initial&Final&Reference&Initial&Final&Reference\vspace*{-1mm}
\end{tabular}
\caption{\label{fig:pose}%
Two example cases from the cube pose optimization test.
With trivial noise-based regularization, we obtain an average error of 48.62$^\circ$ which is an improvement over the 63.57$^\circ$ of Liu et al.~\protect\shortcite{softras}, indicating that the blur and transparency offered by SoftRas are not necessary in this task.
The average error is dominated by local minima where the pose looks correct but the colors are wrong (see example on the right with $\sim$180$^\circ$ final pose error\,---\,the yellow face should be on top instead of white).
Customizing the optimization method to suit the task better lowers the average error to 2.61$^\circ$.
Some cases are impossible due to only one face of the cube being visible in the reference image, but they are rare enough to not contribute significantly to the averages.
}
\end{figure}
}

\newcommand{\earthimg}[1]{\includegraphics[width=0.095\linewidth]{figures/earth/#1}}
\newcommand{\earthhbo}[1]{\raisebox{0.0475\linewidth}{\rotatebox[origin=b]{90}{\smash{\footnotesize{#1}}}}\hspace*{0.1mm}}
\newcommand{\earthbox}[1]{\raisebox{-.25mm}{\makebox[0.095\linewidth][c]{#1}}}
\newcommand{\espa}{\hspace*{.25em}}
\newcommand{\espb}{\hspace*{1em}}
\newcommand{\espc}{\hspace*{2em}}
\newcommand{\espd}{\hspace*{1.5em}}
\newcommand{\figearth}{
\begin{figure*}
\footnotesize
\begin{tabular}{@{}ll@{}}
\earthbox{}\espa%
\earthbox{}\espc%
\noraise{\earthhbo{}}\espa%
\earthbox{1k iterations}\espa%
\earthbox{5k iterations}\espa%
\earthbox{20k iterations}\espa%
\earthbox{Ground truth}\\
\earthimg{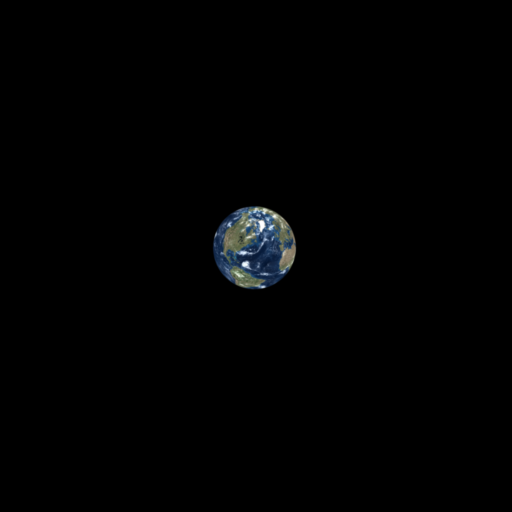}\espa%
\earthimg{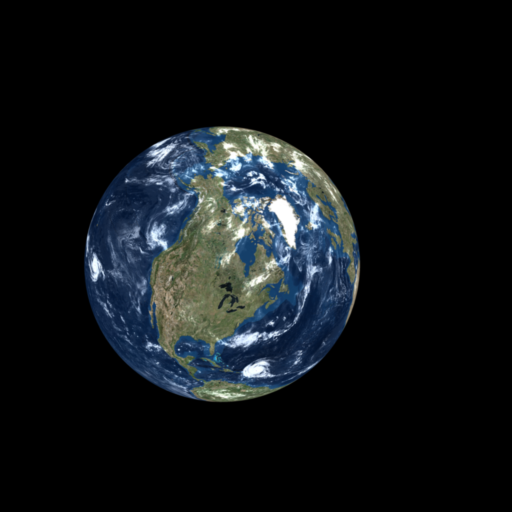}\espc%
\earthhbo{With mipmaps}\espa%
\earthimg{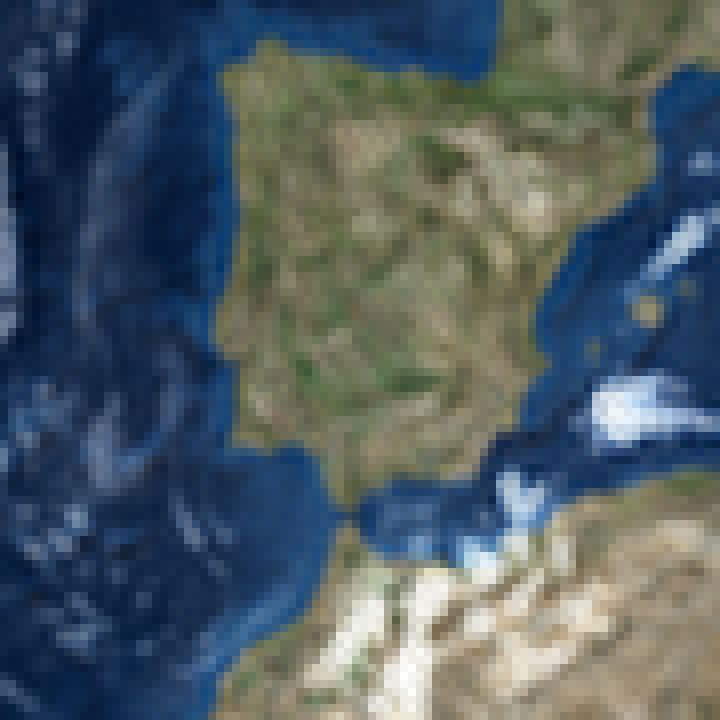}\espa%
\earthimg{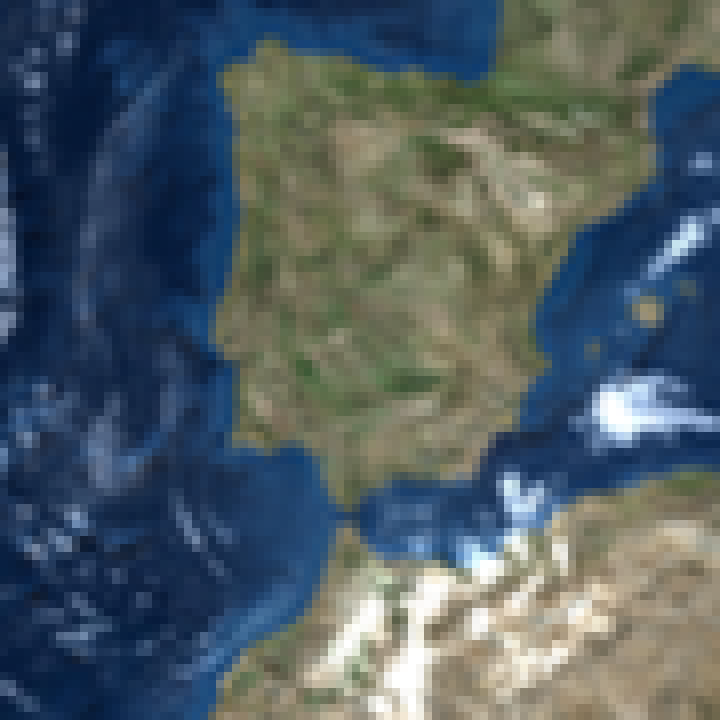}\espa%
\earthimg{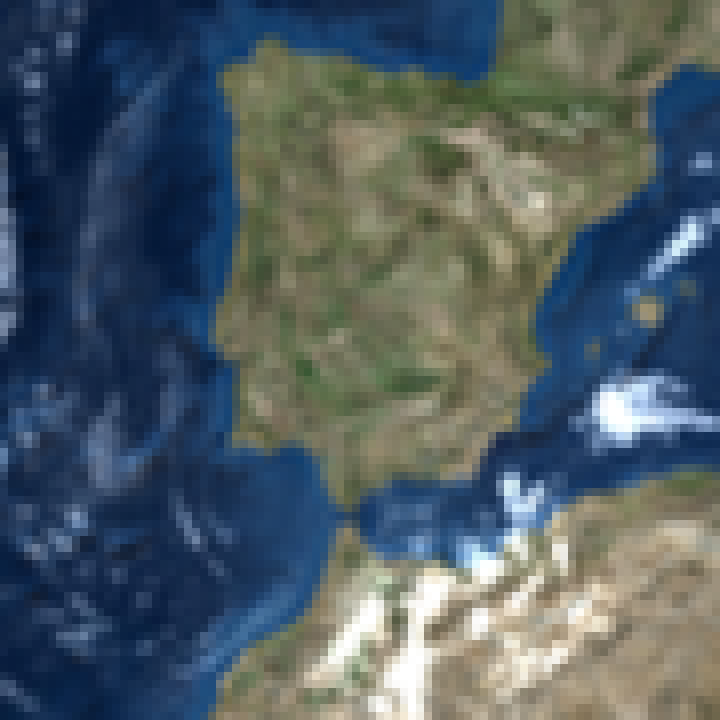}\espa%
\earthimg{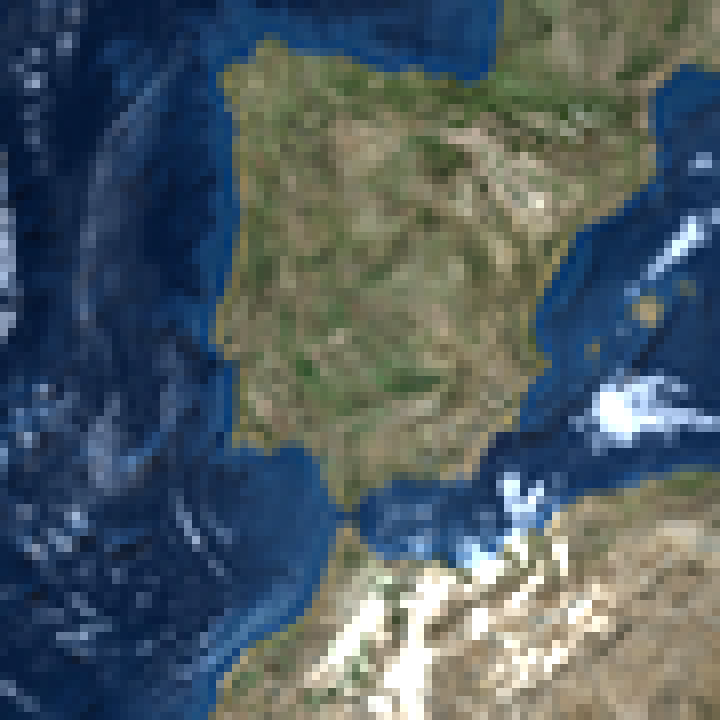}\espd%
\multirow{2}{*}{\raisebox{-16mm}[0mm][0mm]{\includegraphics[width=0.365\linewidth]{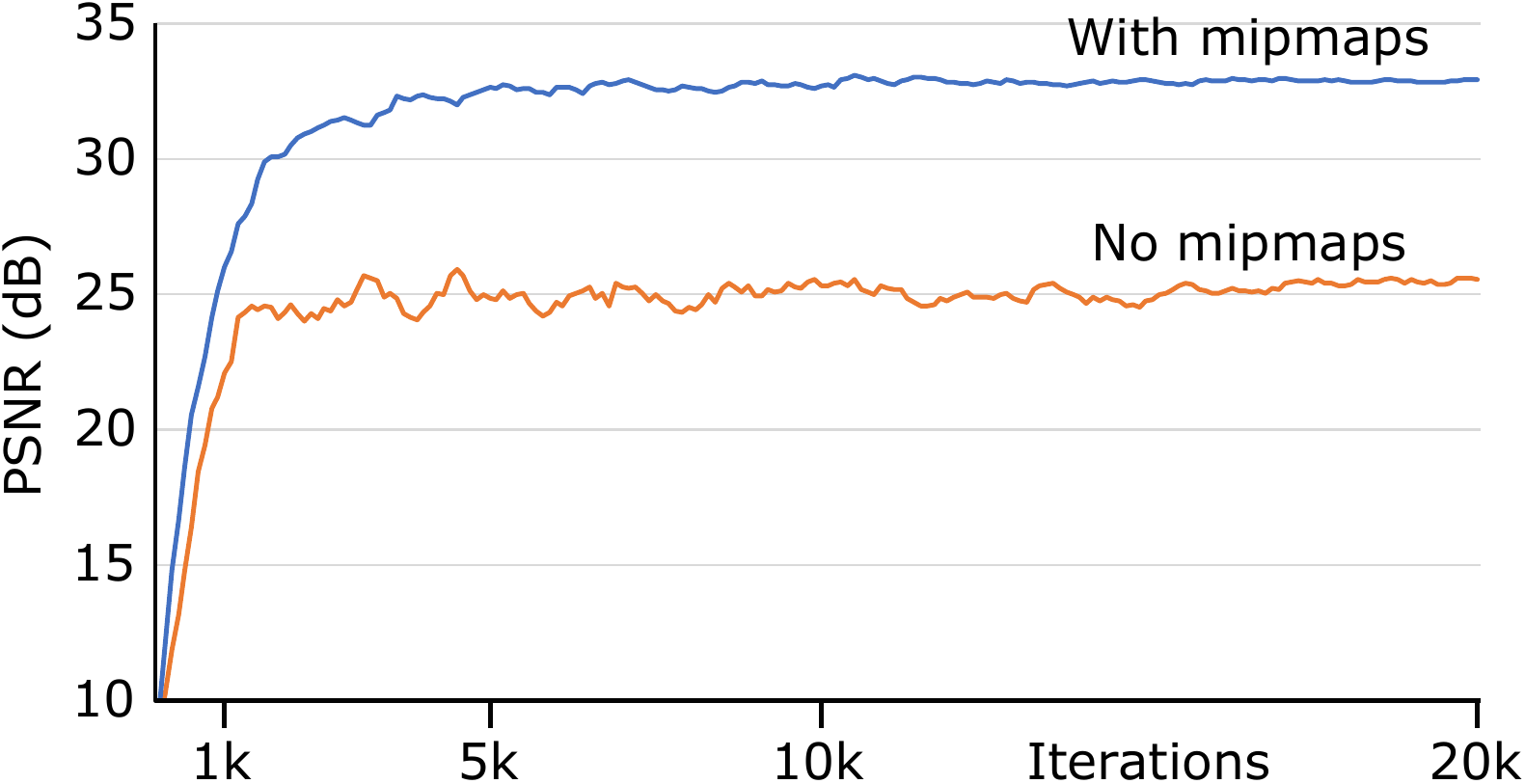}}}\vspace*{-0.25mm}\\
\earthimg{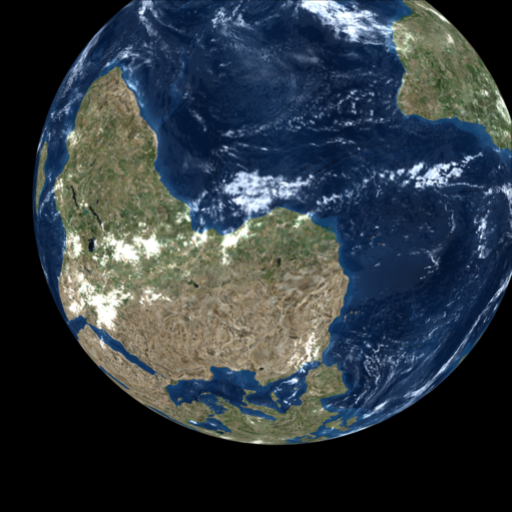}\espa%
\earthimg{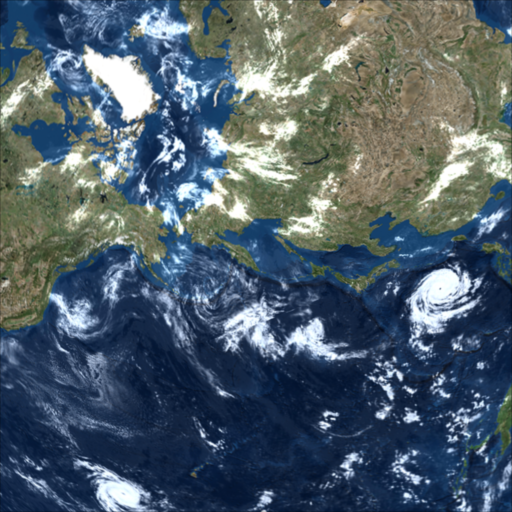}\espc%
\earthhbo{No mipmaps}\espa%
\earthimg{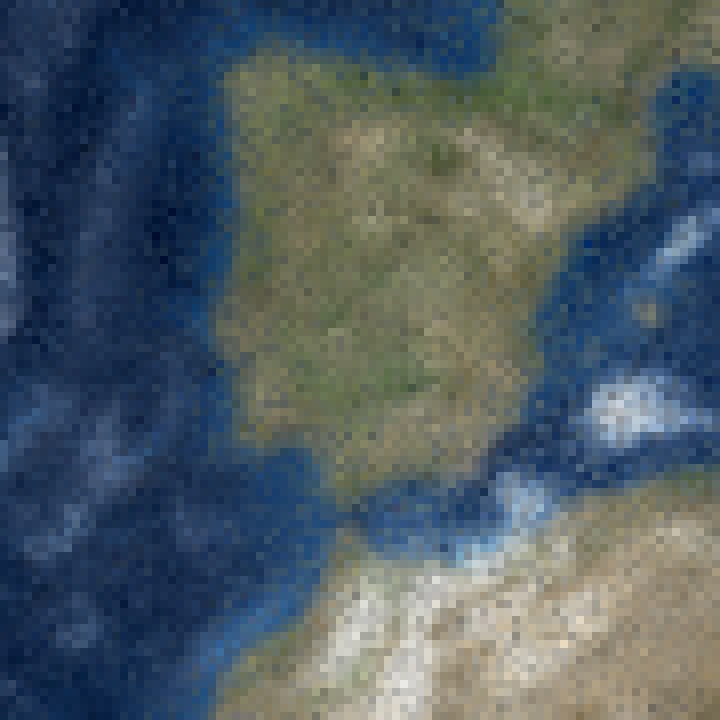}\espa%
\earthimg{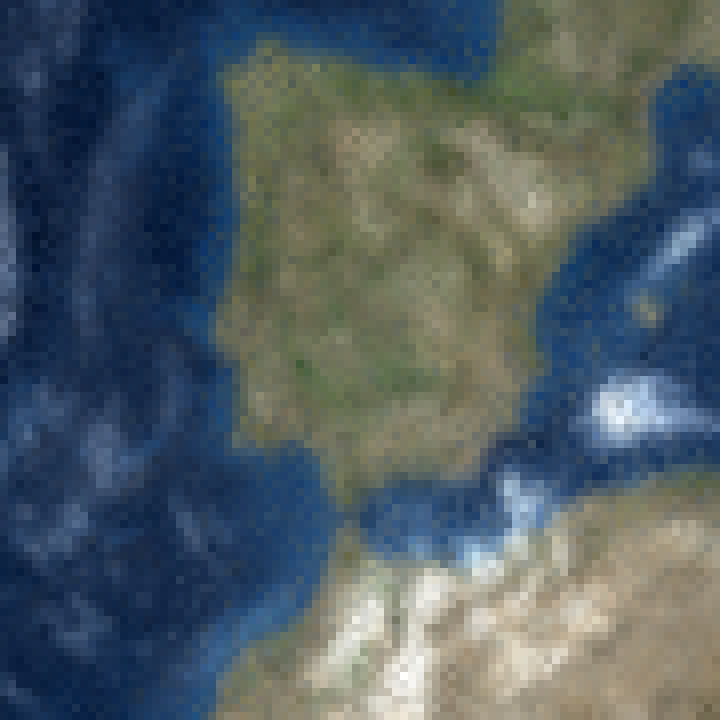}\espa%
\earthimg{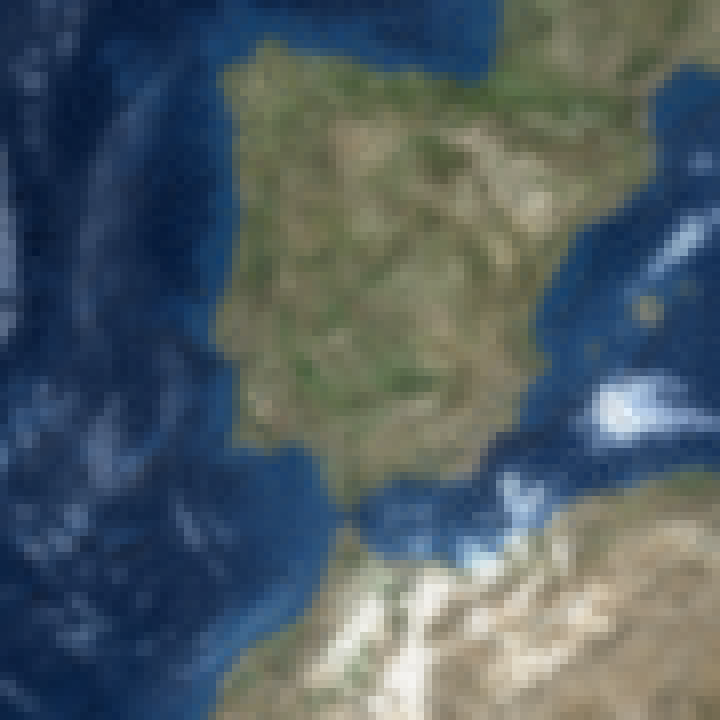}\espa%
\earthimg{ref.png}\\
\end{tabular}\\
\small%
\makebox[0.195\linewidth][c]{(a) Example reference images}\espc\noraise{\earthhbo{}}%
\makebox[0.39\linewidth][c]{(b) Closeups of learned texture}\espd%
\makebox[0.365\linewidth][c]{(c) Texture convergence vs.~reference}\\
\caption{\label{fig:earth}%
Filtered texture sampling via mipmaps helps considerably when learning a texture in a difficult geometric setup.
See text for full description.
(a)~Example synthetic reference images in \res{512} resolution.
(b)~With mipmapping enabled, sampling is prefiltered and gradients are routed to the correct detail levels. 
Without mipmaps, faraway views yield badly filtered samples and spurious, noisy texture updates that do not converge to the correct solution.
(c)~Convergence of the learned texture compared to the reference texture.
Learning rate schedule was optimized for the ``no mipmaps'' case, and the same schedule was used with mipmaps.
}
\end{figure*}
}

\newcommand{\psp}{\hspace*{.2em}}
\newcommand{\psq}{\hspace*{.7em}}
\newcommand{\perfimg}[1]{\includegraphics[width=0.060\linewidth]{figures/perf/small#1.png}}
\newcommand{\perfhlabel}[2]{\multirow{#1}{*}{#2}}
\newcommand{\perfhlabell}[3]{\perfhlabel{#1}{\begin{tabular}{@{}c@{}}#2\\#3\end{tabular}}}
\newcommand{\perfhlabelll}[3]{\perfhlabel{#1}{\begin{tabular}{@{}c@{}}Speedup\\#2\\#3\end{tabular}}}
\newcommand{\hulk}[1]{\makebox[0mm][l]{#1}}
\newcommand{\lhulk}[1]{\makebox[0mm][r]{#1}}
\newcommand{\tblperf}{
\begin{table*}
\caption{\label{tbl:perf}%
Performance comparison between our method, the official implementation of Soft Rasterizer (SoftRas) \protect\cite{softras}, and PyTorch3D \protect\cite{pytorch3d}.%
}%
\vspace*{-2mm}%
\small\centering%
\begin{tabu}{@{\psp}c@{\psq}c@{}c@{}c@{}c@{}c@{}c@{}c@{}c@{}c@{}c@{}c@{}c@{}c@{}c@{}c@{}}%
&&%
\perfimg{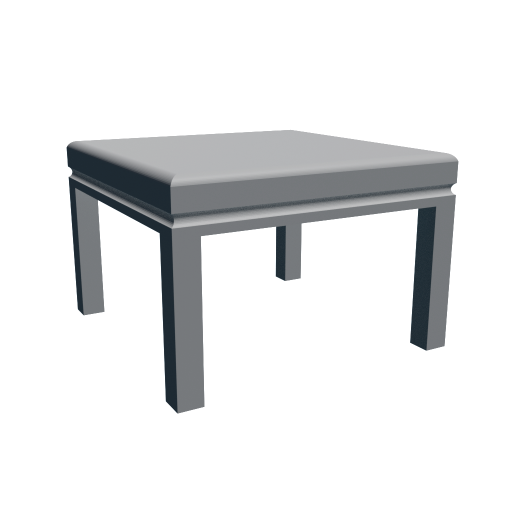}&%
\perfimg{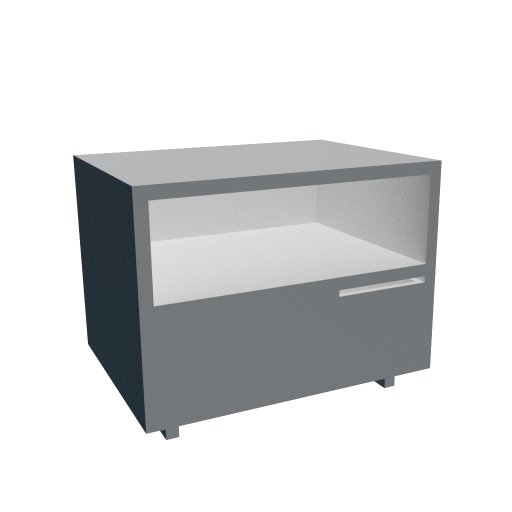}&%
\perfimg{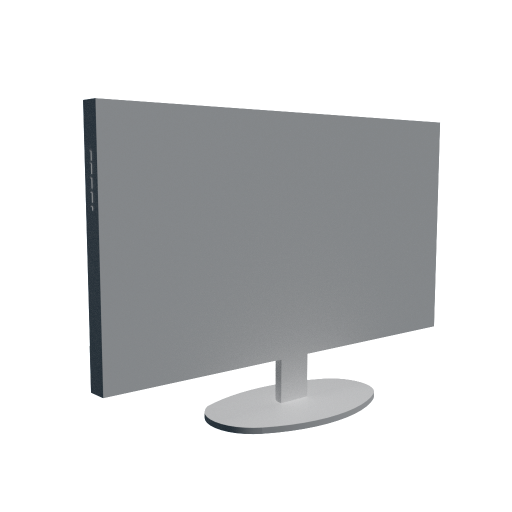}&%
\perfimg{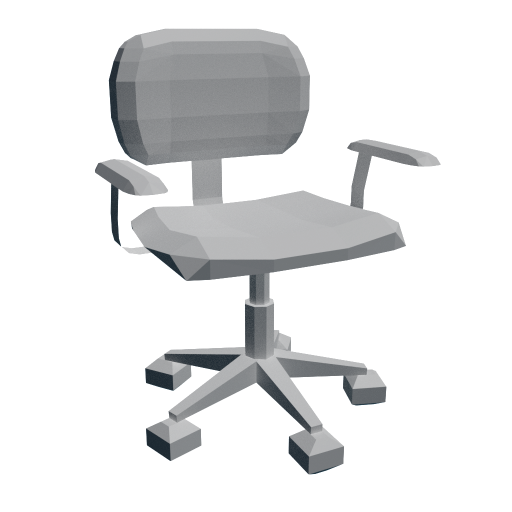}&%
\perfimg{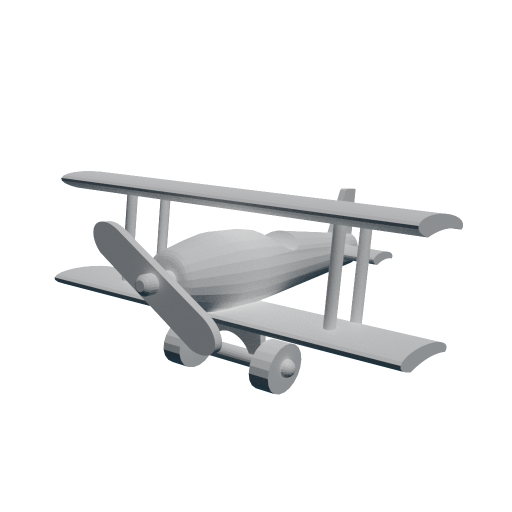}&%
\perfimg{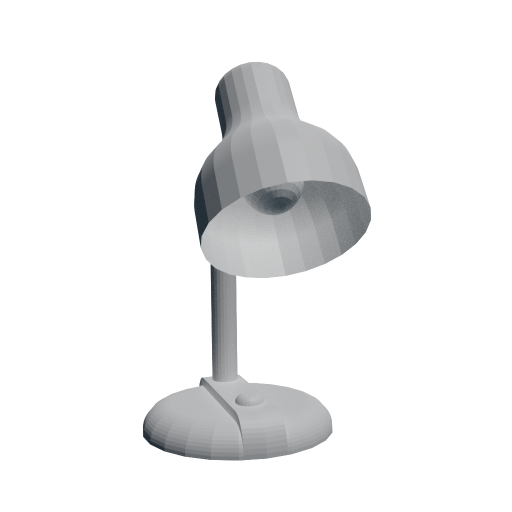}&%
\perfimg{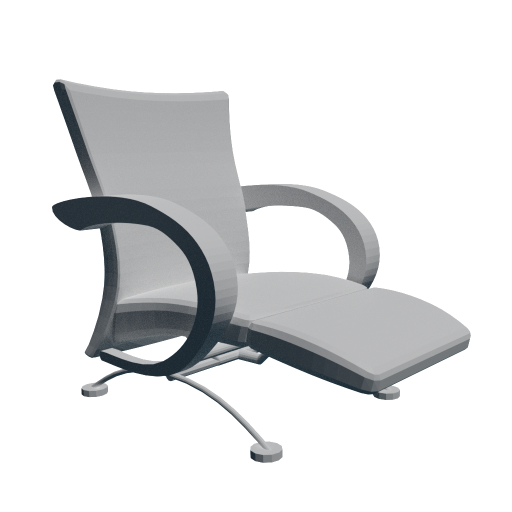}&%
\perfimg{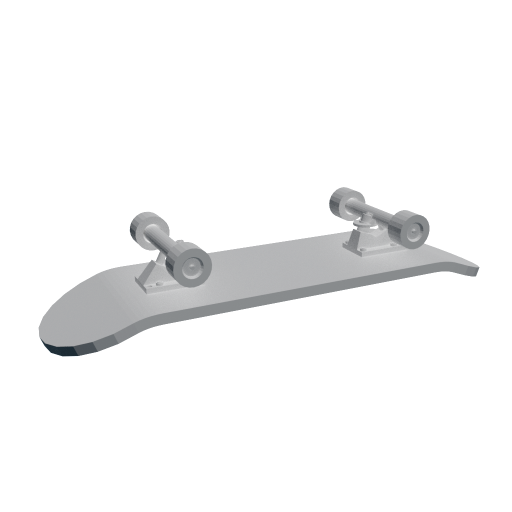}&%
\perfimg{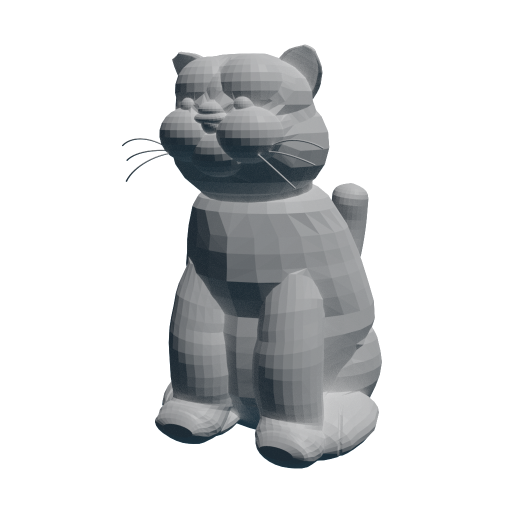}&%
\perfimg{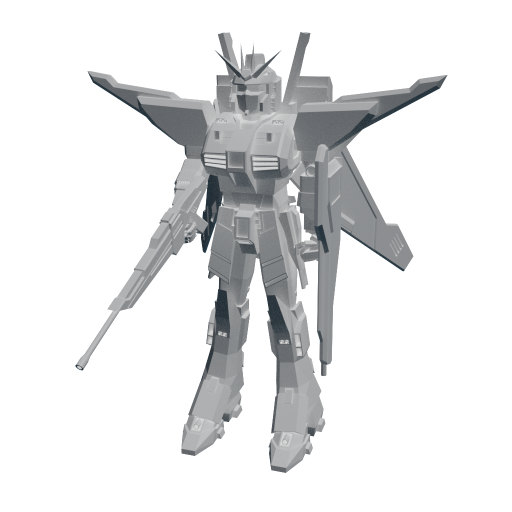}&%
\perfimg{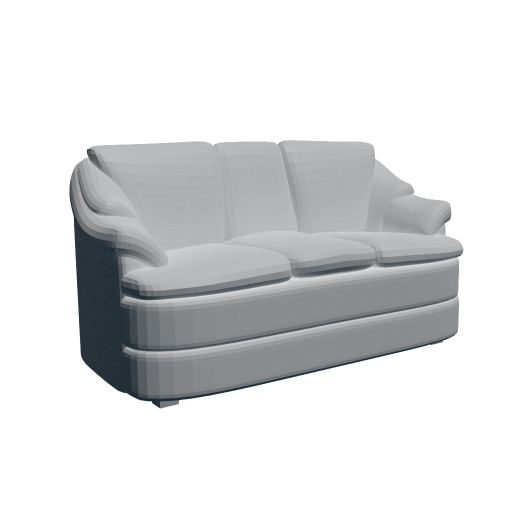}&%
\perfimg{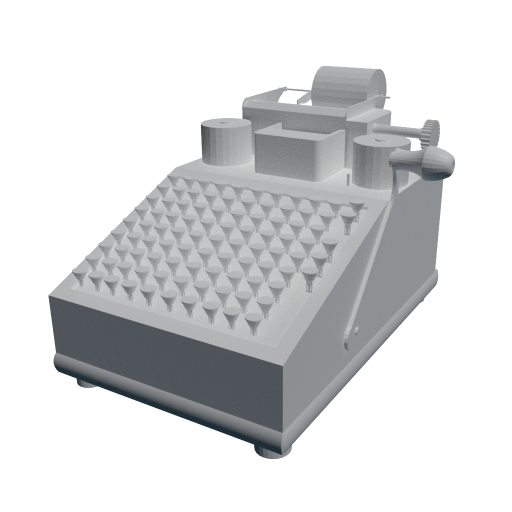}&%
\perfimg{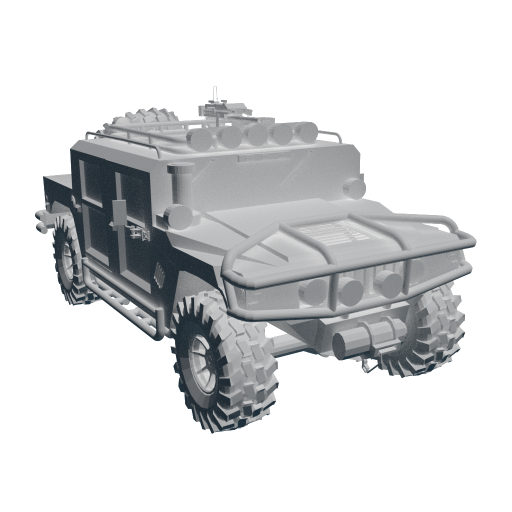}&%
\perfimg{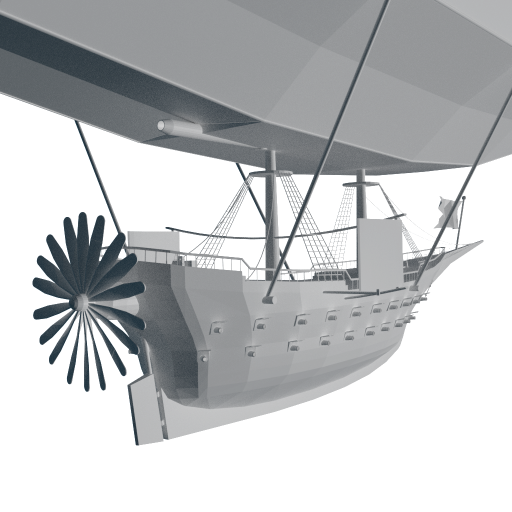}\vspace*{-1mm}\\
&Triangles \hulk{$\rightarrow$}& 284 & 352 & 514 & 1236 & 4474 & 5216 & 5344 & 10908 & 21695 & 25643 & 43448 & 91145 & 196179 & 308170\\
\arrayrulecolor{black}\hline
&Resolution&\multicolumn{11}{c}{{\bf Rendering + gradients time$^*$ (ms)}}\\
\arrayrulecolor{black}\hline
\perfhlabell{5}{\bf Our}{\bf method}%
&\res{256}  & 2.13 & 2.03 & 1.95 & 1.89 & 1.95 & 2.02 & 2.02 & 2.08 & 2.02 & 2.05 & 2.00 & 2.18 & 2.97 & 3.12 \\
&\res{512}  & 2.26 & 2.29 & 2.07 & 2.08 & 2.23 & 2.12 & 2.08 & 2.11 & 2.14 & 2.20 & 2.27 & 2.44 & 2.66 & 3.33 \\
&\res{1024} & 2.70 & 2.94 & 2.60 & 2.56 & 2.52 & 2.56 & 2.61 & 2.59 & 2.61 & 2.64 & 2.66 & 3.03 & 3.36 & 4.13 \\
&\res{2048} & 6.21 & 6.72 & 4.31 & 4.95 & 4.53 & 4.58 & 5.31 & 4.35 & 4.95 & 4.46 & 4.82 & 5.64 & 6.46 & 6.21 \\
&\res{4096} & 17.73 & 20.11 & 12.12 & 13.18 & 12.40 & 12.49 & 13.29 & 12.05 & 12.92 & 12.39 & 12.73 & 14.70 & 17.73 & 15.64 \\
\arrayrulecolor{black}\hline
\perfhlabell{5}{SoftRas}{\protect\shortcite{softras}}%
&\res{256}  & 7.48 & 6.76 & 7.84 & 7.73 & 18.11 & 14.29 & 10.91 & 29.21 & 30.93 & 49.06 & 65.21 & 144.57 & 331.30 & 788.92 \\
&\res{512}  & 10.01 & 10.42 & 8.78 & 10.68 & 24.05 & 24.33 & 24.40 & 50.09 & 82.48 & 99.76 & 163.51 & 375.39 & 865.63 & 1430.17 \\
&\res{1024} & 20.73 & 24.55 & 15.74 & 22.00 & 69.43 & 73.56 & 74.07 & 153.14 & 277.36 & 336.54 & 556.92 & 1250.74 & 3012.63 & 4856.97 \\
&\res{2048} & 66.25 & 86.49 & 46.66 & 68.43 & 250.65 & 276.06 & 280.30 & 557.96 & 1039.49 & 1234.61 & 2044.79 & 4602.11 & 11487.70 & 18402.19 \\
&\res{4096} & 223.88 & 332.33 & 163.87 & 240.71 & 946.76 & 1055.14 & 1082.35 & 2104.77 & 4036.47 & 4768.91 & 7958.07 & 17992.60 & 45499.74 & 72277.20 \\
\arrayrulecolor{black}\hline
\perfhlabell{5}{PyTorch3D}{\protect\shortcite{pytorch3d}}%
&\res{256}  & 27.12 & 27.19 & 26.81 & 27.95 & 27.67 & 27.14 & 26.94 & 28.62 & 27.93 & 30.08 & 32.03 & 37.52 & 54.82 & 115.87 \\
&\res{512}  & 31.83 & 30.93 & 31.08 & 31.19 & 31.70 & 31.82 & 30.83 & 32.06 & 34.84 & 38.41 & 41.50 & 55.53 & 95.21 & 158.71 \\
&\res{1024} & 53.70 & 53.24 & 52.03 & 51.38 & 52.44 & 52.17 & 53.02 & 53.99 & 58.98 & 64.87 & 83.04 & 140.34 & 267.45 & 438.82 \\
&\res{2048} & 156.31 & 153.17 & 145.34 & 141.66 & 144.91 & 145.48 & 141.49 & 148.75 & 165.77 & 182.21 & 259.04 & 456.07 & 930.77 & 1435.20 \\
&\res{4096} & 571.53 & 553.69 & 525.49 & 513.69 & 524.58 & 521.84 & 508.87 & 528.99 & 604.20 & 677.21 & 966.76 & 1754.35 & 3527.62 & 5567.26 \\
\arrayrulecolor{black}\hline
&&\multicolumn{11}{c}{\bf Speedup factor}\\
\arrayrulecolor{black}\hline
\perfhlabell{5}{Our vs}{SoftRas}%
&\res{256}  & 3.51 & 3.33 & 4.02 & 4.09 & 9.29 & 7.07 & 5.40 & 14.04 & 15.31 & 23.93 & 32.60 & 66.32 & 111.55 & 252.86 \\
&\res{512}  & 4.43 & 4.55 & 4.24 & 5.13 & 10.78 & 11.48 & 11.73 & 23.74 & 38.54 & 45.35 & 72.03 & 153.85 & 325.42 & 429.48 \\
&\res{1024} & 7.68 & 8.35 & 6.05 & 8.59 & 27.55 & 28.73 & 28.38 & 59.13 & 106.27 & 127.48 & 209.37 & 412.79 & 896.62 & 1176.02 \\
&\res{2048} & 10.67 & 12.87 & 10.83 & 13.82 & 55.33 & 60.28 & 52.79 & 128.27 & 210.00 & 276.82 & 424.23 & 815.98 & 1778.28 & 2963.32 \\
&\res{4096} & 12.63 & 16.53 & 13.52 & 18.26 & 76.35 & 84.48 & 81.44 & 174.67 & 312.42 & 384.90 & 625.14 & 1223.99 & 2566.26 & 4621.30 \\
\arrayrulecolor{black}\hline
\perfhlabell{5}{Our vs}{PyTorch3D}%
&\res{256}  & 12.73 & 13.39 & 13.75 & 14.79 & 14.19 & 13.44 & 13.34 & 13.76 & 13.83 & 14.67 & 16.02 & 17.21 & 18.46 & 37.14 \\
&\res{512}  & 14.08 & 13.51 & 15.01 & 15.00 & 14.22 & 15.01 & 14.82 & 15.19 & 16.28 & 17.46 & 18.28 & 22.76 & 35.79 & 47.66 \\
&\res{1024} & 19.89 & 18.11 & 20.01 & 20.07 & 20.81 & 20.38 & 20.31 & 20.85 & 22.60 & 24.57 & 31.22 & 46.32 & 79.60 & 106.25 \\
&\res{2048} & 25.17 & 22.79 & 33.72 & 28.62 & 31.99 & 31.76 & 26.65 & 34.20 & 33.49 & 40.85 & 53.74 & 80.86 & 144.08 & 231.11 \\
&\res{4096} & 32.24 & 27.53 & 43.36 & 38.97 & 42.30 & 41.78 & 38.29 & 43.90 & 46.76 & 54.66 & 75.94 & 119.34 & 198.96 & 355.96 \\
\arrayrulecolor{black}\hline
\end{tabu}
\parbox{\linewidth}{\small%
$^*$%
Execution times include both forward and gradient evaluations for rendering one frame.
Each mesh was rendered several times from multiple angles and the results were averaged to reduce random variation.
The exact same meshes, viewpoints, and camera parameters were used for all methods.
For our method, we perform rasterization, attribute interpolation, and antialiasing, but no texturing.
For Soft Rasterizer, rasterization with default lighting is computed.
PyTorch3D was set up to perform Gouraud shading, i.e., attribute interpolation.
}
\end{table*}
}

\newcommand{\figoccluplot}{
\begin{figure}
\centering\footnotesize
\includegraphics[width=0.9\linewidth]{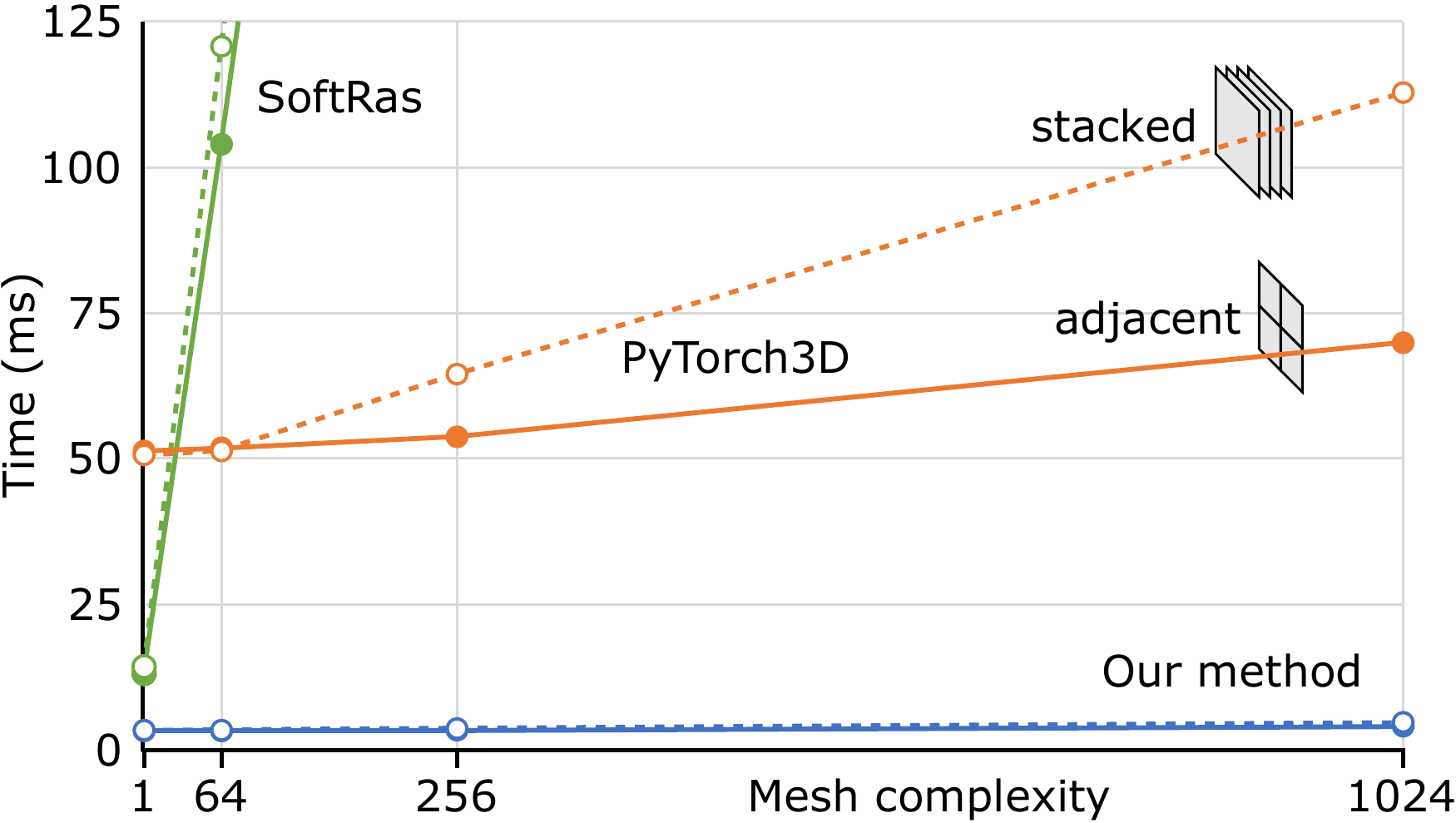}\hphantom{xx}\vspace*{-.5em}\\
\caption{\label{fig:occluplot}%
The effect of geometric configuration on forward~+~gradient pass execution times in \res{1024} resolution.
A \res{8} grid base mesh (128~triangles) is repeated 1, 64, 256, or 1024 times (horizontal axis).
The repeated meshes are either stacked in depth direction (dashed lines) or placed adjacent to each other on the same plane (solid lines). 
The meshes are scaled so that the output image always has the same number of pixels covered.
PyTorch3D~\protect\cite{pytorch3d} scales mostly with the total area of geometry, occluded or not, whereas our method is not slowed down by hidden geometry.
SoftRas~\protect\cite{softras} has approximately constant cost per triangle, and cannot keep up with the other two methods.
}
\end{figure}
}